\pdfoutput=1
\documentclass[a4paper,11pt]{article}

\newcommand\papertitle{Young Researchers School 2024 Maynooth:\\ Lectures on CFT, BCFT and DCFT}

\usepackage{jheppub}

\usepackage[dvipsnames]{xcolor}

\usepackage{tikz}
\usetikzlibrary{arrows, backgrounds, calc, decorations.pathreplacing, decorations.markings, fadings, quotes, positioning, shadings, intersections, angles}

\usepackage{amsmath,amsfonts,amssymb,mathrsfs, dsfont}
\usepackage{mathtools}
\usepackage{bbold}
\usepackage{bbm}
\usepackage[inline]{enumitem}
\usepackage{multirow}
\usepackage{subcaption}
\usepackage{nicefrac}
\usepackage{xspace}
\usepackage{MnSymbol}
\usepackage{simpler-wick}

\usepackage[colorlinks=true]{hyperref}
\hypersetup{
    bookmarks=true,         
    unicode=false,          
    pdftoolbar=true,        
    pdfmenubar=true,        
    pdffitwindow=false,     
    pdfstartview={FitH},    
    pdftitle={\papertitle}, 
    pdfauthor={Christian Northe},
    pdfnewwindow=true,      
    colorlinks=true,        
    linkcolor=teal,         
    citecolor=Maroon,          
    filecolor=magenta,      
    urlcolor=teal          
}

\makeatletter

\@addtoreset{equation}{section}
\makeatother

\newcommand{\p}	{\partial}
\newcommand{\bp}{\bar{\partial}}

\newcommand{\R}	{\mathds{R}}
\newcommand{\C}	{\mathbb{C}}
\newcommand{\Z}	{\mathbb{Z}}
\newcommand{\N}	{\mathbb{N}}

\newcommand{\cB}{\mathcal{B}}
\newcommand{\cC}{\mathcal{C}}

\newcommand{\cE}{\mathcal{E}}
\newcommand{\cF}{\mathcal{F}}

\newcommand{\cH}{\mathcal{H}}
\newcommand{\cI}{\mathcal{I}}
\newcommand{\cJ}{\mathcal{J}}

\newcommand{\cM}{\mathcal{M}}

\newcommand{\cS}{\mathcal{S}}
\newcommand{\cT}{\mathcal{T}}

\newcommand{\cZ}{\mathcal{Z}}

\DeclareMathOperator{\Tr}{Tr}
\DeclareMathOperator{\tr}{\tr}

\newcommand{\bra}[1]	{\langle{#1}\vert}
\newcommand{\ket}[1]	{\vert{#1}\rangle}
\newcommand{\braket}[2]	{\langle{#1}\vert{#2}\rangle}
\newcommand{\ketbra}[2]	{\ket{#1}\bra{#2}}
\newcommand{\corr}[1]   {\left\langle{#1}\right\rangle}
\newcommand{\CPT}       {\Theta}

\newcommand{\bT}    {\bar{T}}
\newcommand{\bz}    {\bar{z}}
\newcommand{\bw}    {\bar{w}}
\newcommand{\bphi}  {\bar{\phi}}

\newcommand{\bL}    {\bar{L}}
\newcommand{\bh}    {\bar{h}}

\newcommand{\bq}    {\bar{q}}
\newcommand{\btau}  {\bar{\tau}}
\newcommand{\bb}    {\bar{b}}
\newcommand{\bi}    {\bar{\imath}}
\newcommand{\bj}    {\bar{\jmath}}

\newcommand{\tq}    {\tilde{q}}

\renewcommand{\Tr}	{\mathrm{Tr}}
\renewcommand{\tr}	{\mathrm{tr}}

\newcommand{\diag}  {\mathrm{diag}}

\newcommand{\End}	{\mathrm{End}}
\newcommand{\Hom}   {\mathrm{Hom}}

\newcommand\Algebra[1]	{\mathfrak{#1}}
\newcommand\Group[1]	{\mathrm{#1}}
\newcommand{\SO}	{\Group{SO}}

\newcommand{\SL}	{\Group{SL}}

\newcommand{\SU}	{\Group{SU}}

\newcommand{\PSL}	{\Group{PSL}}

\newcommand{\so}	{\Algebra{so}}
\newcommand{\su}	{\Algebra{su}}

\newcommand{\vir}   {\mathsf{Vir}}

\newcommand{\bbra}[1]	{\langle\!\langle{#1}\lVert}
\newcommand{\bket}[1]	{\lVert{#1}\rangle\!\rangle}
\newcommand{\ibra}[1]	{\langle\!\langle{#1}\lvert}
\newcommand{\iket}[1]	{\lvert{#1}\rangle\!\rangle}

\newcommand{\proj}[1]   {|\hspace{-0.1em}|\,{#1}\,|\hspace{-0.1em}|}

\newcommand{\fus}       {\mathsf{N}}
\newcommand{\fuse}      {\star}
\newcommand{\qdim}      {\mathsf{d}}
\newcommand{\multZ}     {\cZ}
\newcommand{\multZtwist}[3]{\mathsf{n}^{#1}{}_{#2#3}}
\newcommand{\modS}      {\cS}
\newcommand{\modT}      {\cT}

\newcommand{\cc}        {\mathsf{c}}
\newcommand{\spa}       {\sigma}
\newcommand{\transpose}{\mathsf{T}}
\newcommand{\iu}{\mathsf{i}}
\newcommand{\pl}        {(P)}
\newcommand{\hp}        {(H)}
\newcommand{\confFam}   {\cI}
\newcommand{\nia}       {\mathsf{n}^i_\alpha}
\newcommand{\niaa}      {\mathsf{n}^i_{\alpha\alpha}}
\newcommand{\niab}      {\mathsf{n}^i_{\alpha\beta}}
\newcommand{\ihp}       {\confFam^{\hp}}
\newcommand{\ipl}       {\confFam^{\pl}}
\newcommand{\defect}    {D}

\newcommand\id		{\mathbf{1}}

\newcommand\quotes[1]  {``{#1}"}
\newcommand\emphasize[1] {\textbf{\textit{{#1}}}}

\newcommand\Secref[1]	{Section~\ref{#1}\xspace}

\newcommand\secref[1]	{section~\ref{#1}\xspace}

\newcommand\figref[1]	{figure~\ref{#1}\xspace}

\DeclareMathAlphabet{\mathdutchcal}{U}{dutchcal}{m}{n}
\SetMathAlphabet{\mathdutchcal}{bold}{U}{dutchcal}{b}{n}
\DeclareMathAlphabet{\mathdutchbcal}{U}{dutchcal}{b}{n}

\newtheorem{theorem}{Theorem}[section]
\newtheorem{exercise}[theorem]{Exercise}
\newcommand{\solution}[2]{\textbf{Solution \ref{#1}. }\textit{#2} }

\usepackage[most]{tcolorbox}
\tcbset{colback=SkyBlue!35!white,
        colframe=black!50!black, 
        highlight math style= {enhanced, 
            colframe=black,colback=black!20!white,boxsep=0pt}
        }

\tikzset{
    partial ellipse/.style args={#1:#2:#3}{
        insert path={+ (#1:#3) arc (#1:#2:#3)}
    }
}

\definecolor{boxlinkcolor}{rgb}{0.92, 0.51, 0.45} 
\newcommand{\boxlinkcolor}{Maroon}

\newif\ifsol
\soltrue 

\newcommand\new[1]   {\textcolor{black}{ {#1}}}

\begin{document}
\title{\papertitle}

\author[a,b]{Christian Northe\footnote{\new{This work was performed while being at Ben-Gurion University, whereas the submission and revision process was conducted at the Czech Academy of Sciences.}}}


\affiliation[a]{Department of Theoretical Physics,\\Ben-Gurion University of the Negev,Be'er Sheva, Israel} 

\affiliation[b]{\new{CEICO, Institute of Physics of the Czech Academy of Sciences,\\
Na Slovance 2, 182 00 Prague 8, Czech Republic}}

\abstract{%
\new{These notes were} presented at the Young Researchers School (YRS) in Maynooth in April 2024 and \new{provide} an introduction to Conformal Field Theory CFT, Boundary Conformal Field Theory (BCFT) and Defect Conformal Field Theory (DCFT). This class is mostly self-contained and includes exercises with solutions. The first part of these notes is concerned with the basics of CFT, and was taught by the author during the pre-school for the YRS 2024. Here the aim is to convey the notion of conformal families, their fusion and the construction of partition functions. The second part of these notes is dedicated to boundaries and defects in CFT and was presented by the author at the main school. As far as boundaries are concerned, emphasis is placed on boundary operators and their state spaces, as well as the boundary state formalism with the Cardy constraint. Topological defects are discussed in analogy, i.e. defect state spaces and the relevant consistency constraint are derived. Verlinde lines are constructed as their simplest solution and their properties are inspected.
}

\maketitle
\flushbottom
\section{Before the Movie}\label{secIntroCFT}
In the realm of theoretical physics, few concepts rival the elegance and ubiquity of conformal symmetry. Conformal Field Theory (CFT) stands as a testament to the profound insights gained through the study of symmetry. Indeed, conformal symmetry enables us to solve examples of interacting quantum field theories -- a feat achieved only with few tools in theoretical physics.

Many areas of research have profited from CFT, including the study of (quantum) critical phenomena, deep in-elastic scattering and also string theory. More importantly for this school, conformal invariance plays a major role in topological phases of matter. Quantum Hall devices for instance carry anyonic edge modes, which are described by CFT \cite{tong2016lectures}. In this year's edition of the Young Researchers School (YRS), we learn more about topological insulators in Flore Kunst's class and more about anyonic excitations in Pieralberto Marchetti's class. Concepts which are well known in quantum computation such as braiding and fusion, as we hear about in Joost Slingerland's class, were discussed early on within the context of CFT \cite{Moore:1988uz, moore1989classical}. A modern approach to symmetry is presently developed and goes by the name of non-invertible symmetry, as is taught by Ho Tat Lam at this school. Many of the ideas derived there are very naturally implemented in CFT, in particular those dealing with boundaries and defects.

It is the purpose of this note to provide a \quotes{soft} introduction to the study of CFT leading to the study of boundaries in CFT,  see \secref{secBCFT}, and topological defects, see \secref{secDCFT}. \new{Along the way}, emphasis is also placed on concepts appearing in the other lectures of this school. The author had the pleasure to teach an introduction to CFT at the pre-school of the 2024 YRS and has taken the liberty to include this material in this manuscript spanning sections \ref{secConformalGroup}-\ref{secHilbertSpace}. Hence, for novices, this set of notes provides a self-contained text leading them from the very start to an understanding of boundaries and defects in CFT. Of course, this text is not complete here and there so that pointers to the literature are included. Veterans of CFT, who are interested only in the advanced material on boundaries and defects, are invited to skim the introductory sections to familiarize themselves with the notation.

This lecture is split into two parts. The first part, sections \ref{secConformalGroup}-\ref{secHilbertSpace}, provide standard CFT lore, and were presented at the pre-school. The \emphasize{aim} of this first part is to explain \emphasize{representations of the Virasoro algebra}, their \emphasize{fusion} and how \emphasize{partition functions} are constructed in CFT.
Excellent introductory resources on CFT are already available at several levels. A small selection thereof is loosely followed here. It consists of 
\begin{itemize}
 \item Ginsparg presented a set lectures titled \quotes{Applied Conformal Field Theory} at the Les Houches in summer of 1988 \cite{Ginsparg:1988ui}. By now it has become a classic introductory text to CFT. 
 \item The book \quotes{Conformal Field Theory} by Di Francesco, Mathieu, Senechal \cite{DiFrancesco} is the \quotes{CFT bible} of its day and the holy grail of any aspiring practitioner of the conformal arts. Its size is as impressive as its depth. Even though long, the clear and extensive explanations often save time rather than consuming it.
 \item \new{Vintage fans and/or purists may actually cut directly to the source: The groundbreaking paper \quotes{Infinite Conformal Symmetry In Two-Dimensional Quantufm Field Theory} by Belavin, Polyakov and Zamolodchikov \cite{belavin1984infinite} not only initiated the study two-dimensional CFT but also developed many core elements of CFT on the plane, i.e. without boundaries or defects. Many introductory texts in fact follow this paper or are inspired by it -- the first part in the present set of notes is no exception.}
 \item The book \quotes{Introduction to Conformal Field Theory} by Blumenhagen and Plauschinn \cite{Blumenhagen} covers the vast field of CFT mostly in the operator language using commutators  instead of OPEs as in the previous book. 
 \item Gaberdiel is a leading expert in CFT and has contributed many reviews on CFT, out of which this one \cite{Gaberdiel:1999mc} is reflected in the present text. 
 \item \new{Aficionados of statistical mechanics and the renormalization group are invited to explore Cardy's lecture notes \cite{cardy1988conformal} and his book \cite{cardy1996scaling}.}
\end{itemize}

The second part, which was presented in the main school, deals with boundaries and defects, respectively \secref{secBCFT} and \secref{secDCFT}. The former has already found its way into textbooks and lecture notes, while, to the authors knowledge, the material presented in the topological defect section has not yet been presented in uniform pedagogical manner. \Secref{secBCFT} and \secref{secDCFT} contain their own introduction and pointers to useful literature.

We set out in \secref{secConformalGroup} with a brief introduction to the conformal group in spacetime dimensions larger than two. After that we restrict to two dimensions of spacetime for the remainder of this introduction to CFT. In \secref{secConfSym2d} we discuss the conformal algebra in two spacetime dimensions. In particular, we become acquainted with the Witt algebra and encounter the Virasoro algebra. Afterward we turn to the fields and states in a CFT in \secref{secFieldsStates}. This presents the jumping board to understand the Virasoro algebra in-depth in \secref{secVirasoro} meaning that we study its representations and also its fusion rules. Finally, in \secref{secHilbertSpace} we learn how representations are gathered in order to build a physical CFT state space. Finally, we arrive at the main event, namely the study of boundaries in CFT, \secref{secBCFT}, and topological defects, see \secref{secDCFT}.   

Familiarity with QFT, notions of symmetry and complex analysis are assumed. However, one need not be an absolute expert on QFT, as CFT oftentimes employs its own very sophisticated toolkit, which we develop here or point to useful resources when space and time are scarce. 

\subsection*{Acknowledgements}
I am very greatful to have had the great pleasure and opportunity to teach at the YRS 2024 in Maynooth. Hence I express my gratitude to its organizers: Marius de Leeuw, Saskia Demulder, Graham Kells, Alessandro Sfondrini, Joost Slingerland and Jiri Vala.

Having visited an earlier edition of this school as a student with great joy, I prepared this set of notes with enthusiam in the hopes of sharing my fascination for CFT with the students. Looking back, I am happy to have presented this material to a curious and enthusiastic crowd. 

I wish to thank Ho Tat Lam and Tobias Hössel for pleasant discussions, which lead to a sharpening of my presentation on defects and boundary states, respectively. Regarding the notes themselves, I thank Evgenii Zheltonozhskii for pointing out typos in the CFT pre-lecture manuscript, and I am particularly indebted to Saskia Demulder for a careful reading of the draft and for valuable comments on the structure of the presentation. 

My work is supported by the Israel Science Foundation (grant No.~1417/21), the German Research Foundation through a German-Israeli Project Cooperation (DIP) grant ``Holography and the Swampland'', by  Carole and Marcus Weinstein through the BGU Presidential Faculty Recruitment Fund and by the ISF Center of Excellence for theoretical high-energy physics. \new{As of November 2024, my work is funded by the European Union’s Horizon Europe Research and Innovation Program under the Marie Skłodowska-Curie Actions COFUND, Physics for Future, grant agreement No 101081515.}

The YRS 2024 in Maynooth was financially supported by the COST action \quotes{Fundamental challenges in theoretical physics - CA22113}, the University of Padova, the University of Maynooth, Trinity College, the Science Foundation Ireland, the European Union and the European Research Council, Horizon Quantum Computing, World Quantum Day, Theory Challenges, arQus European University Alliance, the Dublin Institute for Advanced Studies and the Hamilton Mathematics Institute.

\section{The Conformal Group and Algebra}\label{secConformalGroup}
Like with any other introductory text on CFT, our journey starts with the exposition of the conformal group in $d$ spacetime dimensions. We begin with a general exposition of the conformal transformations encountering the four types of \emphasize{global conformal transformations} innate to any CFT, namely translations, rotations, dilations and special conformal transformations and announce the \emphasize{conformal group} in $d\geq3$.

We will not delve into their representations at this point. Interested readers are referred to chapter 4 of \cite{DiFrancesco} or Rychkov's lectures \cite{Rychkov:2016iqz} dedicated entirely to the topic.

\subsection{Conformal Symmetry in $d\geq3$}
Conformal transformations are those which locally preserve angles between vectors, but not necessarily their lengths. Be reminded that the Poincar\'e group preserves only the latter and as such the conformal group is an extension of the Poincar\'e group.
Given two spacetime manifolds $\cM$ and $\tilde\cM$ equipped with metrics $g$ and $\tilde g$, a conformal transformation $f:\cM\mapsto\tilde\cM$ is such that 
\begin{equation}
 f_*\tilde g=\Lambda g
\end{equation}
for a positive smooth function $\Lambda$ on $\cM$. In other words, the metric tensor on $\tilde\cM$, pulled back to $\cM$ by virtue of $f$, coincides with the metric on $\cM$ up to a local rescaling $\Lambda$. 

Using coordinate patches $x^\mu$ on $\cM$ and $y^\alpha$ on $\tilde\cM$ this takes the more practical form
\begin{equation}\label{ConfMetric}
 \tilde g_{\alpha\beta}(y(x))\frac{\p y^\alpha}{\p x^\mu}\frac{\p y^\beta}{\p x^\nu}
 =
 \Lambda(x)g_{\mu\nu}(x)
\end{equation}
\begin{exercise}\label{exPreservedAngles}
  Given two vectors $X=X^\mu\frac{\p}{\p x^\mu},\,Y=Y^\mu\frac{\p}{\p x^\mu}\in T_p\cM$ at a point $p\in\cM$, show that conformal transformations preserve their enclosed angle
  \begin{equation}
   \cos(\theta(X,Y))
   =
   \frac{g_{\mu\nu}X^\mu Y^\nu}{\sqrt{g_{\zeta\eta}X^\zeta X^\eta\,g_{\kappa\lambda}Y^\kappa Y^\lambda}}
  \end{equation}
What happens to lengths $|X|=\sqrt{g_{\mu\nu}X^\mu X^\nu}$?
\end{exercise}
\ifsol
\solution{exPreservedAngles}{After a conformal transformation the angle is
\begin{align}
 \cos(\theta(X,Y)')
   &=
   \frac{\Lambda(x(p))g_p(X,Y)}{\sqrt{\Lambda(x(p))g_p(X,X)\,\Lambda(x(p))g_(Y,Y)}}\notag\\
   &=
   \frac{\Lambda(x)g_{\mu\nu}X^\mu Y^\nu}{\sqrt{\Lambda(x)g_{\zeta\eta}X^\zeta X^\eta\,\Lambda(x)g_{\kappa\lambda}Y^\kappa Y^\lambda}}
   \qquad=
   \cos(\theta(X,Y))
\end{align}
Lengths rescale $|X|\to\sqrt{\Lambda}|X|$.
}
\fi

Our interest lies with conformal maps of $\cM=\R^{p,q}$ -- actually extensions thereof called \textbf{\textit{conformal compactifications}} -- to itself, so that henceforth we use 
\begin{equation}
 g_{\mu\nu}=\eta_{\mu\nu}=\text{diag}(-1,\dots,+1,\dots)\,.
\end{equation}
Obviously, the Poincar\'e transformations are those with $\Lambda(x)=1$ in \eqref{ConfMetric}.

\begin{exercise}\label{exInversion}
 Check that the inversion $\cI: x^\mu\mapsto y^\mu=\frac{x^\mu}{|x|^2}$ is a conformal transformation. $\cI$ maps the exterior of the unit sphere to the interior of the unit sphere and vice versa. 
\end{exercise}
\ifsol
\solution{exInversion}{
\begin{equation}
 \frac{\p y^\alpha}{\p x^\mu}=\frac{\delta_\mu^\alpha}{|x|^2}-\frac{2x_\mu x^\alpha}{|x|^4}
 \qquad
 \Rightarrow
 \qquad
 \eta_{\alpha\beta}\frac{\p y^\alpha}{\p x^\mu}\frac{\p y^\beta}{\p x^\nu}
 =
 \frac{\eta_{\mu\nu}}{|x|^4}
\end{equation}
}
\fi

Note that the inversion $\cI$ is not well-defined at the origin. Hence, conformal compactifications of $\R^{p,q}$ need to be considered. In two-dimensional Euclidean space $\R^{0,2}=\R^2$ the point at infinity is added to arrive at the conformal compactification of $\R^2$, i.e. the Riemann sphere $S^2=\C\cup\{\infty\}$.

Any conformal transformation can be written as a sequence of the following four transformations\medskip

\noindent
\renewcommand{\arraystretch}{1.5}
 \begin{tabular}{c p{3cm} c p{4.4cm}}
 \hline\hline
 Transformation& Action & \#& Remarks\\
 \hline\hline
 Translation $T$ & $x^\mu\mapsto x^\mu+a^\mu$ & $d=p+q$ &$\Lambda(x)=1$, Poincar\'e group\\
 \hline
 Rotation/Boost $R$ & $x^\mu\mapsto R^\mu{}_\nu x^\nu$ & $\frac{d(d-1)}{2}$ & $R^{\transpose}\eta R=\eta$, Poincar\'e group\\
 \hline
 Dilation $D$ & $x^\mu\mapsto \lambda x^\mu$ & 1 & $\Lambda(x)=\lambda^2$\\
 \hline
 Special conformal $SC$ & $x^\mu\mapsto \frac{x^\mu+|x|^2b^\mu}{1+2\,b\cdot x+|b|^2|x|^2}$ & $d$ &$\cI\circ T\circ\cI$\\
 \hline
 \end{tabular}\medskip
 
\noindent where $\lambda>0$ as well as $a^\mu, b^\mu$ are constants. A proof that these are indeed all globally defined conformal transformations can be found in any of the textbooks mentioned in \secref{secIntroCFT}. 

The first three types of transformations are well-defined everywhere on $\R^{p,q}$. Special conformal transformations require conformal compactifications however. Note that inversions by themselves do not have a continuous parameter, and are thus not desirable, if we wish to have a Lie group description of conformal symmetry. Hence the concatenation including a translation by the vector $b^\mu$ leading to the special conformal transformation. \footnote{The reason the inversion needs to be applied twice here is that the inversion is not part of the identity component of the conformal group. After applying it twice, one winds up in the identity component once more.}

The total number of distinct conformal transformations is $d+\frac{d(d-1)}{2}+1+d=\frac{(d+1)(d+2)}{2}$. It can be shown that, and once more the reader is referred to the textbooks in \secref{secIntroCFT}, the generators of the above transformations satisfy the commutation relations of $\so(p+1,q+1)$. The conclusion is that
\begin{tcolorbox}
\begin{center}
 The conformal group in $d=p+q\geq3$ on $\R^{p,q}$ is $\SO(p+1,q+1)$.  
\end{center}

\end{tcolorbox}
On the other hand, $d=2$ is \quotes{where the magic happens}. In this case, as we will see in the following, the conformal group is not just $\SO(2,2)$ in Lorentzian signature or $SO(1,3)$ in Euclidean signature, but is infinitely bigger.

\section{Conformal Symmetry in Two Dimensions}\label{secConfSym2d}
It is time to restrict to two spacetime dimensions. In this section we describe the conformal algebra, classically and quantum mechanically. The former is the given by the \textbf{\textit{Witt Algebra}}, while the latter is given by the \textbf{\textit{Virasoro Algebra}}. Both of these algebrae are infinite-dimensional, implying that conformal symmetry imposes an equally large number of constraints on a system. The success of two-dimensional CFT hinges on this fact, as it allows us to solve these QFT -- a generally very difficult task! At the section's end, we catch a first glimpse of the \emphasize{energy-momentum tensor}, whose role will be developed more and more as we move further into the territory of CFT. Below, whenever we write CFT, we always imply two-dimensional CFTs.

\subsection{Global Conformal Transformations}
From now on we work with Euclidean signature, i.e. on $\R^2$ with coordinates $x^\mu=(x^0,x^1)$. It pays off, however, to view this as the complex plane $\C$, coordinatized by $z=x^0+\iu x^1$. Let us develop some intuition on what the two-dimensional versions of the above transformations are. 

\medskip

\noindent
\renewcommand{\arraystretch}{1.5}
 \begin{tabular}{p{4.5cm} p{4.5cm} c}
 \hline\hline
 Transformation& Action & \#(real parameters)\\
 \hline\hline
 Translation $T$ & $z\mapsto z+a,\, a\in\C$  &2\\
 \hline
 Rotation $R$ & $z\mapsto e^{\iu\phi}z,\, \phi\in[0,2\pi)$ & 1 \\
 \hline
 Dilation $D$ & $z\mapsto r z,\,r>0$ & 1 \\
 \hline
 Special conformal $SC$ & $z\mapsto \frac{z}{\bb z+1}$ & 2\\
 \hline
 \end{tabular}\medskip

\noindent Rotations and dilations have the same nature showing in that they can be combined into a single complex dilation $z\mapsto (r\,e^{\iu\phi})z$. To reach the expression for the special conformal transformation, one employs the following expression for the scalar product $x\cdot b=x^\mu b_\mu=(\bz b+z\bb)/2$ with $b=b^0+\iu b^1$.

Taken together, the transformations in the table generate the Möbius group, which is characterized by
\begin{equation}\label{moebius}
 z\mapsto\frac{a\,z+b}{c\,z+d},
 \quad
 \text{where}
 \quad
 \begin{pmatrix}
  \,a\, & \,b\,\\
  c & d
 \end{pmatrix}
\in \PSL(2,\C)\simeq \SL(2,\C)/\Z_2
\end{equation}
This group is in fact very natural in our context as it furnishes the automorphism\footnote{A reminder: Automorphisms are isomorphisms which are also endomorphisms. The former means these are invertible maps and the latter that they map a space onto itself.} group of the Riemann sphere $S^2=\C\cup\{\infty\}$, which we already identified as the conformal compactification of $\R^2$. Note two things. Firstly, the Möbius transformations have no poles on the Riemann sphere, in the sense that, e.g. the point $z=-d/c$ is mapped to $\infty\in S^2$. Secondly, $\PSL(2,\C)$ has three free complex parameters, agreeing with the amount of generators in the table. The conformal transformations in the table are thus represented by the following matrices ($\lambda=re^{\iu\phi}\in\C$)
\begin{equation}\label{globalConfTransf}
 T\cong\begin{pmatrix}
      \,1\, & \,a\,\\
      0 & 1
     \end{pmatrix},
\quad
R/D\cong\begin{pmatrix}
      \lambda^{1/2} & 0\\
      0 & \lambda^{-1/2}
     \end{pmatrix},
     \quad
SC\cong\begin{pmatrix}
      \,1\, & \,0\,\\
      b & 1
     \end{pmatrix},
\end{equation}
Combined conformal transformations are thus represented by the matrix product of these three matrices; a perk of this formalism! The reader may check this as an optional exercise. The transformations \eqref{globalConfTransf} are called \emphasize{global conformal transformations} because these transformations are well-defined everywhere on the Riemann sphere $S^2$. 

So far we have succeeded in understanding the two-dimensional analog of the conformal group in $d\geq3$. However, magic was promised and here it comes...

\subsection{The Conformal Group in Two Dimensions}
In two dimensions the set of conformal transformations is extended to be infinite-dimensional. It is in fact the set of all holomorphic and anti-holomorphic functions.

To proof this claim, we show that first that a holomorphic function of a complex coordinate is indeed a conformal transformation, and turn to the inverse direction afterward. Indeed, given the following complex coordinates
\begin{equation}\label{complexCoord}
 z=x^0+\iu x^1,
 \quad
 \bz=x^0-\iu x^1
 \quad
 \p:=\p_z=\frac{1}{2}(\p_0-\iu \p_1)
 \quad
 \bp:=\p_{\bz}=\frac{1}{2}(\p_0+\iu \p_1)
\end{equation}
the Cauchy-Riemann differential equations assume the following simple form for functions $w(z,\bz)=w^0(z,\bz)+\iu w^1(z,\bz)$ and $\bw(z,\bz)=w^0(z,\bz)-\iu w^1(z,\bz)$
\begin{equation}
 \p \bw(z,\bz)=0
 \qquad
 \bp w(z,\bz)=0
\end{equation}
so that a (anti-)holomorphic mapping takes $z\mapsto w(z)$ ($\bz\mapsto\bw(\bz)$). A holomorphic differential transforms as follows $dz=\frac{\p z}{\p w}dw$ and the metric in turn picks up a scale factor,
\begin{equation}
 ds^2=\frac{1}{2}(dz\otimes d\bz+d\bz\otimes dz)=\frac{\p z}{\p w}\frac{\p \bz}{\p \bw}\frac{1}{2}(dw\otimes d\bw+d\bw\otimes dw)
\end{equation}
Hence, (anti-)holomorphic maps are conformal, as claimed. The inverse direction is left for the reader as an exercise. 

\begin{exercise}\label{exCauchyRiemann} Show that \eqref{ConfMetric} reduces to the Cauchy-Riemann differential equations on the transformation $y=y(x)$. Use the Euclidean metric $\eta_{\mu\nu}=\diag(1,1)$
\end{exercise}
\ifsol
\solution{exCauchyRiemann}{In the case at hand, \eqref{ConfMetric} becomes
\begin{equation}
 \eta_{\alpha\beta}(y)\frac{\p y^\alpha}{\p x^\mu}\frac{\p y^\beta}{\p x^\nu}
 =
 \Lambda(x)\eta_{\mu\nu}(x)
\end{equation}
Evaluating this for the two cases $\mu=\nu=0$ and $\mu=\nu=1$ gives $\Lambda$ on their rhs so the respective lhs' can be equated, yielding
\begin{equation}
 \left(\frac{\p y^0}{\p x^0}\right)^2+\left(\frac{\p y^1}{\p x^0}\right)^2
 =
 \left(\frac{\p y^0}{\p x^1}\right)^2+\left(\frac{\p y^1}{\p x^1}\right)^2
\end{equation}
Due to symmetry in $(\mu\leftrightarrow\nu)$ we only need to evaluate one remaining case. Pick $\mu=0,\,\nu=1$
\begin{equation}
 \frac{\p y^0}{\p x^0}\frac{\p y^0}{\p x^1}+\frac{\p y^1}{\p x^0}\frac{\p y^1}{\p x^1}
 =
 0
\end{equation}
These condition are equivalent to either
\begin{equation}
 \frac{\p y^1}{\p x^0}=\frac{\p y^0}{\p x^1} 
\qquad \text{and} \qquad
 \frac{\p y^0}{\p x^0}=-\frac{\p y^1}{\p x^1}
\end{equation}
which are the Cauchy-Riemann equations for holomorphic functions, or to
\begin{equation}
 \frac{\p y^1}{\p x^0}=-\frac{\p y^0}{\p x^1} 
\qquad \text{and} \qquad
 \frac{\p y^0}{\p x^0}=\frac{\p y^1}{\p x^1}
\end{equation}
which are the Cauchy-Riemann equations for anti-holomorphic functions.
}
\fi

We arrive at the foreboded conclusion giving rise to the super powers of CFT in two dimensions:
\begin{tcolorbox}
 \textit{In two dimensions, the conformal group is the set of all holomorphic and anti-holomorphic maps. Group multiplication is the standard composition of maps. This set is infinite dimensional.}
\end{tcolorbox}

A few remarks are in order:
\begin{itemize}
 \item Clearly, the global conformal transformations \eqref{moebius}, i.e. the Möbius transformations, are a subset of these.
 \item Strictly speaking we will be dealing with meromorphic functions, as we will allow countably many isolated singularities leading to Laurent (instead of Taylor) series. This subtlety is usually glossed over in the physics literature. This implies that some of the allowed transformations are not well-defined everywhere on the Riemann sphere $S^2$. Tranformations that are ill-defined at at least one isolated point on the Riemann sphere are called \emphasize{local conformal transformations}.
 \item The coordinates $z$ and $\bz$ are treated as independent, rather than complex conjugates. The proper way of thinking about this is to analytically continue the Cartesian coodinates $x^0$ and $x^1$ to the complex plane. In this way, the relations \eqref{complexCoord} define a simple coordinate transformation between independent coordinates. After any calculation one reverts to the \textit{physical surface} by setting 
 \begin{equation}\label{physSurface}
  \bz=z^*
 \end{equation}
\end{itemize}

\subsection{Classical Conformal Symmetry and the Witt Algebra}\label{secClassCFTWitt}
Any holomorphic transformation is a conformal transformation, also the infinitesimal ones, which we restrict to next. Locally, analytic and meromorphic maps can be expanded into a power series. As mentioned above, these will be Laurent series. Consider the following one 
\begin{equation}\label{InfConfTransf}
 g_w:
 \quad
 w(z)
 =
 z+\epsilon(z)
 =
 z+\sum_{n\in\Z}\epsilon_nz^{n+1}, 
 \qquad 
 |\epsilon_n|\ll1
\end{equation}
where $g_w$ is the group element of the conformal group coding for $w(z)$. To obtain a feeling for this structure, let us restrict to a single $\epsilon_n$ and declare a generator $\ell_n$ responsible for this transformation
\begin{equation}\label{EllnTransf}
 \ell_n:
 \quad
 z\mapsto w_n(z)=z+\epsilon_n z^{n+1}
 \qquad
 |\epsilon_n|\ll1
\end{equation}
The global conformal transformations \eqref{globalConfTransf}, for instance, correspond to
\begin{equation}
T\cong\ell_{-1} 
\qquad
R/D\cong\ell_0
\qquad
SC\cong\ell_1
\end{equation}
For translations this is seen trivially. For $R/D$ this comes about because $z\mapsto\lambda z\simeq(1+\epsilon)z$ and for $SC$ we have $z\mapsto z/(\epsilon z+1)\simeq1+\epsilon z^2$. These are indeed global since they have no poles on the Riemann sphere $\C\cup\{\infty\}$. For all $n\notin{0,\pm1}$, the transformations are ill-defined somewhere on the Riemann sphere, hence their characterization as \textit{local} conformal transformations. 

Group elements responsible for transformations are represented on functions as $\pi(g)[f(z)]=f(g^{-1}(z))$ \footnote{The inverse appears to secure that this representation respects the group multiplication of the group, as it should.}. To see how the generator $\ell_n$ is represented on functions, introduce $g_n\simeq \id+\epsilon\ell_n$. The calculation to be performed is
\begin{align}
 \pi(g_n)[f(z)]&=f(w_n^{-1}(z))\notag\\
 \pi(\id+\epsilon\ell_n)[f(z)]&=f(z-\epsilon z^{n+1})\notag\\
 \simeq\qquad
 f(z)+\epsilon\pi(\ell_n)[f(z)]&=f(z)-\epsilon z^{n+1}\p f(z)
\end{align}
From now on the presence of the representation symbol $\pi$ is understood from context and not written explicitly. Hence we arrive at the following conclusion
\begin{equation}\label{WittGenerator}
 \ell_n=-z^{n+1}\p
\end{equation}
Observe that this expression is non-singular at $z=0$ only for $n\geq-1$. The other troublesome point is $z=\infty$, which is studied best after the inversion $z=-1/w$ by contemplating $w\to0$. Using $\p_z=w^2\p_w$ we obtain
\begin{equation}
 \ell_n=-\left(-\frac{1}{w}\right)^{n-1}\p_w
\end{equation}
This is non-singular at $w=0$ only for $n\leq1$. Note that the global conformal transformations form precisely the intersection of the allowed sets at $0$ and $\infty$. This demonstrates clearly that the $\ell_{0,\pm1}$ generate the only conformal transformations, which are well-behaved everywhere on the Riemann sphere $S^2=\C\cup\{\infty\}$. In contrast, all transformations with $n\neq0,\pm1$ are ill-behaved at isolated points on $S^2$, and hence called \emphasize{local conformal transformations}.
\begin{exercise}\label{exWittAlgebra}
 Show that the generators of conformal transformations satisfy the \textbf{Witt Algebra}
 \begin{equation}\label{Witt}
  [\ell_n,\ell_m]=(n-m)\ell_{n+m}
 \end{equation}

\end{exercise}
\ifsol
\solution{exWittAlgebra}{Applying the commutator to a test function $f$ gives
\begin{equation}
 \left[z^{n+1}\p,z^{m+1}\p\right]f
 =
 \left(z^{n+1}\p(z^{m+1})\p-(n\leftrightarrow m)\right)f
 =
 (m-n)z^{n+m+1}\p f
 =
 (n-m)\ell_{n+m}
\end{equation}
where the piece proportional to $\p^2$ drops out due to the commutator.}
\fi

This is the Lie algebra of \textit{classical} conformal symmetry in two dimensions... well, frankly, it is almost said algebra.

The attentive reader may have noticed a disparity between the number of global conformal transformations represented by the $\ell_n$ and their total number. Indeed, $\ell_n$ for $n=0,\pm1$ are only three, while the conformal algebra $\so(2,2)$ in two dimensions  has six generators. The missing three generators are the anti-holomorphic counterparts of $\ell_n$, namely $\bar{\ell}_n=-\bz^{n+1}\p_{\bz}$. Since $z$ and $\bz$ are taken to be completely independent barred generators commute with the unbarred ones. In total we now have
\begin{itemize}
 \item Two translations $\ell_{-1}, \bar{\ell}_{-1}$
 \item Two special conformal transformations $\ell_{1}, \bar{\ell}_{1}$
 \item One dilation $\ell_0+\bar{\ell}_0$
 \item One rotation $\iu(\ell_0-\bar{\ell}_0)$
\end{itemize}
Of course, antiholomorphic $\bar{\ell}_n$ exist for any $n\in\Z$ and they ovbiously also satisfy the Witt algebra \eqref{Witt}.
\begin{exercise}\label{ExDilationRotation}
 By going to polar coordinates $z=re^{\iu \spa}$, convince yourself that the linear combinations $\ell_0\pm\bar{\ell}_0$ indeed correspond to dilations and rotations. 
\end{exercise}
\ifsol
\solution{ExDilationRotation}{
In complex coordinates we have
\begin{equation}
 \ell_0
 =
 -\frac{r}{2}\p_r+\frac{\iu}{2}\p_\spa\,,
 \qquad
 \bar{\ell}_0
 =
 -\frac{r}{2}\p_r-\frac{\iu}{2}\p_\spa\,
\end{equation}
These can be recombined into
\begin{equation}
 \ell_0+\bar{\ell}_0=r\p_r\,
 \qquad
 \iu(\ell_0-\bar{\ell}_0)=-\p_\spa
\end{equation}
which are the generators of dilations and rotations, respectively.
}
\fi

\subsection{Quantum Conformal Symmetry and the Virasoro Algebra}

Upon quantization of a system with classical conformal symmetry, the Witt Algebra is in fact deformed to the celebrated \textbf{\textit{Virasoro Algebra}} 
\begin{tcolorbox}
\begin{equation}\label{Virasoro}
 [L_n,L_m]=(n-m)L_{n+m}+\frac{\cc}{12}n(n^2-1)\delta_{n+m,0}
\end{equation}
\end{tcolorbox}
\noindent The numbers $n$ and $m$ are integer\footnote{Though there are situations where they are fractional numbers; we will not encounter such situations.}. The additional piece is the \textbf{\textit{central charge}} $\cc$ of the system. In full rigor, this is treated as an operator which commutes with all $L_n$, hence it is \textit{central} and the deformation is called a \textit{central extension}. By Schur's Lemma, this acts as a multiple of the identity on irreducible representations of the Virasoro algebra. In practice it is thus a number, and arises typically due to normal ordering of operators or double Wick contractions. Examples will be presented, see exercise \ref{exfreeFieldsTTOPE}, once we have some more technology at our disposal. Note that the central term disappears on the global conformal generators with $n\in\{0,\pm1\}$, implying that $L_{\pm1}$ still generate translation and special conformal transformations respectively, and that $L_0$ still generates dilations and rotations.

In broad terms, the central charge is an anomaly. Anomalies are known to spoil a classical symmetry upon quantization. In our case the anomaly leaves conformal symmetry largely intact, evident by the resemblance of the Virasoro and the Witt algebra. We say that \emphasize{conformal symmetry is broken \quotes{softly}}. A quantum system with conformal symmetry will react to the introduction of a macroscopic scale, such as temperature or a finite extend of the system, through the central charge. We will see this a little more explicitly when getting acquainted with the energy-momentum tensor. The central charge indicates the \quotes{amount} by which conformal symmetry is broken by quantum effects and as it is a feature of the CFT. It can be shown that the above central extension is unique up to normalization, the reader is referred to the textbooks in \secref{secIntroCFT}.

As in the classical case, the symmetry algebra of a CFT contains one copy of the Virasoro algebra for the $L_n$ and one for their antiholomorphic counterparts $\bL_n$. These two copies commute. For completeness we gather the \textbf{\textit{symmetry algebra of a quantum CFT}} 
\begin{subequations}\label{SymAlgebra}
\begin{align}
 [L_n,L_m]&=(n-m)L_{n+m}+\frac{\cc}{12}n(n^2-1)\delta_{n+m,0}\\
 [\bL_n,\bL_m]&=(n-m)\bL_{n+m}+\frac{\bar{\cc}}{12}n(n^2-1)\delta_{n+m,0}\\
 [L_n,\bL_m]&=0
\end{align}
\end{subequations}
The antiholomorphic central charge $\bar{\cc}$ is completely independent of the holomorphic one $\cc$. The examples we encounter in these notes, however, all have $\bar{\cc}=\cc$.

\subsection{Enter the Protagonist: The Energy-Momentum Tensor}
The energy-momentum tensor $T_{\mu\nu}$ can be derived from the variation of the action with respect to infinitesimal transformations of the metric, $g_{\mu\nu}\mapsto g_{\mu\nu}+\delta g_{\mu\nu}$. Conformal transformations are included in such transformations, and additionally, in a conformally invariant system, they impose an infinite number of constraints.

Emmy Noether taught us that any continuous symmetry leads to a conserved current $\p_\mu j^\mu=0$. Our interest lies with transformations of the type $x^\mu\mapsto x^\mu+\epsilon^\mu(x)$, for which the conserved current turns out to be 
\begin{equation}
 j_\mu=T_{\mu\nu}\epsilon^\nu(x)\,.
\end{equation}
Checking its conservation $\p_\mu j^\mu=0$ for various types of transformations leads to the following properties of the energy-momentum tensor
\begin{itemize}
 \item Translations: $\epsilon^\mu=$constant $\Rightarrow$ $\p_\mu T^{\mu\nu}=0$
 \item Rotations: $\epsilon^\mu=\omega^{\mu\nu}x_\nu$ for a constant antisymmetric tensor $\omega_{\mu\nu}$ $\Rightarrow$ $T^{\mu\nu}=T^{\nu\mu}$
 \item Dilations: $\epsilon^\mu=\varepsilon x^\mu$ $\Rightarrow$ $T_\mu{}^\mu=0$
\end{itemize}
The first two are obviously familiar from any Poincar\'e invariant scalar theory. The third is new and indeed a hallmark of conformal symmetry -- well actually a hallmark of scaling symmetry by itself -- namely that the \emphasize{energy-momentum tensor is traceless in a CFT}. It is this last conservation law that is violated upon quantization by the central charge leading to the terminology of an anomaly.

In two dimensions, the energy-momentum tensor has thus only two independent real components. Indeed, symmetry forces $T^{01}=T^{10}$ and tracelessness imposes $T^{00}=-T^{11}$. This is found using a flat Euclidean metric $\delta_{\mu\nu}=\diag(1,1)$. Switching to complex coordinates $z=x^0+ix^1$ results in the two components
\begin{equation}
 T_{zz}=\frac{1}{2}(T_{00}-\iu T_{01}),
 \qquad 
 T_{\bz\bz}=\frac{1}{2}(T_{00}+\iu T_{01})
\end{equation}
These are both, in principle, functions of $z$ and $\bz$. Employing the conservation of the energy-momentum tensor leads however to the following desirable properties
\begin{equation}
 \bp T_{zz}=\frac{1}{2}(\p_0+\iu\p_1)\frac{1}{2}(T_{00}-\iu T_{01})=0
 \qquad
 \p T_{\bz\bz}=\frac{1}{2}(\p_0-\iu\p_1)\frac{1}{2}(T_{00}+\iu T_{01})=0
\end{equation}
Hence, these two components are holomorphic and antiholomorphic, respectively. It is customary to define
 \begin{equation}\label{ETcomponents}
  T(z):=T_{zz},
  \qquad \qquad
  \bT(\bz):=T_{\bz\bz}
 \end{equation}

\section{Fields and States in a CFT}\label{secFieldsStates}
Now that we have an idea of the symmetry algebra, we want to take a closer look at the fields of a CFT. In this section we get acquainted with the notions of \textbf{\textit{primary}} and \textbf{\textit{descendant fields}}, the \textbf{\textit{operator-state correspondence}}. Moreover, we catch a glimpse at \textbf{\textit{correlators}} and the \textbf{\textit{operator product expansion}}. In this section

\subsection{Primary Fields}
An important piece of information when dealing with field theories endowed with symmetries is how the fields transform. In regular QFT one encounters free scalar fields, which transform in the trivial representation of the Lorentz group, while fermions transform non-trivially. Similarly, we need to describe the representations of the conformal group. An extensive discussion thereof is found in \cite{DiFrancesco}. Only the results are stated here. 

First, we pick a maximally commuting set of generators amongst the symmetry algebra, i.e. a Cartan torus. Canonically, this is the pair $(L_0,\bL_0)$. Physically, these correspond to the dilation operator $D=L_0+\bL_0$, whose eigenvalue is the scaling dimension $\Delta=h+\bh$, and the rotation operator $R=L_0-\bL_0$ with spin $s=h-\bh$ as eigenvalue. It will be most convenient to work with the (anti-)holomorphic conformal weight $h\,(\bh)$, which is the eigenvalue of $L_0\,(\bL_0)$.

For starters, fields transforming under scalings $z\mapsto \lambda z$ according to
\begin{equation}
 \phi(z,\bz)\mapsto\phi'(z,\bz)=\lambda^h\bar{\lambda}^{\bh}\phi(\lambda z,\bar{\lambda}\bz)
\end{equation}
are said to have conformal weights $(h,\bh)$. Fields which mimick this behavior for an arbitrary conformal transformation $z\mapsto f(z)$
\begin{equation}\label{primary}
 \phi(z,\bz)
 \mapsto
 \phi'(z,\bz)
 =
 U_f\phi(z,\bz)U_f^{-1}
 =
 \left(\frac{\p f}{\p z}\right)^h
 \left(\frac{\bp \bar{f}}{\p \bz}\right)^{\bh}
 \phi\bigl(f(z),\bar{f}(\bz)\bigr)
\end{equation}
are called \textbf{\textit{primary fields}} of conformal weights $(h,\bh)$. Fields which satisfy this transformation behavior only for the global conformal group $\SL(2,\C)/\Z_2$ are called \textbf{\textit{quasi-primary fields}}. \new{$U_f$ implements\footnote{\new{Actually, $U_f$ is the product of two copies of a projective representation of the conformal group on Hilbert space, one generated by $T$, the other by $\bT$.}} the conformal transformation $f$ unitarily on the Hilbert space $\cH$ of the theory, while the field $\phi$ is to be viewed as an operator on $\cH$. In this section and the next, we lay the groundwork for understanding the structure of this Hilbert space, and discuss it in-depth only in \secref{secHilbertSpace}.}  Aficionados of differential geometry will appreciate that the transformation behavior \eqref{primary} is that of a tensor of rank $\Delta=h+\bh$ with $h$ \quotes{$z$} indices and $\bh$ \quotes{$\bz$} indices. 

Note that there is a distinguished primary field which remains inert under conformal transformations: the \textbf{\textit{identity field $\id$}}, which is constant throughout the plane, and has $h=\bh=0$. It is reasonable to assume that a CFT has only one such field, and indeed we shall do so\footnote{However, looking ahead to \secref{secDCFT}, when introducing defects to the theory, they come with their own excitations, some of which may have $h=\bh=0$ as well. These will be topological excitations! Since we are looking ahead already, let us also foreshadow that the primary fields \eqref{primary} for particular CFTs are synonymous with anyons in the fractional quantum Hall effect.}

Because the global conformal group is a subgroup of the full conformal symmetry, a primary is always quasi-primary. Clearly, the reverse is not true. Primary fields are a very special class of fields in a CFT and, as is discussed below, these furnish the highest weights in representations of the Virasoro algebra. All other fields are called \textbf{\textit{descendant fields}}. They transform as
\begin{equation}\label{descendants}
 \phi(z,\bz)
 \mapsto
 \phi'(z,\bz)
 =
 \left(\frac{\p f}{\p z}\right)^h
 \left(\frac{\bp \bar{f}}{\p \bz}\right)^{\bh}
 \phi\bigl(f(z),\bar{f}(\bz)\bigr)+\textrm{extra terms}
\end{equation}
More on descendants later, once we have developed better tools to describe these matters. 

\begin{exercise}\label{exInfConfTransfPrimary}
 Resorting to infinitesimal conformal transformations $f(z)=z+\epsilon(z)$ and $\bar{f}(\bz)=\bz+\bar{\epsilon}(\bz)$ with $\epsilon(z),\bar{\epsilon}(\bz)\ll1$, show that the transformation behavior \eqref{primary} implies 
 \begin{equation}\label{InfConfTransfPrimary}
  \delta_{\epsilon,\bar{\epsilon}}\phi(z,\bz)
  :=
  \phi'(z,\bz)-\phi(z,\bz)
  =
  \bigl[h(\p\epsilon)(z)+\epsilon(z)\p
  +
  \bh(\bp\bar{\epsilon})(\bz)+\bar{\epsilon}(\bz)\bp\,\bigr]\phi(z,\bz)
 \end{equation} 
 
\end{exercise}
\ifsol
\solution{exInfConfTransfPrimary}{
Use the Taylor expansions
\begin{equation}
 \left(\frac{\p f}{\p z}\right)^h
 =
 \left(1+\p \epsilon\right)^h\approx1+h\p\epsilon\,,
 \quad
 \left(\frac{\bp \bar{f}}{\p \bz}\right)^{\bh}
 =
 \left(1+\p \bar\epsilon\right)^{\bh}\approx1+\bh\bp\bar\epsilon\,,
\end{equation}
and 
\begin{equation}
 \phi(f(z),\bar f(z))\approx \phi(z,\bz)+\epsilon(z)\p\phi(z,\bz)+\bar\epsilon(\bz)\bp\phi(z,\bz)
\end{equation}
Plugging this into \eqref{primary} and rearranging provides \eqref{InfConfTransfPrimary}}.
\fi

Using $\epsilon(z)=\sum_n\epsilon_nz^{n+1}$ as in \eqref{InfConfTransf}, we recognize that the second term becomes $\epsilon(z)\p\phi=-\sum_n\epsilon_n\ell_n\phi$ and similarly for the antiholomorphic fourth term. This shows that these two terms implement the transformations for the spacetime argument, while the terms proportional to $h$ or $\bh$ indicate the transformation due to the field's conformal representation. We will turn to representations of the conformal algebra in detail in \secref{secVermaConfFam}.
\begin{exercise}\label{exFreeTheories}
 The actions of free massless real bosons and Majorana fermions can be expressed in complex coordinates (in two dimensions) as follows
 \begin{subequations}\label{freeTheories}
   \begin{align}
  S_\text{boson}&=\int\textrm{d}z\textrm{d}\bz\,(\p\varphi)(\bp\varphi)\label{actionBoson}\\
  S_\text{fermion}&=\int\textrm{d}z\textrm{d}\bz\,\left[\psi(\bp\psi)+\bar{\psi}(\p\bar{\psi})\right]\label{actionFermion}
 \end{align}
 \end{subequations}
where $\varphi=\varphi(z,\bz)$ and $\bar{\psi}$ is not meant to be the Dirac adjoint, but rather $\Psi=(\psi(z,\bz),\bar{\psi}(z,\bz))$ form the conventional Majorana spinor. Argue from the equations of motion in both cases that 
\begin{equation}\label{freeFieldsPrimaries}
 \p\varphi=\p\varphi(z),
 \qquad  
 \bp\varphi=\bp\varphi(\bz),
 \qquad
 \psi=\psi(z),
 \qquad
 \bar{\psi}=\bar{\psi}(\bz).
\end{equation}
What can you conclude for the holomorphic and anti-holomorphic conformal weights of these fields from here? Assume that these fields are primaries and use the invariance of the action under conformal transformations $(z,\bz)\to (f(z),\bar{f}(\bz))$ to derive the tuple of conformal weights $(h,\bh)$ for all four fields \eqref{freeFieldsPrimaries}.
\end{exercise}
\ifsol
\solution{exFreeTheories}{
The bosonic equation of motion is $\p\bp\varphi=0$ implying that $\varphi(z,\bz)=\phi(z)+\bar\phi(\bz)$. Thus $\p\varphi=\p\phi(z)$ and thus $\p\varphi$ has $\bh=0$. Similarly $\bp\varphi$ has $h=0$. Transforming the action \eqref{actionBoson} amounts to
\begin{equation}
 S_\text{boson}
 \to
 \int \textrm{d}f\textrm{d}\bar f\,
 \frac{\p z}{\p f}\frac{\p \bz}{\p \bar f}
 \left(\frac{\p f}{\p z}\right)^h
 \left(\frac{\p \bar f}{\p \bz}\right)^{\bh}
 (\p_f\varphi)(\bp_{\bar f}\varphi)
\end{equation}
Demanding invariance leads to $(h,\bh)=(1,0)$ for $\p\varphi$ and $(h,\bh)=(0,1)$ for $\bp\varphi$. The equations of motion for the fermion are $\p\bar\psi=0$ and $\bp\psi=0$. Thus $\psi$ has $\bh=0$ and $\bar\psi$ has $h=0$. Demanding invariance of the action shows that $\psi$ has $(h,\bh)=(1/2,0)$ and $\bar\psi$ has $(h,\bh)=(0,1/2)$. 
}
\fi

\subsection{A Few Words on Correlators and the Operator Product Expansion}
Most of the richness of QFTs relies in how fields \quotes{communicate} with each other in correlation functions. Here, we only have time to state important results on correlators in CFT leaving all their intricacies and derivations to any of the excellent resources mentioned in the introduction. 

\textbf{\textit{One-point functions}} of any field with non-vanishing $\Delta\neq0$ vanish as a consequence of scale symmetry. Because of holomorphic-antiholomorphic factorization this property can be already formulated at the level of a (anti-) holomorphic field 
\begin{equation}\label{1pt}
 \corr{\phi(z)}=\delta_{h,0},
 \qquad\qquad
 \corr{\bphi(\bz)}=\delta_{\bh,0}
\end{equation}
This means that only the identity field $\id$ is allowed to have a non-vanishing expectation value!

\textbf{\textit{Two-point functions}} are likewise totally constrained by conformal symmetry, up to normalization. It is instructive to derive this. Define the shorthand $g(z,w)=\corr{\phi_1(z)\phi_2(w)}$. Invariance under translations, given by the conformal map $f(z)=z+a$ imposes that $g(z,w)=g(z-w)$. Invariance under scalings $f(z)=\lambda z$ demand that
\begin{equation}
 \corr{\phi_1(z)\phi_2(w)}
 \to
 \corr{\lambda^{h_1}\phi_1(\lambda z)\lambda^{h_2}\phi_2(\lambda w)}
 =
 \lambda^{h_1+h_2}g\bigl(\lambda(z-w)\bigr)
 \overset{!}{=}
 g(z-w)
\end{equation}
which is fulfilled by $g(z-w)=d_{12}(z-w)^{-h_1-h_2}$ for a constant $d_{12}$. Lastly, consider the effect of the inversion $f(z)=-1/z$,
\begin{align}
 \corr{\phi_1(z)\phi_2(w)}
 &\to
 \frac{1}{z^{2h_1}w^{2h_2}}\corr{\phi_1\left(-\frac{1}{z}\right)\phi_2\left(-\frac{1}{w}\right)}\notag\\
 &=
 \frac{1}{z^{2h_1}w^{2h_2}}\frac{d_{12}}{\left(-\frac{1}{z}+\frac{1}{w}\right)^{h_1+h_2}}
 \overset{!}{=}
 \frac{d_{12}}{(z-w)^{h_1+h_2}}
\end{align}
This requirement can only be met if $h_1=h_2$. In conclusion, the correlators of two (anti-)holomorphic fields $\phi_i(z)$ ($\bphi_i(\bz)$) with conformal dimensions $h_i$ ($\bh_i$) are 
\begin{align}\label{2pt}
 \corr{\phi_1(z_1)\phi_2(z_2)}
 =
 d_{12}\frac{\delta_{h_1,h_2}}{(z_1-z_2)^{2h_1}}
 \qquad\qquad
 \corr{\bar\phi_1(\bz_1)\bar\phi_2(\bz_2)}
 =
 \bar{d}_{12}\frac{\delta_{\bh_1,\bh_2}}{(\bz_1-\bz_2)^{2\bh_1}}
\end{align}
where the $d_{12}$ ($\bar{d}_{12}$) are normalizations for the fields. Two characteristics are paramount here. 
\begin{itemize}
 \item The appearance of the delta function signifies that two fields with differing conformal weights stand orthogonal to each other. This notion is made more precise shortly once we unlock the operator-state correspondence. Note that one recovers \eqref{1pt} from \eqref{2pt} by fixing either of the two involved fields to be the unit field $\id$.
 \item The correlator follows a power-law behavior instead of decaying exponentially. When encountering a system in the wilderness, this hallmark can be used to diagnose conformal symmetry upon measuring its correlators.
\end{itemize}
Finally, correlators of \quotes{full} fields, i.e. those with holomorphic and antiholomorphic weights are products of the expressions provided for the individual chiralities, i.e.
\begin{equation}
 \corr{\phi_1(z_1,\bz_1)\phi_2(z_2,\bz_2)}
 =
 d_{12}\bar{d}_{12}\frac{\delta_{h_1,h_2}}{(z_1-z_2)^{2h_1}}
 \frac{\delta_{\bh_1,\bh_2}}{(\bz_1-\bz_2)^{2\bh_1}}
\end{equation}
This is evidently not only true of two-point functions but more generally for multi-point functions.

\begin{exercise}\label{exFreeField2pt}
 Evaluate the chiral two-point function of a real free boson and fermion
 \begin{equation}
  \corr{\p\varphi(z)\p\varphi(w)}, \qquad \corr{\psi(z)\psi(w)}
 \end{equation}
and confirm that these expressions satisfy bosonic and fermionic statistics, respectively.
\end{exercise}
\ifsol
\solution{exFreeField2pt}{
\begin{equation}
 \corr{\p\varphi(z)\p\varphi(w)}
 =
 \frac{d_{\p\varphi\p\varphi}}{(z-w)^2},
 \qquad 
 \corr{\psi(z)\psi(w)}
 =
 \frac{d_{\psi\psi}}{z-w}
\end{equation}
The bosonic correlator is invariant under $(z\leftrightarrow w)$, while the fermionic correlator picks up a minus.}
\fi

Historically, the great success of two-dimensional CFT is due to its predictions for critical behavior\footnote{Back then the focus lied heavily on second order phase transitions. Nowadays, CFT is also used to describe quantum phase transitions and many other phenomena!} of statistical models. In the next exercise we catch a glimpse at how one proceeds in matching a CFT to a statistical model.  
\begin{exercise}\label{exIsing2pt}
 The study of representations of the Virasoro algebra instructs us that at central charge $\cc=1/2$ there is a CFT with a field content
 \begin{equation}
  \phi_{1,1}\,:(h,\bh)=(0,0),\qquad
  \phi_{2,1}\,:(h,\bh)=(1/2,1/2),\qquad
  \phi_{1,2}\,:(h,\bh)=(1/16,1/16).
 \end{equation}
The Ising model has a local spin operator $\sigma_n$ and local energy operator $\epsilon_n$, where $n$ denotes a site. Their correlators are 
\begin{equation}
 \corr{\varepsilon_n\varepsilon_0}\propto\frac{1}{|n|^2},\qquad
 \corr{\sigma_n\sigma_0}\propto\frac{1}{|n|^{1/4}} 
\end{equation}
Associate the non-trivial CFT fields with these Ising operators. This CFT is referred to in following as the \emphasize{Ising CFT} and its primaries are labeled as $\id,\,\varepsilon(z,\bz),\,\sigma(z,\bz)$.
\end{exercise}
\ifsol
\solution{exIsing2pt}{
\begin{equation}
  \phi_{1,1}=\id
  \qquad
  \phi_{2,1}=\varepsilon(z,\bz)
  \qquad
  \phi_{1,2}=\sigma(z,\bz)
 \end{equation}
}
\fi

\textbf{\textit{Three-point functions}} are almost fixed entirely by conformal symmetry
\begin{align}
 \corr{\phi_1(z_1)\phi_2(z_2)\phi_3(z_3)}
 =
 \frac{C_{123}}{(z_{12})^{h_1+h_2-h_3}(z_{23})^{h_2+h_3-h_1}(z_{13})^{h_1+h_3-h_2}}\label{3pt}\\
 \corr{\bphi_1(\bz_1)\bphi_2(\bz_2)\bphi_3(\bz_3)}
 =
 \frac{\bar{C}_{123}}{(\bz_{12})^{\bh_1+\bh_2-\bh_3}(\bz_{23})^{\bh_2+\bh_3-\bh_1}(\bz_{13})^{\bh_1+\bh_3-\bh_2}}
\end{align}
where we have introduced the short-hand notation $z_{ij}=z_i-z_j$. We will not derive this result here -- readers are once more referred to the literature in \secref{secIntroCFT} -- but only point out its features. Once more, a characeristic power-law appears. More importantly, this time a constant $C_{123}$ appears, which cannot be normalized as they will reappear in $d_{ij}=1$. Hence the coefficients $C_{123}$ carry physical meaning; they are in fact very important. Schematically, we can think of having field $\phi_1$ and $\phi_2$ collide and decompose into a sum of other fields in the theory. This procedure is called
\begin{tcolorbox}
The \emphasize{Operator Product Expansion (OPE)} 
 \begin{equation}\label{OPE}
 \phi_1(z_1)\phi_2(z_2)=\sum_{k}C_{12}{}^{k} \,z_{12}^{h_k-h_1-h_2}\, \phi_k(z_2) 
\end{equation}
The label $k$ runs over all fields in the theory. Because \new{an expression, which is} quadratic in fields becomes linear, we regard this expression as an \emphasize{Operator Algebra} \new{and the $C_{12}^k$ are its \emphasize{structure constants}}.
\end{tcolorbox}
The way the operator product expansion is presented in \eqref{OPE} is not \new{enirely} standard. Typically one splits the sum over $k$ into primaries and descendants, in which case $C_{12k}=\sum_lC_{12}{}^ld_{lk}$ holds up to functions of $h_1,h_2,h_k$. The interested reader is referred to the literature in \secref{secIntroCFT}.

\subsubsection*{\new{A Remark on Consistent Structure Constants}}\label{remPlaneSewing}
Employing the OPE in \eqref{3pt}, a three-point function reduces to a two-point function, which in turn carries $C_{12}{}^{k}\delta_{h_k,h_3}$. We obtain the relation $C_{12k}=\sum_lC_{12}{}^ld_{lk}$. \new{Therefore}, the physical meaning of $C_{123}$ is to indicate \textit{how} fields $\phi_1$ and $\phi_2$ combine into $\phi_3$. \new{Evidently, the structure constants thus encode the dynamical data and are of paramount importance for the theory. They turn out be subject to powerful consistency constraints called \emphasize{sewing relations}, which are consequences of two important requirements:
\begin{itemize}
 \item[1)] \emphasize{Locality} requires correlators to be analytic functions of its insertion points $z_i$ with $z_j\neq z_i$ for $j\neq i$ after returning to the physical surface, $\bz_i=z_i^*$.
 \item[2)] \emphasize{Associativity of the OPE}  imposes indepence of the order in which several fields are contracted by use of \eqref{OPE} in a correlator.
\end{itemize}
A proper discussion of these important concepts leads us too far afield, but the reader is encouraged to consult the work of Moore and Seiberg on this matter \cite{moore1989classical,moore1990lectures}.}

\subsubsection*{\new{Examples of OPEs}} 
When spelling out an OPE, by convention, one usually only writes those terms of an OPE, which are singular in their spacetime argument. They suffice for many important purposes; non-singular terms can be very important too, and we encounter one such example. Here are three examples. Ellipsis stand for non-singular terms in $(z-w)$
\begin{itemize}
 \item Real free massless boson 
 \begin{equation}\label{bosonOPE}
  (\p\varphi)(z)(\p\varphi)(w)
  =
  \frac{d_{\p\varphi\p\varphi}}{(z-w)^2}\id+\dots
 \end{equation}
and similarly for the antiholomorphic field $(\bp\bar\varphi)(\bz)$. Note that a first order pole is forbidden by bosonic statistics.
\item Real free massless fermion 
 \begin{equation}\label{fermionOPE}
  \psi(z)\psi(w)
  =
  \frac{d_{\psi\psi}}{z-w}\id
  +\dots
 \end{equation}
and similarly for the antiholomorphic field $\bar\psi(\bz)$. 
\item Ising CFT
\begin{subequations}\label{IsingOPE}
\begin{align}
 \varepsilon(z,\bz)\varepsilon(w,\bw)
 &=
 \frac{\id}{|z-w|^2}+\dots\\
 \varepsilon(z,\bz)\sigma(w,\bw)
 &=
 \frac{C_{\varepsilon\sigma}{}^\sigma}{|z-w|}\sigma(w,\bw)+\dots\\
 \sigma(z,\bz)\sigma(w,\bw)
 &=
 \frac{\id}{|z-w|^{\frac{1}{4}}}+C_{\sigma\sigma}{}^{\varepsilon}|z-w|^{\frac{3}{4}}\varepsilon(w,\bw)+\dots
\end{align}
\end{subequations}
In the last line we have kept a non-singular term. The utility of this term will be justified once we have become acquainted with conformal families and fusion rules, see \secref{secVermaConfFam} and \secref{secFusion}. The computation of the three-point coefficients is usually more involved, and we refer the reader to chapter 12 of \cite{DiFrancesco}. Their values are $C_{\varepsilon\sigma}{}^\sigma=\frac{1}{2}=C_{\sigma\sigma}{}^{\varepsilon}$ and $C_{\sigma\sigma}{}^{\id}=1$
\end{itemize}
Observe that the OPE truncates at some negative power. Suppose the contrary was true and we had an infinity of negative powers. This would require fields of negative conformal weights, which exist in non-unitary theories (but we actually stay clear of these). If there were an infinite number of these, it would mean that energy is unbounded below. Hence, the truncation of the OPE at some negative power reflects the boundedness of energy. Their is no truncation toward positive powers of $(z-w)$. These are an infinity's worth of descendant fields.

While these are OPEs specific to particular models, we will come across two OPEs which are fundamental to conformal symmetry in \secref{secTandVir}.

\begin{exercise}\label{exVevOPE}
 Take the vacuum expectation value of all these OPEs. How many orders in $(z-w)$ does one need for the result to be exact?
\end{exercise}
\ifsol
\solution{exVevOPE}{Due to \eqref{1pt}, one only needs the terms proportional to the unit field $\id$ to recover the two-point functions.}
\fi

\subsection{Radial Quantization}
\begin{figure}[t]
\centering
 \includegraphics[scale=.3]{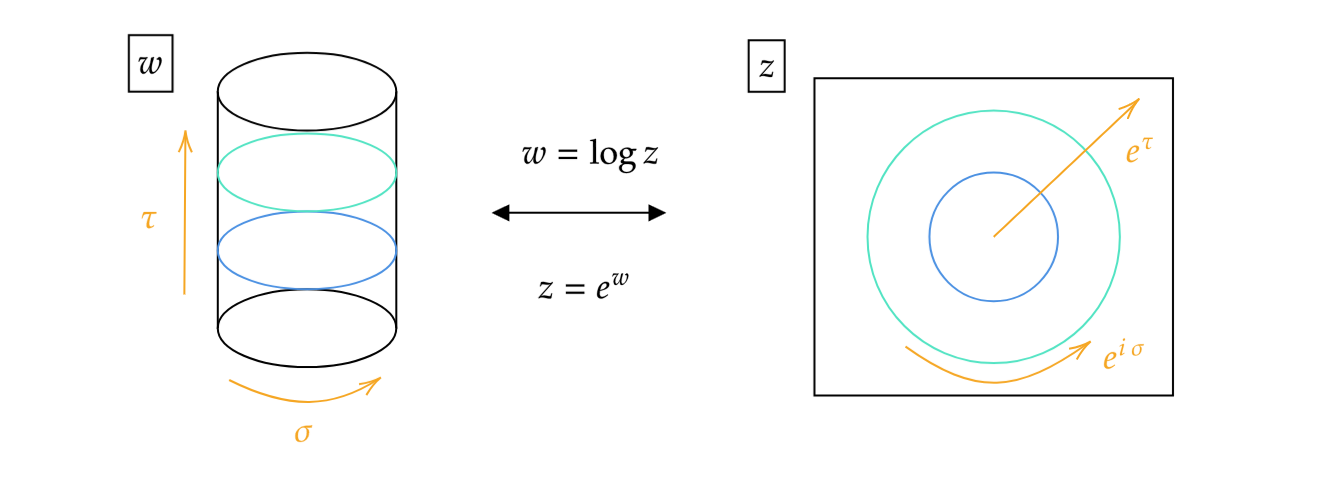}
 \caption{The Exponential map \eqref{ExpMap} maps a cylinder onto the plane. Constant time slices on the cylinder are mapped into concentric circles on the plane and time flows radially outward.}
 \label{figExpMap}
\end{figure}

In CFT a field is in fact synonymous with a state in the Hilbert space of states. To understand this, we make contact with QFT on a Minkowskian cylinder $\R\times S^1$, where $t\in\R$ parametrizes time and $\spa\in [0,2\pi)$ parametrizes space, where $\spa+2\pi\simeq \spa$. Wick-rotating to Euclidean space via $\tau=\iu t$, we form a complex coordinate $w=\tau+\iu \spa$. The exponential map 
\begin{equation}\label{ExpMap}
 z=e^w=e^{\tau+\iu \spa}
\end{equation}
explodes the cylinder onto the plane, as seen in \figref{figExpMap}. As evident from the picture, a constant time slice on the cylinder amounts to a concentric circle on the plane. Hence temporal evolution corresponds to the radial direction, which, as seen in exercise \ref{ExDilationRotation}, are generated by $L_0+\bL_0$. Because the Hamiltonian generates temporal evolution we can identify $H=L_0+\bL_0$. Note that the distant past (future), i.e. $\tau\to-\infty$ ($\tau\to+\infty$) lies at $z=0$ ($z=\infty$). 

We always quantize a QFT with respect to a constant time slice, meaning that states are associated with this time slice. In what follows, we associate states with the constant time slice in the distant past (which is more like a dot really), but states may be promoted to any other constant time slice by Hamiltonian evolution. Because constant time slices are concentric circles in our case, this proceedure is called \emphasize{radial quantization}. Let us see how this is handled.

\subsubsection*{Asymptotic In-States}
First we have to argue asymptotic in-states into existence. In order to do so, a few assumptions are customary
\begin{itemize}
 \item Taking inspiration from free field theory, we demand the existence of a vacuum state $\ket{0}$ upon which the CFT Hilbert space is constructed in a Fock-space-like manner, but more on that in \secref{secVermaConfFam}. The vacuum state $\ket{0}$ is considered a generic asymptotic in-state defined at the constant time slice at $\tau\to-\infty$ ($z\to0$).
 \item For interacting CFT, we assume that the structure of Hilbert space is the same as for free theories, but their eigenstates are allowed to be different.  
 \item Interactions fall off at $\tau=\pm\infty$ so that the asymptotic field $\lim_{\tau\to\pm\infty}\phi(\tau,\spa)$ is free.
\end{itemize}
Altogether, these assumptions provide us with a definition of arbitrary in-states within radial quantization:

\begin{tcolorbox}
The \textbf{\textit{operator-state correspondence}} relates field operators of the CFT to states in its Hilbert space
 \begin{equation}\label{instate}
 \ket{\phi_\textrm{in}}=\lim_{z,\bz\to0}\phi(z,\bz)\ket{0} \in\cH
\end{equation}
\end{tcolorbox}

\noindent \new{As the following} exercise shows, this definition indeed reduces to a single operator acting on $\ket{0}$.
\begin{exercise}\label{exFreeFieldsInState}
 When working on the cylinder, the free boson and fermion fields admit a Fourier expansion. Mapping this onto the plane via \eqref{ExpMap}, this becomes a Laurent series expansion 
 \begin{equation}
  \iu \p\varphi(z)=\sum_{n\in\Z}a_nz^{-n-1}\,,
  \qquad
  \psi(z)=\sum_{r\in\Z+1/2}\psi_rz^{-r-1/2}
 \end{equation}
Evaluate $\ket{\iu\p\varphi}:=\iu\p\varphi(0)\ket{0}$ and $\ket{\psi}:=\psi(0)\ket{0}$ and derive constraints on subsets of the $a_n$ and $\psi_r$. This naturally splits the mode operators into annihilators and creators. The remaining operator exemplifies the operators-state correspondence.  
\end{exercise}
\ifsol
\solution{exFreeFieldsInState}{
One finds $a_n\ket{0}=0$ for $n\geq0$ and $\psi_r\ket{0}=0$ for $r\geq1/2$. The states corresponding to the fields are $\ket{\iu\p\varphi}:=\iu\p\varphi(0)\ket{0}=a_{-1}\ket{0}$ and $\ket{\psi}:=\psi(0)\ket{0}=\psi_{-1/2}\ket{0}$.
}
\fi
 
\subsubsection*{Asymptotic Out-States} 
An out-vacuum is created easily by Hermitian conjugation of the vacuum state, $\bra{0}=\ket{0}^\dagger$. We would like to understand what it means to Hermitian conjugate \eqref{instate}, which requires in particular a field $\phi^\dagger(z,\bz)$. First of all, because Hermitian conjugation does not affect the temporal coordinate $t$ in Minkowski space, it indeed affects Euclidean time by reversing it $\tau=\iu t\to -\tau$. Glancing back at \eqref{ExpMap}, this induces $z\to1/z^*$. On the physical surface (see \eqref{physSurface} for an explanation) the following definition of a Hermitian conjugate field is thus sensible
\begin{equation}\label{dagger}
 \phi^\dagger(z,\bz)=\bz^{-2h}z^{2\bh}\,\phi(1/\bz,1/z)
\end{equation}
The ominous prefactors are justified by demanding that the inner product $\braket{\phi_\textrm{out}}{\phi_\textrm{in}}$ be finite and hence well-defined. Indeed,
\begin{align}
 \braket{\phi_\textrm{out}}{\phi_\textrm{in}}
 &=
 \lim_{z,\bz,w,\bw\to0}\bra{0}\phi^\dagger(z,\bz)\phi(w,\bw)\ket{0}\notag\\
 &=
 \lim_{z,\bz,w,\bw\to0}\bz^{-2h}z^{2\bh}\corr{\phi(1/\bz,1/z)\phi(w,\bw)}\notag\\
 &=
 \lim_{\xi,\bar{\xi}\to\infty}\lim_{w,\bw\to0}\xi^{2h}\bar{\xi}^{2\bh}\corr{\phi(\xi,\bar{\xi})\phi(w,\bw)}\\
 &=
 \lim_{\xi,\bar{\xi}\to\infty}\lim_{w,\bw\to0}d_{\phi\phi}\frac{\xi^{2h}}{(\xi-w)^{2h}}\frac{\bar{\xi}^{2\bh}}{(\bar{\xi}-\bw)^{2\bh}}
 \quad=
 d_{\phi\phi}
\end{align}
where we relabeled $z^{-1}=\bar{\xi}$ and $\bz^{-1}=\xi$ and employed \eqref{2pt}. Thus, the following definition of the  
\begin{tcolorbox}
\emphasize{asymptotic out-state} \new{applies}
 \begin{equation}\label{outstate}
  \bra{\phi_\textrm{out}}
  =
  \lim_{z,\bz\to0}\bra{0}\,\phi^\dagger(z,\bz)
  =
  \lim_{\xi,\bar{\xi}\to\infty}\xi^{2h}\bar{\xi}^{2\bh}\bra{0}\,\phi(\xi,\bar{\xi})
 \end{equation}
\end{tcolorbox}
Back on the cylinder this is a state placed in the distant future, i.e. $\tau\to\infty$, just opposed to the \new{in-state} which is placed at the distant past $\tau\to-\infty$.

There is in fact another way to see that \eqref{outstate} is the correct definition of an out-state based on the inversion, which is arguably more physical. It may have already struck the reader that the prefactors in \eqref{dagger} show resemblance with the prefactors $(\p\cI/\p z)^h$ a primary field picks up upon transforming under $\cI:z\mapsto1/z$. Let walk through this observation with care.

Following Ginsparg \cite{Ginsparg:1988ui}, our in-states are placed at the origin of the plane, $z\to0$, i.e. the distant past. The inversion $\cI:z\mapsto w=1/z$ maps a neighborhood of the origin on the Riemann sphere to a neighborhood of the point at $z=\infty$. Let us now declare the operator $\phi'(w,\bw)$ to be the one for which $w=0$ corresponds to $z=\infty$. This naturally suggests $\bra{\phi}=\lim_{w,\bw\to0}\bra{0}\phi'(w,\bw)$. Employing \eqref{primary} for $f(w)=1/w$ yields
\begin{equation}
 \phi'(w,\bw)=(-w^{-2})^{h}(-\bw^{-2})^{\bh}\phi(1/w,1/\bw)
\end{equation}
Combining what we have gathered, we obtain
\begin{align}
 \bra{\phi_\textrm{out}}
 &=
 \lim_{z,\bz\to0}\bra{0}\phi'(z,\bz)\notag\\
 &=
 \lim_{z,\bz\to0}\bra{0}z^{-2h}\bz^{-2\bh}\phi(1/z,1/\bz)\notag\\
 &=
 \lim_{z,\bz\to0}\bra{0}\phi^\dagger(\bz,z)\notag\\
 &=
 \bigl[\lim_{z,\bz\to0}\phi(\bz,z)\ket{0}\bigr]^\dagger\notag\\
 &=
 \ket{\phi_\textrm{in}}^\dagger\label{phiInDagger}
\end{align}
This confirms \eqref{outstate}. From now on, we drop the subscripts \quotes{in} and \quotes{out}.

\begin{exercise}\label{exFreeFieldOutStates}
 Use what you have learned in exercise \ref{exFreeFieldsInState} to evaluate  $\bra{\iu\p\varphi}$ and $\bra{\psi}$. Restrict to the physical surface \eqref{physSurface} and use the field conjugation \eqref{dagger} on $\p\varphi(z)$ and $\psi(z)$ to derive $a_n^\dagger=a_{-n}$ and $\psi_r^\dagger=\psi_{-r}$, respectively. Use this to confirm \eqref{phiInDagger} for the free boson and fermion.
\end{exercise}
\ifsol
\solution{exFreeFieldOutStates}{
$\bra{\iu\p\varphi}=\bra{0}a_1$ and $\bra{\psi}=\bra{0}\psi_{1/2}$.
For the second part we have
\begin{equation}
 (\iu\p\varphi(z))^\dagger=\sum_{n\in\Z}a_n^\dagger\bz^{-n-1}
\end{equation}
By virtue of \eqref{dagger} and $h_{\p\varphi}=1$, this equals 
\begin{equation}
 \bz^{-2}\,\iu\p\varphi(1/\bz)=\sum_{n\in\Z}a_n\bz^{n-1}\,\,\overset{n\to-n}{=}\,\,\sum_{n\in\Z}a_{-n}\bz^{-n-1}
\end{equation}
By comparison, we find $a_n^\dagger=a_{-n}$. Finally, $\bra{\iu\p\varphi}=\ket{\iu\p\varphi}^\dagger$ is obvious from the result of exercise \ref{exFreeFieldsInState}. The analysis runs through similarly for the fermion, where now $h_\psi=1/2$. }
\fi

\section{Virasoro Algebra}\label{secVirasoro}

In this section we finally dedicate some well-deserved time on the object that lies at the heart of CFT, namely the \textbf{\textit{Virasoro algebra}}. Its representations, mainly \textbf{\textit{Verma modules}}, are discussed and we learn about \textbf{\textit{fusion rules}}. 
\medskip

Before we begin though, we state a few commutator identities without derivation. Interested readers may consult section 6.1.2 of \cite{DiFrancesco}. Consider two operators which can be expressed as a contour integrals on the Riemann sphere as follows
\begin{align}
 A=\oint_0dz\,a(z),
 \qquad
 \qquad
 B=\oint_0dz\,b(z)
\end{align}
where the contour encloses the origin $z=0$. In the cases of interest to us, the operator-valued functions $a(z),b(z)$ are holomorphic in $z$, and as such the contour may be deformed arbitrarily so long as no other operator insertions are crossed. For physical interpretation the contours are taken to be circles at some radius around the origin, since, as we have seen, these correspond to constant time slices in radial quantization. We will see that operators of this type furnish Noether-like charges.

Commutators of such objects can then be evaluated according to (see for instance chapter 6 in \cite{DiFrancesco})
\begin{align}
 [A,b(w)]
 &=
 \oint_wdz\,R\bigl(a(z)b(w)\bigr)\label{commutatorAb}\\
 [A,B]
 &=
 \oint_0dw\,\oint_wdz\,R\bigl(a(z)b(w)\bigr)\label{commutatorAB}
\end{align}
The operators under the integral are \textbf{\textit{radial ordered}},
\begin{equation}
 R(a(z)b(w))
 =
 \begin{cases}
  a(z)b(w),\quad |z|>|w|\\
  b(w)a(z),\quad |w|>|z|
 \end{cases}
\end{equation}
Therefore, if $a(z)$ lies further away from the origin than $b(w)$ then it is evaluated last. In radial quantization this is equivalent to saying that $a$ is placed later in time than $b$, and thus this is nothing other than the analog of time ordering in Minkowski space.

As we will see, expressions \eqref{commutatorAb} and \eqref{commutatorAB} relate commutators to OPEs \eqref{OPE}. We will see this explicitly in examples. This is important since this permits us to translate the dynamical data and constraints following from conformal symmetry into operator language.

\subsection{The Energy-Momentum Tensor and the Virasoro Algebra}\label{secTandVir}
At long last, we are finally here! In this subsection we encounter the hallmarks of CFT in two dimensions, so pay close attention! For brevity and clarity, our discussion focuses only on the holomorphic sector. Halt your fear of missing out right there though; since the anti-holomorphic sector is treated in exact analogy, there is no need to panic. Instead, get comfortable and enjoy the ride!

First of all, recall that the energy-momentum-tensor is holomorphic, see \eqref{ETcomponents}. Hence we can expand it in terms of Laurent modes
\begin{equation}\label{Texpansion}
 T(z)=\sum_{n\in\Z}L_n\,z^{-n-2},
 \qquad
 L_n=\oint_0\frac{dz}{2\pi \iu}z^{n+1}T(z)
\end{equation}
In a quantum theory, the Laurent coefficients $L_n$ are operators on Hilbert space. While we do not know anything about their commutation relations yet, we will soon find these modes to satisfy the Virasoro algebra \eqref{Virasoro}, hence the notation. The 2 in the exponent on the LHS is actually the conformal weight $h_T=2$ of $T_{zz}=T(z)$. Indeed, recall that the conformal weight $h$ counts the number of $z$ indices. Inserting the conformal weight here secures that the expansion \eqref{Texpansion} becomes a Fourier series after mapping the plane to the cylinder via \eqref{ExpMap}. 
\begin{exercise}\label{exVirModeDagger}
 Restrict to the physical surface \eqref{physSurface} and use the field conjugation \eqref{dagger} on $T(z)$ to derive $L_n^\dagger=L_{-n}$.
\end{exercise}
\ifsol
\solution{exVirModeDagger}{On the one hand we have
\begin{equation}
 T(z)^\dagger=\sum_{n\in\Z}L_n^\dagger\bz^{-n-2}
\end{equation}
By virtue of \eqref{dagger} and $h_T=2$, this equals 
\begin{equation}
 \bz^{-4}T(1/\bz)=\sum_{n\in\Z}L_n\bz^{n-2}\,\,\overset{n\to-n}{=}\,\,\sum_{n\in\Z}L_{-n}\bz^{-n-2}
\end{equation}
The claim follows by comparison. }
\fi

The RHS of \eqref{Texpansion} is just an inversion using Cauchy's residue theorem. It is reminiscent of Noether charges familiar from QFT $Q=\int dx\, j^0$, which is evaluated at constant time for some current $j^\mu$. Let us define a general \textbf{\textit{conformal charge}}
\begin{equation}
 Q_\epsilon=\oint_0\frac{dz}{2\pi\iu}\epsilon(z)T(z)=\sum_n\epsilon_nL_n
\end{equation}
responsible for an infinitesimal coordinate change \eqref{InfConfTransf} where $\epsilon(z)=\sum_n\epsilon_nz^{n+1}$. Clearly, the $L_n$ in \eqref{Texpansion} are of the same type with $\epsilon(z)=z^{n+1}$. Thus they are responsible for transformations of the type \eqref{EllnTransf}. Naively, we may thus conclude that $L_n=\ell_n$. While this is true classically, we are interested in a quantum theory here, so let us not jump to conclusions. To make progress, we wish to derive the commutation relations of the $L_n$.

Before we can do so, however, a little more technology is required. Speaking of Noether charges, recall that they generate infinitesimal transformation of observable $A$ via $\delta A=[Q,A]$. Furtunately, you have already worked out what an infinitesimal conformal transformations looks like on a primary field, see \eqref{InfConfTransfPrimary}. This is repeated here for convenience for a purely holomorphic field ($w$ is a coordinate on the plane in the following, not the cylinder as before),
\begin{equation}\label{InfConfTransfHolPrimary}
 \delta_\epsilon\phi(w)
 =
 \bigl(h\epsilon'(w)+\epsilon(w)\p\bigr)\phi(w)
\end{equation}
On the other hand, this is supposed to be
\begin{align}\label{InfConfTransfCommutator}
 \delta_\epsilon\phi(w)
 =
 [Q_\epsilon,\phi(w)]
 =
 \oint_w\frac{dz}{2\pi\iu}\epsilon(z)\, R(\,T(z)\phi(w)\,)
\end{align}
where \eqref{commutatorAb} has been used. In order for the last two expressions to be compatible the following must hold:
\begin{tcolorbox}
 The energy-momentum tensor and a primary field satisfy the OPE
 \begin{equation}\label{TphiOPE}
  T(z)\phi(w)=\frac{h\phi(w)}{(z-w)^2}+\frac{\p_w\phi(w)}{z-w}+\dots
 \end{equation}
where the ellipses represent non-singular orders in $(z-w)$. Descendant fields have terms with higher singularities in $(z-w)$. 

The second order pole codes for the behavior of conformal fields under scaling and the first order pole codes for behavior under translations.
\end{tcolorbox}
Here and in the following, we omit the radial ordering symbol $R$, as is customary in the literature. It is stressed though that it is always implied in OPEs. Moreover, and this is not evident from our derivation here, OPEs are meant as operator identities \textit{valid inside correlation functions}. This ties in with the fact that, ultimately, physics is read off from correlators, not operators. 

\begin{exercise}\label{exFreeFieldTphiOPE}
 The result of exercise \ref{exFreeField2pt} yields the following Wick contractions
 \begin{equation}\label{contractions}
 \wick{
 \p\c 1\varphi(z) \p\c1\varphi(w)}
 =
 \frac{d_{\p\varphi}}{(z-w)^2}
\qquad
\wick{\c1\psi(z)\c1\psi(w)}
=
\frac{d_\psi}{z-w}
 \end{equation}
where the constants $d_{\p\varphi}$ and $d_{\psi}$ are the normalization constants appearing in \eqref{2pt}. Employ these to demonstrate that $\p\varphi$ and $\psi$ are primary fields and derive their conformal weights by use of the OPE \eqref{TphiOPE}. The energy-momentum tensors are, respectively,
\begin{equation}\label{freeFieldsT}
 T_\varphi(z)=\frac{1}{2d_{\p\varphi}}:\p\varphi\p\varphi:(z),
 \qquad
 T_\psi(z)=-\frac{1}{2d_{\psi}}:\psi\p\psi:(z),
\end{equation}
where the colons denote normal ordering. Be aware of the Pauli principle in the fermionic case. 
\end{exercise}
\ifsol
\solution{exFreeFieldTphiOPE}{
\begin{align}
 \wick{
\c1T(z)\p\c1\varphi(w)
}
&=
\frac{1}{2d_{\p\varphi}}2\frac{\p\varphi(z)d_{\p\varphi}}{(z-w)^2}
=
\frac{\p\varphi(w)}{(z-w)^2}+\frac{\p^2_w\varphi(w)}{z-w}\\
\wick{
\c1T(z)\c1\psi(w)
}
&=
-\frac{1}{2d_{\psi}}\left[-\frac{\p\psi(z)d_\psi}{z-w}-\frac{\psi(z)d_\psi}{(z-w)^2}\right]
=
\frac{\frac{1}{2}\psi(w)}{(z-w)^2}+\frac{\p_w\psi(w)}{z-w}
\end{align}
In the bosonic, to reach the first equality, two identic Wick contractions are carried out, leading to the 2 in the numerator. In going to the first equality in the fermionic case, the first term required moving $\psi$ past $\p\psi$, thereby picking up a sign. The second term here requires a derivative $\p_z$ of the fermionic contraction in \eqref{contractions}, leading to another sign. In reaching the last equalities in both cases, fields depending on $z$ were Taylor expanded around $w$.}
\fi

If the field $\phi$ is taken to be quasi-primary, the OPE $T(z)\phi(w)$ can have higher order singularities, except for a third order pole. Indeed, for global conformal transformations the function $\epsilon(w)$ is at most quadratic in $w$, thus we obtain information on the poles of first, second and third order only. 

Equation \eqref{TphiOPE} is amongst the most important OPEs in CFT. Commit it to memory. The single most important OPE is
\begin{tcolorbox}
 The OPE of the energy-momentum tensor with itself is
 \begin{equation}\label{TTOPE}
  T(z)T(w)=\frac{\cc/2}{(z-w)^4}+\frac{2T(w)}{(z-w)^2}+\frac{\p T(w)}{z-w}+\dots
 \end{equation}
where the ellipses represent non-singular orders in $(z-w)$.
\end{tcolorbox}
That this is indeed correct is argued as follows. The last two terms are, as above, just the behavior of a conformal field with weight $h=2$. The fourth order pole is allowed because of the existence of a field with $h=0$, namely $\id$. The particular choice of the proportionality factor $\cc/2$ is such that the free boson theory has $\cc=1$. For higher order poles, we would require fields of negative conformal weight, which are absent in unitary theories. This leaves us with a potential third order pole, which is forbidden however by Bose symmetry $T(z)T(w)=T(w)T(z)$. Observe that the energy-momentum tensor is not a primary field itself, unless $\cc=0$. It is quasi-primary as seen by the absence of a third order pole.

\begin{exercise}\label{exVirTTOPE}
 Using \eqref{commutatorAB} and \eqref{Texpansion} show that \eqref{TTOPE} is equivalent to the Virasoro algebra
 \begin{equation}\label{VirasoroAgain}
 [L_n,L_m]=(n-m)L_{n+m}+\frac{\cc}{12}n(n^2-1)\delta_{n+m,0}
\end{equation}
\end{exercise}
\ifsol
\solution{exVirTTOPE}{
Application of the rhs in \eqref{Texpansion} and \eqref{commutatorAB} yields
\begin{align}
 [L_n,L_m]
 &=
 \oint_0\frac{dw}{2\pi\iu}\oint_w\frac{dz}{2\pi\iu}z^{n+1}w^{m+1}T(z)T(w)\notag\\
 &=
 \oint_0\frac{dw}{2\pi\iu}\oint_w\frac{dz}{2\pi\iu}z^{n+1}w^{m+1}\left[\frac{\cc/2}{(z-w)^4}+\frac{2T(w)}{(z-w)^2}+\frac{\p T(w)}{z-w}\right]
\end{align}
Regular terms in $(z-w)$ drop out, because they have no residue. Hence, the calculation is exact, despite our ignorance of regular terms in \eqref{TTOPE}. Employ the expansion
\begin{align}\label{ZatW}
 z^{n+1}
 &=
 (w+(z-w))^{n+1}\\
 &=
 w^{n+1}+(n+1)w^n(z-w)
 +\frac{(n+1)n}{2}w^{n-1}(z-w)^2
 +\frac{(n+1)n(n-1)}{6}w^{n-2}(z-w)^3\notag
\end{align}
in the commutator and integrate out $z$ to find
\begin{align}
 [L_n,L_m]
 &=
 \oint_0\frac{dw}{2\pi\iu}
 \left[
 \frac{\cc\, n(n^2-1)}{12}w^{m+n-1}
 +
 2(n+1)w^{m+n+1}T(w)
 +
 w^{n+m+2}\p T(w)
 \right]
\end{align}
The last term requires a partial integration, $w^{n+m+2}\p T(w)\to-(n+m+2)w^{n+m+1}T(w)$. Integrating out $w$ returns the Virasoro algebra \eqref{VirasoroAgain}.}
\fi

Hence, the $L_n$ cannot simply be the generators $\ell_n$ of the Witt algebra, as naively hypothesized above. You may now wonder if $\cc$ is indeed non-vanishing. It turns out that one need not look far to find such examples. Even real free bosons and real free fermions have have non-vanishing $\cc$, as you will check shortly. This gives weight to the claim made above, that the central charge is not some pecularity of some particular CFT, but a very generic feature of a QFT with conformal symmetry.  
\begin{exercise}\label{exfreeFieldsTTOPE}
 Evaluate the $TT$ OPE for the real massless free boson $\varphi$ and fermion $\psi$, whose energy-momentum tensors are given in \eqref{freeFieldsT} and show that their central charges are $\cc_\varphi=1$ and $\cc_\psi=1/2$, respectively. Convince yourself that the central term in the OPE \eqref{TTOPE} stems from double Wick contractions. The central charge is thus an inherently quantum effect, as expected of an anomaly.
\end{exercise}
\ifsol
\solution{exfreeFieldsTTOPE}{
\begin{align}
 \wick{\c1T_\varphi(z)\c1T_\varphi(w)}
 &=
 \frac{1}{(2d_{\p\varphi})^2}\left[2\left(\frac{d_{\p\varphi}}{(z-w)^2}\right)^2+4\frac{:\p\varphi(z)\p\varphi(w):\,d_{\p\varphi}}{(z-w)^2}\right]\notag\\
 &=
 \frac{1/2}{(z-w)^4}+\frac{2T_\varphi(w)}{(z-w)^2}+\frac{\p T_\varphi(w)}{z-w}
\end{align}
The first term after the first equality results from two double Wick contractions and the second term from 4 single contractions. In going to the last line, $\p\varphi(z)=\p\varphi(w)+\frac{1}{2}\p\varphi(w)(z-w)+\dots$ and $\p^2\varphi(w)\p\varphi(w)=\frac{1}{2}\p(\p\varphi(w)\p\varphi(w))$ were employed. The free boson central charge is thus $\cc=1$. The following are useful for the free fermion case
\begin{equation}
 \wick{\p\c1\psi(z)\c1\psi(w)}
 =
 -\frac{d_\psi}{(z-w)^2}\,,
 \quad
 \wick{\c1\psi(z)\p\c1\psi(w)}
 =
 \frac{d_\psi}{(z-w)^2}\,,
 \quad
 \wick{\p\c1\psi(z)\p\c1\psi(w)}
 =
 -\frac{2d_\psi}{(z-w)^3}\,.
\end{equation}
When contracting, one has to watch out for minus signs picked up when passing fermions passed each other,
\begin{align}
 \wick{\c1T_\psi(z)\c1T_\psi(w)}
 &=
 \frac{1}{4d_\psi^2}\biggl[
 \frac{(2-1)d_\psi^2}{(z-w)^4}
 +
 \frac{2d_\psi:\psi(z)\psi(w):}{(z-w)^3}\notag\\
 &\qquad\qquad+
 \frac{d_\psi:(\p\psi(z)\psi(w)-\psi(z)\p\psi(w)):}{(z-w)^2}
 -
 \frac{d_\psi:\p\psi(z)\p\psi(w):}{z-w}
 \biggr]\notag\\
 &=
 \frac{1/4}{(z-w)^4}+\frac{2T_\psi(w)}{(z-w)^2}+\frac{\p T_\psi(w)}{z-w}
\end{align}
After the first equality, the first term is due to the double contractions, while the remaining terms are single contractions. We have proceeded as for the boson afterward. Keep in mind, the Pauli principle forces $\psi(w)^2=(\p\psi(w))^2=0$ and $\p(\psi(w)\p\psi(w))=\psi(w)\p^2\psi(w)$. The free fermion has therefore a central charge $\cc=1/2$.}
\fi

While \eqref{TTOPE} encodes the behavior of $T(z)$ under infinitesimal conformal transformations, it is useful to infer its behavior under a \quotes{normal sized} conformal transformation. To this end consider
\begin{align}
 \delta_\epsilon T(w)
 =
 [Q_\epsilon,T(w)]
 &=
 \oint_w\frac{dz}{2\pi\iu}\epsilon(z)T(z)T(w)\notag\\
 &=
 \frac{\cc}{12}\p^3\epsilon(w)+2T(w)\p\epsilon(w)+\epsilon(w)\p T(w)
\end{align}
This is integrated for large $\epsilon$ to
\begin{equation}\label{Ttransform}
T'(z)
=
U_fT(z)U_f^{-1}
=
\left(\frac{\p f}{\p z}\right)^2T(f(z))+\frac{\cc}{12}S(f,z)
\end{equation}
where the Schwarzian derivative has been introduced,
\begin{equation}\label{Schwarzian}
 S(f,z)=\frac{f'f'''-\frac{3}{2}(f'')^2}{(f')^2}, 
 \qquad
 S(f,z)=-\left(\frac{\p f}{\p z}\right)^2S(z,f)
\end{equation}
and primes indicate derivates with respect to $fz$. It is the unique derivative vanishing on Möbius transformations. Hence, $T(z)$ is quasi-primary, as for $f\in\PSL(2,\C)$ it transforms like a primary \eqref{primary}.

\subsection{Verma Modules and Conformal Families}\label{secVermaConfFam}
Now that we have access to the symmetry algebra governing our physical system, we are in a position to discuss some general aspects of the ensuing structure of Hilbert space. 

Global conformal invariance imposes that 
\begin{equation}
 L_{0}\ket{0}=L_{\pm1}\ket{0}=0
\end{equation}
Given that the energy-momentum tensor contains these three modes -- recall \eqref{Texpansion} -- we see however that this does not suffice to have $T(z)$ produce a well-behaved state via the operator-state correspondence \eqref{instate}. Indeed, for $\ket{T}=T(0)\ket{0}$ to be well-defined, we are forced to accept a much larger set of constraints on the vacuum, namely
\begin{equation}
 L_n\ket{0}=0
 \qquad
 \textrm{for }n\geq-1 
\end{equation}
This secures $\bra{0}T(z)\ket{0}=0$, which had been stated already in \eqref{1pt} in CFT for any field besides the identity. The identity field is but one of many primary fields however and in order to find out how they behave under application of the $L_n$s we need the following exercise
\begin{exercise}\label{exVirHWS}
 Show that 
 \begin{equation}\label{LnPrimaryCommutator}
  [L_n,\phi(z)]=h(n+1)z^n\phi(z)+z^{n+1}\p\phi(z)
 \end{equation}
What is the second term reminiscent of? What does the first term thus encode? Apply this relation to the asymptotic state $\ket{\phi}=\lim_{z\to0}\phi(z)\ket{0}$ to derive
\begin{equation}\label{primaryState}
 L_0\ket{\phi}=h_\phi\ket{\phi},
 \qquad
 L_n\ket{\phi}=0,
 \quad \textrm{for }n\geq1
\end{equation}
This establishes primaries as highest weight states of the Virasoro algebra. Note that if $\phi$ was not primary, additional singular terms in the OPE with $T$ would make the action of some of the $L_{n>0}$ non-vanishing.
\end{exercise}
\ifsol
\solution{exVirHWS}{
Using \eqref{commutatorAb} and \eqref{TphiOPE} we have
\begin{align}
 [L_n,\phi(w)]
 &=
 \oint_w\frac{dz}{2\pi\iu}z^{n+1}T(z)\phi(w)\notag\\
 &=
 \oint_w\frac{dz}{2\pi\iu}\left(w^{n+1}+(n+1)w^n(z-w)\right)\left(\frac{h\phi(w)}{(z-w)^2}+\frac{\p_w\phi(w)}{z-w}\right)\notag\\
 &=
 h(n+1)w^n\phi(w)+w^{n+1}\p\phi(w)
\end{align}
where the expansion \eqref{ZatW} was used to second order. We recognize the action of the Witt generators \eqref{WittGenerator} in the second term. This term is therefore responsible for the spacetime transformation. The first term reflects thus the representation of the conformal algebra that $\phi$ sits in, labeled by $h$. This is made more precise by
\begin{align}
 L_0\ket{\phi}
 &=
 \lim_{z\to0}[L_0,\phi(z)]\new{\ket{0}}
 =
 \lim_{z\to0}h\phi(z)\new{\ket{0}}
 =
 h\ket{\phi}\\
 n>0:\quad
 L_n\ket{\phi}
 &=
 \lim_{z\to0}[L_n,\phi(z)]\new{\ket{0}}
 =
 \lim_{z\to0}\left(z^nh\phi(z)+z^{n+1}\p\phi(z)\right)\new{\ket{0}}
 =
 0
\end{align}
}
\fi

These primary states lend themselves now for the construction of \quotes{Fock spaces} by application of Virasoro modes with negative index,
\begin{equation}\label{descendant}
 L_{-k_1}L_{-k_2}\dots L_{-k_n}\ket{\phi},
 \qquad
 1\leq k_1\leq\dots\leq k_n\,.
\end{equation}
The ordering of increasing $k_i$ is a convention. These states are the \textbf{\textit{descendant states}} that we have come across above already. Here we have finally encountered their concrete form. Observe that $\ket{T}\equiv\lim_{z\to0}T(z)\ket{0}=L_{-2}\ket{0}$, and so the energy-momentum tensor is confirmed to be a descendant. In a scenario where we only have access to the global conformal group, which excludes $L_{-2}$, the energy-momentum tensor is seen to be its own (quasi-)primary however.

Like any QFT, CFTs have an infinite amount of states. The organizing principle for CFTs is simple however. Any state that is not primary is descendant, and so we need (mostly) only know the primaries in a model. These correspond to highest weight representations, which in turn bounds the energy from below. 

Given that $[L_0,L_n]=-nL_n$, such a descendant state has $L_0$ eigenvalue
\begin{equation}
 h=k_1+k_2+\dots k_n+h_\phi=N+h_\phi
\end{equation}
where $N$ is called the level of the descendant. We already know that the $L_n$ generate infinitesimal conformal transformations. Hence the set of all descendants is the entire orbit of a primary state under conformal transformations. Note that this never transforms a given primary into a distinct primary. Hence these are representations, or more precisely, a \textit{module} of the Virasoro algebra. These will not be irreducible representations in general however. Reducibility is indicated by the presence of null vectors, but we leave that to \secref{secCharactersNullVec}. In general, the space spanned by all possible descendant states \eqref{descendant}, reducible or not, is called a \textbf{\textit{Verma module}} $V_{h_i}$.

For now, let us observe that the number of linearly independent states at level $N$ is given by the partitions $p(N)$ of $N$, i.e. the way an integer $N$ can be split into sum of positive integers. For instance, 3 can be split into 1+1+1, 1+2 and 3 itself, giving $p(3)=3$. Bases for the first five levels are for instance given by 
\medskip

\noindent
\renewcommand{\arraystretch}{1.5}
 \begin{tabular}{p{3.5cm} p{8.5cm} c}
 \hline\hline
 $L_0$ eigenvalue& Basis vectors & $p(N)$\\
 \hline\hline
 $h$  & $\ket{\phi}$  &1\\
 \hline
 $h+1$ & $L_{-1}\ket{\phi} $  & 1 \\
 \hline
 $h+2$ & $L_{-1}^2\ket{\phi},\,L_{-2}\ket{\phi}$ & 2 \\
 \hline
 $h+3$ & $L_{-1}^3\ket{\phi},\,L_{-1}L_{-2}\ket{\phi},\,L_{-3}\ket{\phi}$ & 3\\
 \hline
 $h+4$ & $L_{-1}^4\ket{\phi},\,L_{-1}^2L_{-2}\ket{\phi},\,L_{-2}^2\ket{\phi},\,L_{-1}L_{-3}\ket{\phi},\,L_{-4}\ket{\phi}$ & 5\\
 \hline
 \end{tabular}\medskip
 
 In order to count all possible descendant states in a Verma module, it is convenient to introduce a book-keeping parameter $q^{N}$ and count all states at level $N$
 \begin{equation}\label{countingStates}
  \sum_{N=0}^\infty p(N)q^N=\prod_{n=1}^\infty\frac{1}{1-q^n}
 \end{equation}
The equality can be verified by Taylor expansion. The subspace of the entire Hilbert space spanned by a primary state and all its descendant states is often called a \textbf{\textit{conformal tower}} or
\begin{tcolorbox}
\hypersetup{linkcolor=\boxlinkcolor}
 A \textbf{\textit{conformal family $[\phi]$}} encompasses a primary $\phi$ and all of its descendants \eqref{descendant}. Members of a single family transform amongst themselves under conformal transformations. In other words, the OPE of $T(z)$ with $\phi(w)$ consists only of members of $[\phi]$. We label the set of all conformal families in the full Hilbert space by $\confFam$. Given some $i\in\confFam$ we denote a family by $[\phi_i]$ and when no confusion can arise we simply write $[i]$.
\end{tcolorbox}
In \new{these lectures} we restrict to conformal symmetry. When extended symmetries are present in the system, such as Kac-Moody symmetry or supersymmetry, it is useful use their families instead. The idea is the same. The extended symmetry algebra will have a set of generators, some of which are annihilators and some of which are creators. The ladder create the descendants of the extended symmetry algebra. A family is again a primary and all of its descendants. 

Consider the Ising CFT as an example. It has three primary fields $\id,\,\varepsilon,\,\sigma$. Any other state in this theory is a descendant thereof. The families are thus $\confFam=\{[\id],\,[\varepsilon],\,[\sigma]\}$. When labelling conformal families, the brackets $[\cdot]$ are sometimes omitted when no potential confusion can arise. 

It can happen here that a Virasoro primary is a descendant with respect to the larger symmetry algebra. We have in fact already encountered such an example in exercise \ref{exFreeFieldsInState}, where we found $\ket{\iu\p\varphi}=a_{-1}\ket{0}$. The mode $a_{n<0}$ are the creators of a $U(1)$ Kac-Moody symmetry. While we have seen that $\p\varphi$ is its own Virasoro primary, i.e. $\p\varphi\in[\p\varphi]_{\vir}$, it is a descendant of the identity when working with $U(1)$ Kac-Moody symmetry, , i.e. $\p\varphi\in[\id]_{U(1)}$.

\subsection{Fusion}\label{secFusion}
In general, the OPE of two distinct conformal families contains any number of other conformal families. This is in essence the statement of the OPE \eqref{OPE}. Because the OPE contains the data of the three-point function \eqref{3pt}, it tells us how strongly two primaries couple to a third, which is encoded in the three-point coefficient $C_{ijk}$. For some purposes, it already suffices to know, however, which conformal families simply appear in an OPE, not how strongly they couple. This is encapsulated in the following notion:
\begin{tcolorbox}
\hypersetup{linkcolor=\boxlinkcolor}
Given two conformal families, the \emphasize{fusion rules} encode which conformal families occur in their OPE \eqref{OPE}
\begin{equation}\label{fusionRules}
 [\phi_i]\fuse[\phi_j]=\sum_{k\in \confFam}\fus_{ij}^k\,[\phi_k], \qquad \fus_{ij}^k\in\N_0
\end{equation}
The structure constants $\fus_{ij}^k$ are called the \emphasize{fusion coefficients}. The family of the identity field $[\phi_0]=[\id]$, denoted $i=0$, presents the unique unit element of this multiplication rule, i.e. $\fus_{0i}^k=\delta_i^k$. Fusion is commutative, $\fus_{ij}^k=\fus_{ji}^k$ and associative,
\begin{equation}\label{fusionAssociativity}
 \sum_{m\in \confFam}\fus_{im}^l\fus_{jk}^m
 =
 \sum_{n\in \confFam}\fus_{ij}^n\fus_{nk}^l
\end{equation}
By construction, $C_{ij}{}^k=0\Leftrightarrow\fus_{ij}^k=0$.
\end{tcolorbox}
Since the energy momentum tensor is a descendant of the identity, i.e. $T\in[\id]$, this formalizes our statement that the OPE of $T$ with $\phi$ remains in the family $[\phi]$.

The fusion rules are a version of a tensor product of representations. However, it is not the standard version that we know and love in group theory, as that construction would have central charges add up, and we wish to avoid that. Mathematical details can be found in \cite{Gaberdiel:1993td, moore1990lectures, Recknagel:2013uja} and references therein.
\begin{exercise}\label{exFusionRules}
 Read off the fusion rules underlying the OPEs \eqref{bosonOPE}, \eqref{fermionOPE}, \eqref{IsingOPE}, \eqref{TphiOPE} and \eqref{TTOPE}. The Ising fields in \eqref{IsingOPE} are conformal primaries.
\end{exercise}
\ifsol
\solution{exFusionRules}{The following fusion rules for conformal families are read off.
\begin{itemize}
 \item The free boson OPE \eqref{bosonOPE} gives $[\p\varphi]\fuse[\p\varphi]=[\id]$.
 \item The free fermion OPE \eqref{fermionOPE} gives $[\psi]\fuse[\psi]=[\id]$.
 \item The Ising OPEs \eqref{IsingOPE} give
 \begin{equation}\label{IsingFusionOPEex}
  [\varepsilon]\fuse[\varepsilon]=[\id]\,,
  \qquad 
  [\varepsilon]\fuse[\sigma]=[\sigma]\,,
  \qquad 
  [\sigma]\fuse[\sigma]=[\id]+[\varepsilon]\,,
  \qquad 
 \end{equation}
 \item The OPE \eqref{TphiOPE}, which defines a primary gives $[\id]\fuse[\phi]=[\phi]$.
 \item The conformal algebra OPE \eqref{TTOPE} gives $[\id]\fuse[\id]=[\id]$.
\end{itemize} 
}
\fi

The \emphasize{charge conjugate family} $i^+$ of a family $i$ is singled out by $\fus_{ij}^0=\delta_{j,i^+}$. The family of the identity field is self-conjugate. Charge conjugation provides an involution of the fusion rules, $[\phi_i]\to[\phi_{i^+}]$. The fusion coefficients are affected by charge conjugation in the following way
\begin{equation}\label{FusionChargeConjugation}
 \fus_{i^+j}^k=\fus_{ik}^j
 \qquad 
 \textrm{and}
 \qquad
 \fus_{i^+j^+}^{k^+}=\fus_{ij}^k\,.
\end{equation}
\begin{exercise}\label{exChargeConjFusion}
 Convince yourself that $([i]\fuse[j])^+=[i^+]\fuse[j^+]$.
\end{exercise}
\ifsol
\solution{exChargeConjFusion}{
\begin{align}
 ([i]\fuse[j])^+
 =
 \sum_k\fus_{ij}^k[k^+]
 =
 \sum_k\fus_{ij}^{k^+}[k]
 =
 \sum_k\fus_{i^+j^+}^{k}[k]
 =
 [i^+]\fuse[j^+]
\end{align}
The second step relabels the summation $k\to k^+$ and the third step uses the right entry of \eqref{FusionChargeConjugation}.}
\fi

It is customary to phrase the fusion coefficients via the $\confFam\times\confFam$ \emphasize{fusion matrices} $\fus_i$
\begin{equation}\label{fusionMatrix}
 \left(\fus_i\right)_{jk}=\fus_{ij}^k,
 \qquad
 \fus_{i^+}=\fus_i^\transpose
\end{equation}
where $\transpose$ denotes matrix transposition. 
\begin{exercise}\label{exFusionRulesRegRep}
 Show that the fusion matrices furnish the regular representation of the fusion rules. 
 \begin{equation}\label{fusionRulesRegularRep}
  \fus_i\fus_j=\sum_{k\in\confFam}\fus_{ij}^k\,\fus_k
 \end{equation}
\end{exercise}
\ifsol
\solution{exFusionRulesRegRep}{
The associativity condition \eqref{fusionAssociativity} can directly be written as $(\fus_j\fus_i)_{kl}=\sum_n\fus_{ji}^n(\fus_n)_{kl}$.
}
\fi

Commutativity of fusion, i.e. $\fus_{ij}^k=\fus_{ji}^k$ is naturally passed on to the fusion matrices, in particular $[\fus_i,\fus_{i^+}]=0$. Because of $\fus_{i^+}=\fus_i^\transpose$, this means that $\fus_i$ is normal and is therefore diagonalized by a unitary matrix. Later on in \eqref{Verlinde}, we will learn which matrix this is, and that it is in fact a central player in CFT. For now, this guarantees that all fusion matrices $\{\fus_i\}$ can be diagonalized  simultaneously. Pick an $|\confFam|$-dimensional eigenvector $\textbf{v}=(v_1,\dots,v_{|\confFam|})$,
\begin{equation}\label{FusionEigensystem}
 \fus_i\,\textbf{v}=n_i\,\textbf{v},
 \qquad\Leftrightarrow\qquad
 \sum_k\fus_{ij}^k\,v_k=n_i\,v_j
\end{equation}
\new{By applying \eqref{fusionRulesRegularRep} to the eigenvector $\textbf{v}$ from the left}, we easily find that the eigenvalues $n_i$ for the eigenvector $\textbf{v}$ furnish one-dimensional irreducible representations of the fusion rules \eqref{fusionRules}
\begin{equation}
 n_in_j=\sum_k\fus_{ij}^kn_k
\end{equation}
Clearly this equation holds for the eigenvalues of any eigenvector of $\fus_i$. The eigenbasis is $|\confFam|$-dimensional and obviously it is shared amongst all the fusion matrices $\{\fus_i\}$ since they commute with one another. Hence we could add a label $a=1,\dots,|\confFam|$ to the eigensystem, $\textbf{v}\to\textbf{v}_a$ and $n_i\to n_{ia}$. We will not do this now, but anticipate that this plays a role in the important relation \eqref{Verlinde}.

Invoking the Perron-Frobenius theorem of mathematics, which deals with eigensystems of matrices with non-negative entries, we learn that there exists a unique maximal eigenvalue $\qdim_i\in\R_{>0}$ of $N_i$ belonging to an eigenvector $\hat{\textbf{v}}$ with strictly positive entries. The eigenvalues $\qdim_i$ are called \emphasize{quantum dimensions} and play a central role in the study of topological phases of matter. Being an eigenvalue of $\fus_i$, they satisfy the fusion rules \eqref{PerronFrobeniusVector}, i.e. $\qdim_i\qdim_j=\sum_k\fus_{ij}^k\qdim_k$.

Given the symmetry $\fus_{ij}^k=\fus_{ji}^k$, we can evaluate the entries of $\hat{\textbf{v}}$. Indeed, 
\begin{equation}\label{PerronFrobeniusVector}
 \sum_k\fus_{ij}^k\,\hat{v}_k
 =
 \qdim_i\,\hat{v}_j
 \overset{!}{=}
 \sum_k\fus_{ji}^k\,\hat{v}_k
 =
 \qdim_j\,\hat{v}_i
 \qquad
 \Rightarrow
 \qquad
 \hat{v}_i=\qdim_i
\end{equation}
Hence the Perron-Frobenius vector $\hat{\textbf{v}}=(\qdim_1,\qdim_2,\dots,\qdim_{|\confFam|})$ is filled with the maximal eigenvalues of all fusion matrices $\fus_i$. Because all fusion matrices $\{\fus_i\}$ share an eigenbasis, they evidently also share $\hat{\textbf{v}}$.

\section{The Hilbert Space of a CFT}\label{secHilbertSpace}
In this section we discuss the structure of the state space in a CFT. This will require a discussion of \emphasize{conformal characters, CFTs on a torus} and \emphasize{modular invariance}. We will touch upon \emphasize{null vectors}.

\subsection{Conformal Characters and Null Vectors}\label{secCharactersNullVec}
Just as with any other symmetry, Hilbert space will fall into irreducible representations of the Virasoro algebra. In this subsection we develop an idea of what that means. 

We have already seen that a conformal representation (conformal family) is built upon a primary state $\ket{\phi}$ as in \eqref{descendant}. It may happen that one of its descendant states, let us call it $\ket{\psi}$, usually a linear combination of states \eqref{descendant} at fixed level, is itself primary again, i.e. it satifsfies \eqref{primaryState}. These are called \emphasize{null vectors} or \emphasize{singular vectors}.
\begin{exercise}\label{exNullVecOrthogonal}
 Using $L_n^\dagger=L_{-n}$ to show that null vectors stand orthogonal to all other states in the Verma module $V_{h_\phi}\equiv V_\phi$.
\end{exercise}
\ifsol
\solution{exNullVecOrthogonal}{The claim follows immediately from \eqref{descendant} and Hermitian conjugation,
\begin{equation}
 \bra{\psi}L_{-k_1}L_{-k_2}\dots L_{-k_n}\ket{\phi}
 =
 \bra{\phi}L_{k_1}L_{k_2}\dots L_{k_n}\ket{\psi}^*
 =
 0\,,
 \quad
 k_i>0
\end{equation}
}
\fi

Because such states do not \quotes{talk} to the remainder of the Verma module $V_\phi$, they can be safely ignored. More mathematically rigorously, what is happening is that the null vector $\ket{\psi}$ and all its descendant states form their own invariant subspace, i.e. their own Verma module $V_\psi$, within the Verma module $V_\phi$. We have hence found that the Verma module, as a representation of the Virasoro algebra, is reducible, and in order to reach its irreducible core, the null vector $\ket{\psi}$ and its submodule $V_\psi$ are quotiented out of the Verma module. In practice, we simply set $\ket{\psi}=0$, which also rids us of all descendants of $\ket{\psi}$. Clearly, when several null vectors are present in $V_\phi$, all of them are quotiented away in this manner. In the following, the irreducible core of a Verma module $V_\phi$ is denoted $\cH_\phi$.

From this discussion, null vectors may sound very exotic. They are not difficult to come by, however. In fact, we have already encountered a null vector: the requirement that the ground state $\ket{0}$ be invariant under the conformal group, in particular translations, forces $L_{-1}\ket{0}=0$. Clearly this null vector carries important physical data. More generally, null vectors impose powerful physicality constraints on correlation functions. We will have no time discuss this unfortunately. Any real aspirant of the conformal arts is advised however to delve into this very important feature of CFT in the literature, for instance the resources in \secref{secIntroCFT}.

In order to count states in an irreducible representation $\cH_i$ for a primary $\phi_i$, it is useful to introduce a \quotes{mini partition function},
\begin{equation}\label{character}
 \chi_i(q)=\tr_{\cH_i}\left[q^{L_0-\cc/24}\right],
 \qquad
 i\in\confFam
\end{equation}
This is called the \emphasize{character of the family $i$}. The presence of $\cc/24$ is ad hoc for now, and we justify this convention in hindsight later.
When counting states in a Verma module in \eqref{countingStates}, we have basically already computed the character of a Verma module $V_{h_i}$,
\begin{equation}
 \chi_{V_i}(q)
 =
 \tr_{V_i}\left[q^{L_0-\cc/24}\right]
 =
 q^{h_i-\cc/24}\sum_{N=0}^\infty p(N)q^N
 =
 q^{h_i-\frac{\cc-1}{24}}\,\eta(q)^{-1}
\end{equation}
where the \emphasize{Dedekind $\eta$ function} was introduced,
\begin{equation}\label{Dedekind}
 \eta(q)=q^{\frac{1}{24}}\prod_{n=1}^\infty(1-q^n)
\end{equation}
Rewriting characters so as to include $\eta(q)$ is recommended, since this function transforms conveniently under modular transformations, which we will get to soon.

Let us now investigate the effect of null vectors on characters using $L_{-1}\ket{0}$ as guinea pig. The character of the Verma module built over the vacuum state $\ket{0}$ and $\ket{\psi}=L_{-1}\ket{0}$ are, respectively,
\begin{equation}
 \chi_{V_0}=\frac{q^{-(\cc-1)/24}}{\eta(q)},
 \qquad
 \chi_{V_\psi}=\frac{q^{1-(\cc-1)/24}}{\eta(q)}
\end{equation}
The procedure of quotienting $V_0$ by $V_\psi$ is reflected in the counting of states simply by subtraction 
\begin{equation}
 \chi_{V_0}-\chi_{V_\psi}
 =
 \frac{q^{-(\cc-1)/24}}{\eta(q)}(1-q)
\end{equation}
At central charges $\cc>1$, and admitting only unitary representations of the Virasoro algebra, $L_{-1}\ket{0}$ is the only null vector found in the Verma module of the vacuum $\ket{0}$. In this case, the remaining module is the irreducible, i.e. it is indeed $\cH_0$ and $\chi_0=\chi_{V_0}-\chi_{V_\psi}$ is its character. When $\cc\leq1$ more null vectors may be present in $V_0$, and they all need to be removed to get to the irreducible module $\cH_0$. 

This pattern is followed when quotienting Verma modules of arbitrary conformal weights $h_\phi$ and $h_\psi$, 
\begin{equation}
 \chi_{V_\phi}-\chi_{V_\psi}
 =
 \frac{q^{-(\cc-1)/24}}{\eta(q)}(q^{h_\phi}-q^{h_{\psi}})
\end{equation}
and this has to be carried out with every null vector in a Verma module, so that in spirit
\begin{equation}
 \chi_i
 =
 \chi_{V_i}-\sum_k\chi_{V_k}
\end{equation}
where $k$ runs over all null-vectors in $V_i$. This relation is actually not always the end of the story.  It needs to be refined when descendant spaces of null vectors overlap, \new{which is in fact a common feature. The structure carried by irreducible unitary representations of the Virasoro algebra was unravelled in \cite{kac1979contravariant,feigin1982invariant} and their characters were computed in \cite{rocha1985vacuum}.} A complete discussion of irreducible representations in CFT is found in chapters 7 and 8 of \cite{DiFrancesco}. Here, we content ourselves with presenting the characters of the Ising model as example,
\begin{subequations}\label{IsingCharacters}
\begin{align}
 \chi_0(q)
 &=
 \frac{1}{2}\left(\sqrt{\frac{\new{\vartheta_3(q)}}{\eta(q)}}+\sqrt{\frac{\new{\vartheta_4(q)}}{\eta(q)}}\right)\\
 \chi_\varepsilon(q)
 &=
 \frac{1}{2}\left(\sqrt{\frac{\new{\vartheta_3(q)}}{\eta(q)}}-\sqrt{\frac{\new{\vartheta_4(q)}}{\eta(q)}}\right)\\
 \chi_\sigma(q)
 &=
 \frac{1}{2}\sqrt{\frac{\new{\vartheta_2(q)}}{\eta(q)}}
\end{align}
\end{subequations}
where, as usual, the character of the identity field $\id$ is labeled by $0$ and the \textit{Jacobi theta functions} are
\begin{subequations}\label{JacobiThetaMain}
\begin{align}
 \vartheta_3(q)
 &=
 \sum_{n\in\Z}q^{\frac{n^2}{2}}
 =
 q^{-\frac{1}{24}}\,\eta(q)\,\prod_{n=1}^\infty\left(1+q^{n-\frac{1}{2}}\right)^2\label{JacobiTheta3main}\\
 \vartheta_2(q)
 &=
 \sum_{n\in\Z}q^{\frac{1}{2}\left(n-\frac{1}{2}\right)^2}
 =
 2q^{\frac{1}{12}}\,\eta(q)\,\prod_{n=1}^\infty\left(1+q^{n}\right)^2\label{JacobiTheta2main}\\
 \vartheta_4(q)
 &=
 \sum_{n\in\Z}(-1)^n\,q^{\frac{n^2}{2}}
 =
 q^{-\frac{1}{24}}\,\eta(q)\,\prod_{n=1}^\infty\left(1-q^{n-\frac{1}{2}}\right)^2\label{JacobiTheta4main}\\
 \vartheta_1(q)
 &=
 i\sum_{n\in\Z}(-1)^n\,q^{\frac{1}{2}\left(n-\frac{1}{2}\right)^2}
 =
 \frac{1}{2}q^{\frac{1}{12}}\,\eta(q)\,\prod_{n=0}^\infty\left(1-q^{n}\right)^2=0\label{JacobiTheta1main}
\end{align}
\end{subequations}

\subsection{Structure of the State Space}\label{secStructureStateSpace}
It is time to revive the anti-holomorphic sector. The symmetry algebra of a CFT, as pointed out in \eqref{SymAlgebra}, is $\vir\times\bar{\vir}$. In general, covariance of a QFT under a symmetry algebra means that its space of states carries an action of said symmetry. In our case this means that the full Hilbert space of our CFT decomposes as follows
\begin{equation}\label{CFTstateSpace}
 \cH=\bigoplus_{(i,\bi)\in \confFam\times\confFam}\multZ_{i\bi}\,\cH_i\otimes\cH_{\bi}
\end{equation}
where $\cH_i$ and $\cH_{\bi}$ are irreducible representations of $\vir$ and $\bar{\vir}$, respectively. The tupel $(i,\bi)$ labels the conformal dimensions $(h_i,h_{\bi})$ of a primary field $\phi_{i,\bi}(z,\bz)$, or equivalently its state $\ket{h_i,h_{\bi}}$. Be reminded that these fields have to be primary (highest weight representations), since otherwise the energy spectrum would not be bounded from below. The set $\confFam$ carries all possible irreducible representations of the Virasoro algebra, as before. The $|\confFam|\times|\confFam|$ matrix $\multZ_{i\bi}$ has positive integer entries and counts the multplicity of any primary field. Note that it can pair holomorphic and anti-holomorphic fields in a spinful manner, i.e. $s=h_i-h_{\bi}\neq0$\footnote{Recall from \secref{secClassCFTWitt} that the rotations are generated by $\iu(L_0-\bL_0)$}. Another reasonable physical assumption is that the vacuum be unique, and so we shall demand $Z_{00}=1$. We are now in a position to formally write down the partition function of the CFT,
\begin{equation}\label{partitionFunctionCharacters}
 Z(q,\bq)=\sum_{(i,\bi)\in \confFam\times\confFam}\multZ_{i\bi}\, \chi_i(q)\,\chi_{\bi}(\bq)
\end{equation}
Once more, we introduced book-keeping device $q$ and $\bq=q^*$. In what follows, we will finally discover their physical meaning. Crucially, in doing so, this will naturally lead us to powerful constraints imposed on the matrix $\multZ_{i\bi}$.

\subsubsection{Genus One and the Modular Group}
In QFTs we typically determine the theory's field content by checking which fields run in loops. Given that we consider CFTs on Riemann surfaces, loops correspond to Riemann surfaces of non-trivial genera. It is understood here that CFTs living on different surfaces derive from the same model, if their local properties, such as OPEs, are the same. As \new{argued by Moore and Seiberg} \cite{moore1990lectures, moore1989classical}, no meaningful additional constraints are collected by going to Riemann surfaces of genus larger than one, i.e. we can restrict to the torus.

The torus is reached from the plane as follows. First, map the plane conformally to the cylinder $\R\times S^1$ via \eqref{ExpMap}.
\begin{exercise}\label{ExDilationRotationCylinder}\label{exHamiltonianCyl}
 Show that the Hamiltonian on the plane $H=L_0+\bL_0$ is mapped to $H_{cyl}=L_0+\bL_0-\cc/24-\bar{\cc}/24$ on the cylinder. Similarly, show that the generator of rotations on the plane $P_{pl}=\iu(L_0-\bL_0)$ is mapped to $P_{cyl}=\iu(L_0-\bL_0-\cc/24+\bar{\cc}/24)$ (it generates translations on the cylinder in the compact direction, cf. exercise \ref{ExDilationRotation})\footnote{The exponential map \eqref{ExpMap} actually carries a macroscopic scale, namely the circumference of the cylinder $L$. If the compact direction is time, we rather call this scale $\beta$. This scale leads to \quotes{soft breaking} of scale invariance, as mentioned above. In many cases it is useful to carry this scale around, leading to a modification of the exponential map \eqref{ExpMap} to $z=\exp(2\pi w/L)$ ($z=\exp(2\pi \iu w/\beta)$). Consequently, the spacetime translators are also rescaled on the cylinder 
 \begin{equation}
  H_{cyl}=\frac{2\pi}{L}(L_0+\bL_0-\cc/24-\bar{\cc}/24),
  \qquad
  P_{cyl}=\frac{2\pi}{L}\iu(L_0-\bL_0-\cc/24+\bar{\cc}/24)\notag
 \end{equation}
 Choosing $L=2\pi$ returns us to the case in the main text.} Hint: Use equation \eqref{Ttransform}.  
\end{exercise}
\ifsol
\solution{exHamiltonianCyl}{It suffices to look at the holomorphic half, since the other half follows suit. The energy momentum tensor transforms according to \eqref{Ttransform}. The Schwarzian \eqref{Schwarzian} is $S(e^w,w)=-1/2$, so that
\begin{equation}
 T_{cyl}(w)
 =
 \left(\frac{\p z}{\p w}\right)^2T(z)+\frac{\cc}{12}S(z,w)
 =
 z^2T(z)-\frac{\cc}{24}
\end{equation}
The mode expansion \eqref{Texpansion} turns this into
\begin{equation}
 T_{cyl}(w)
 =
 \sum_nL_nz^{-n}-\frac{\cc}{24}
 =
 \sum_n\left(L_n-\frac{\cc}{24}\delta_{n0}\right)e^{-wn}
\end{equation}
This gives $(L_0)_{cyl}=L_0-\cc/24$ from which the Hamiltonian and angular momentum operator are constructed.}
\fi

\begin{figure}
 \begin{center}
  \includegraphics[scale=.275]{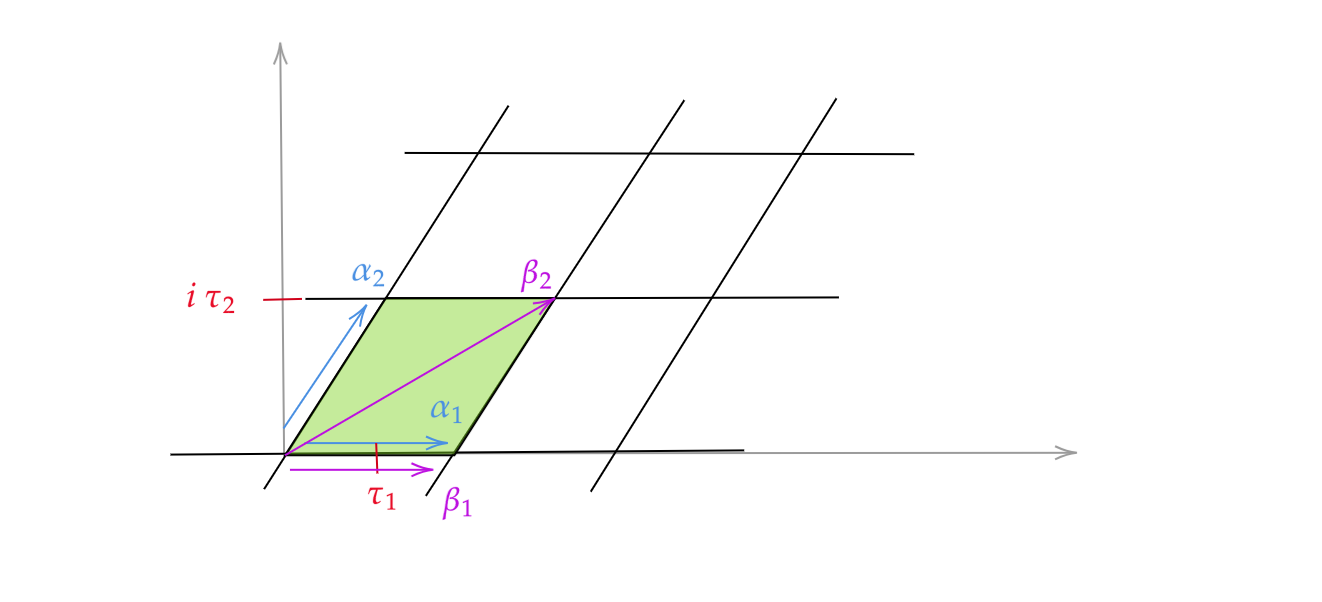}
 \end{center}
 \caption{Two vectors $\alpha_{1,2}$ (red) span the green lattice. Their fundamental domain is shaded in green. Two other vectors (purple), $\beta_{1,2}$ span the same lattice. The modular parameters of $\alpha_{1,2}$ and $\beta_{1,2}$ are related by $\tau\to\tau+\tau_1$, which is a modular transformation.}
 \label{figTorusLattice}
\end{figure}

On the cylinder we cut out a finite slab $[0,\beta]\in\R$ and identify its ends. We are left with a torus $S^1\times S^1$. It turns out that we actually have a choice here. Indeed, before identifying the ends, we have the freedom to twist the ends of the cylinder. There is a general way of analyzing this and it starts on the plane. A torus can also be constructed directly from the plane $\C$ by identifying points
\begin{equation}
 z\sim z+m \alpha_1+n\alpha_2,
 \qquad
 n,m\in\Z,
 \quad
 \alpha_{1},\alpha_2\in\C
\end{equation}
In this way, the tuple $(\alpha_1,\alpha_2)$ is seen to span a lattice, as seen in \figref{figTorusLattice}. Its smallest cell is the fundamental domain of the torus. Crucially, the fundamental domain is in general not a square, but a parallelogram. This skewing of the square to the torus is described by the \emphasize{modular parameter}
\begin{equation}
 \tau=\frac{\alpha_2}{\alpha_1}=\tau_1+\iu\tau_2\in\C
\end{equation}
The modular parameter describes the shape of the torus.


By glancing at \figref{figTorusLattice} it is easy to convince oneself that distinct tuples of $(\alpha_1, \alpha_2)$ and $(\beta_1,\beta_2)$ may construct the same lattice and hence the same torus. This is the case when one tuple of periods is given by an integer multiple of periods of the other tuple
\begin{equation}
 \begin{pmatrix}
  \alpha_1\\
  \alpha_2
 \end{pmatrix}
 =
 \begin{pmatrix}
  \,a\, & \,b\,\\
  \,c\, & \,d\,
 \end{pmatrix}
 \begin{pmatrix}
  \beta_1\\
  \beta_2
 \end{pmatrix}
 \qquad
 a,b,c,d\in\Z,
 \qquad
 ad-bc=1
\end{equation}
That this matrix be unimodular follows from the requirement that the inverse matrix must also have integer entries, thereby rephrasing $(\beta_1,\beta_2)$ as appropriate integer combinations $(\alpha_1,\alpha_2)$. It also implies that the parallelograms spanned by $\alpha_{1,2}$ and $\beta_{1,2}$ have the same area. Furthermore, it makes no difference to choose $(-\alpha_1,-\alpha_2)$ over $(\alpha_1,\alpha_2)$ as basis cell for our lattice, so we can mod out a $\Z_2$. Overall,
\begin{tcolorbox}
 The transformations on $(\alpha_1,\alpha_2)$, which preserve the lattice -- and hence the torus -- are the entirety of 
 \begin{equation}
  \begin{pmatrix}
  \,a\, & \,b\,\\
  \,c\, & \,d\,
 \end{pmatrix}
 \in\PSL(2,\Z)=\SL(2,\Z)/\Z_2\,.
 \end{equation}
 These transformations form the \emphasize{modular group} and describe the isometries of the torus. They allow us to pick a convenient frame for the periods $(\alpha_1,\alpha_2)=(1,\tau)$. The modular group acts as follows on the modular parameter
 \begin{equation}
  \tau\to\frac{a\tau+b}{c\tau+d}\,.
 \end{equation}

\end{tcolorbox}
The modular group is generated by two transformations -- this is non-trivial to see and we will be cavalier about this statement's proof. The first generator is
\begin{equation}
 \modT: \tau\to\tau+1,
 \qquad
 \modT=\begin{pmatrix}
                                      \,1&1\,\\
                                      \,0&1\,
                                     \end{pmatrix}
\end{equation}
and the second generator is
\begin{equation}
 \modS: \tau\to -\frac{1}{\tau}
 \qquad
 \modS=\begin{pmatrix}
                                      \,0&1\,\\
                                      \,-1&0\,
                                     \end{pmatrix}
\end{equation}
Note that the modular $\modS$ transformation corresponds to an interchange of space and time directions on our torus, $\modS:\, (\alpha_1,\alpha_2)\to(-\alpha_2,\alpha_1)$ . These transformations satisfy $\modS^2=-\id\simeq\id$ and $(\modS\modT)^3=\id$ when acting on $\tau$ -- these expressions will be modified once we have the modular group act on characters! 

\subsubsection{The Partition Function and Modular Invariance}
We are now finally in a position to formulate a second expression for the partition function \eqref{partitionFunctionCharacters}. In statistical mechanics we have $Z=\tr_\cH e^{-\beta_0 H}$, where $\beta_0$ is an inverse temperature leading to Boltzmann factors $p_i=e^{-\beta_0 E_i}$ for a state $\ket{\phi_i}$. To interpret this in QFT we rewrite this expression as $Z=\sum_i\bra{\phi_i}e^{-\beta_0 H}\ket{\phi_i}$ and view this as temporal evolution of $\ket{\phi_i}$ via the Hamiltonian $H$ for a distance $\beta_0$. Importantly, time is Euclidean with period $\beta_0$ so this evolution returns us to the same spot that we start out with. Moreover, the overlap with $\bra{\phi_i}$ simply \new{constructs the return amplitude for this evolution.} The trace has us repeat this for all states in the system.

We wish to repeat this for the two-dimensional cylinder, parametrized by $w=\Re(w)+\iu\Im(w)$, whose compact direction is $\Re(w)$ and we interpret as spatial. We have just learned that we can compactify a cylinder into a variety of different tori, distinguished by a modulus $\tau=\tau_1+\iu\tau_2$. Let's start with a rectangular torus which has $\tau=\iu\tau_2$. Since we view $\Im(w)$ as temporal, we can identify $\beta_0=\tau_2$, and plainly adapt the construction above leading to $e^{-\beta_0 H}\to e^{-\tau_2 H_{cyl}}$. 

A torus with $\tau_1\neq0$ on the other hand is skewed by $\tau_1$ and in order for a point to return to itself after the evolution, we have to modify the evolution operator $e^{-\beta_0 H}\to e^{-2\pi\tau_2 H+2\pi\tau_1 P}$. 
Using our findings of exercise \ref{ExDilationRotationCylinder} we find
\begin{tcolorbox}
The \emphasize{partition function} of a CFT is
 \begin{align}
 Z(\tau,\btau)
 =
 \tr_H\left(e^{-2\pi\tau_2 H_{cyl}+2\pi\tau_1 P_{cyl}}\right)
 =
 \tr_H\left(q^{L_0-\cc/24}\bq^{\bL_0-\bar{\cc}/24}\right)
\end{align}
where $q=e^{2\pi\iu\tau}$ and $\bq=e^{-2\pi \iu\btau}=e^{-2\pi \iu(\tau_1-\iu\tau_2)}$. Any physical partition function does not depend on the particular lattice frame $(\alpha_1,\alpha_2)$ we choose to parametrize the torus. Hence we demand that the partition function be \emphasize{modular invariant}
\begin{equation}\label{modInvariance}
 Z(\tau+1,\btau+1)=Z(\tau,\btau)=Z(-1/\tau,-1/\btau)
\end{equation}
\end{tcolorbox}
This can now be connected with our general discussion on representations of the Virasoro algebra leading to \eqref{partitionFunctionCharacters}. The modular group must also act on the characters, and generically, a modular transformation shuffles representations amongst themselves,
\begin{align}
 \modT:\chi(\tau)\to&\chi(\tau+1)=\sum_{i\in\confFam}\modT_{ij}\chi_j(\tau)
 \\
 \modS:\chi(\tau)\to&\chi(-1/\tau)=\sum_{i\in\confFam}\modS_{ij}\chi_j(\tau)\label{modStransformation}
\end{align}
The two matrices $\modT_{ij}$ and $\modS_{ij}$ are $|\confFam|\times|\confFam|$-dimensional. 
It turns out that $\modT$ acts diagonally on a character, $\modT_{ij}=\delta_{ij}e^{2\pi\iu(h_i-\cc/24)}$. Evaluating $Z(\tau+1,\btau+1)$ thus yields the constraint $\multZ_{i\bi}=0$ unless $s=h_i-\bh_{\bi}\in\Z$ for the CFT state space \eqref{CFTstateSpace}. For fermionic degrees of freedom, where we obviously like to have half-integer spin, we therefore need to slightly relax the requirement of modular invariance by demanding invariance only under $\modT^2$ rather than $\modT$.

While the matrix $\modT_{ij}$ is rather innocent, the modular $\modS$ transformation carries much non-trivial information. Unlike $\modT_{ij}$, $\modS_{ij}$ has a non-trivial model-dependent form. For unitary theories, the following properties can nevertheless be derived \cite{fuchs1994fusion}
\begin{align}
 \modS^{-1}=\modS^*,
 \qquad
 \modS^\transpose=\modS,
 \qquad
 \modS^2=\cC\\
 \modS_{ij^+}=\modS_{ij}^*=\modS_{i^+j}, 
 \qquad
 \modS_{i0}\geq\modS_{00}>0
 \qquad \label{modSidentities}
\end{align}
where $\cC_{ij}=\delta_{j,i^+}$ is the charge conjugation matrix. The circumstance that $\modS^2\neq\id$ indicates that we are dealing with a representation of the double cover of the modular group\footnote{This is just as with spin in quantum mechanics, where a $2\pi$ rotation does not lead back to the starting state. This is because $\SU(2)$ is the double cover of $\SO(3)$}. These relations imply in particular $\modS_{i0}=\modS_{i0}^*\in\R$. The most striking property of the modular $\modS$ matrix however is its relation to the fusion rules expressed through the celebrated
\begin{tcolorbox}
 \emphasize{Verlinde formula} \new{\cite{verlinde1988fusion}}
\begin{equation}\label{Verlinde}
 \fus_{ij}^k=\sum_{i\in\confFam}\frac{\modS_{il}\modS_{jl}\modS_{kl}^*}{\modS_{l0}}
\end{equation}
This is the statement that the modular $\modS$ matrix diagonalizes the fusion rules. Concretely, the matrix $\fus_i$ has $|\confFam|$-dimensional eigenvectors $\mathbf{v}_a$ with $a=1,\dots|\confFam|$ and entries $(\mathrm{v}_a)_k=\modS_{ak}$, so that $\fus_i=\modS\Delta_i\modS^\dagger$. 
\end{tcolorbox}

\begin{exercise}\label{exQuantumDimensions}
 Read off the eigenvalues of $\fus_i$ and derive the following expression for the quantum dimensions
 \begin{equation}\label{qdimModS}
  \qdim_i
  =
  \frac{\modS_{i0}}{\modS_{00}},
 \end{equation}
 Show furthermore that this is the same as
 \begin{equation}\label{qdimCharacters}
  \qdim_i=\lim_{q\to1^-}\frac{\chi_i(q)}{\chi_0(q)}
 \end{equation}
In this form, the quantum dimensions acquire a representation theoretic meaning as asymptotic -- this refers to $\lim_{q\to1^-}$ -- measure of size for $\cH_i$ in \quotes{units of $\cH_0$}. Hint: Consider a purely real $q=e^{-\beta_0}$. The limit $\lim_{q\to1^-}$ is the high temperature limit $\beta_0\to0^+$. Which term or terms dominate the sum? 
\end{exercise}
\ifsol
\solution{exQuantumDimensions}{The eigenvalues can simply be read off, but we can also just do it the old-fashioned way by recalling eq. \eqref{FusionEigensystem}, with which
\begin{equation}
 (\fus_i\textbf{v}_a)_j
 =
 \sum_k\fus_{ij}^k\modS_{ka}
 =
 \frac{\modS_{ia}}{\modS_{0a}}\modS_{ja}
 =
 \frac{\modS_{ia}}{\modS_{0a}}(\textbf{v}_a)_j
 \quad
 \Rightarrow
 \quad
 n_{ia}=\frac{\modS_{ia}}{\modS_{0a}}
\end{equation}
where $n_{ia}$ is the eigenvalue of $\fus_i$ for $\textbf{v}_a$. By the Perron-Frobenius theorem, we are seeking the eigenvector with strictly positive entries, since this one pertains to the maximal eigenvalue. This is only guaranteed for $a=0$. Its eigenvalue for $\fus_i$ is $n_{i0}=\qdim_i$ as claimed. The following decomposition is thus clear, $\fus_i=\modS\Delta_i\modS^\dagger$, where $\Delta_i=\diag(n_{i0},\dots n_{ia},\dots n_{i|\confFam|})$. For the second part consider a purely real use that 
\begin{equation}
 \chi_i(q)
 =
 \sum_j(\modS^{-1})_{ij}\chi_j(\tq)
 \overset{q\to1^-}{\approx}
 (\modS^{-1})_{i0}\chi_0(\tq)
\end{equation}
is dominated by the primary with smallest conformal weight, which is the vacuum for unitary theories. This is most easily seen by using a purely imaginary modular parameter, $q=e^{-\beta_0}$ and have $\beta_0\to0^+$. This is the infinite temperature limit, in which the smallest energy (solely) dominates the sum, i.e. $h=0$.}
\fi
\begin{exercise}\label{exIsingVerlindeFusion}
 Given the modular $\modS$ matrix of the Ising CFT,
 \begin{equation}\label{IsingModS}
 \modS=
 \begin{pmatrix}
        \frac{1}{2} & \frac{1}{2} & \frac{1}{\sqrt{2}} \\
        \frac{1}{2} & \frac{1}{2} & -\frac{1}{\sqrt{2}} \\
        \frac{1}{\sqrt{2}} & -\frac{1}{\sqrt{2}} & 0 \\
       \end{pmatrix}
\end{equation}
where rows and columns are ordered as $\{\id,\epsilon,\sigma\}$, work out the model's fusion rules. Compare with exercise \ref{exFusionRules}.
\end{exercise}
\ifsol
\solution{exIsingVerlindeFusion}{The Fusion rules are found in \eqref{IsingFusionOPEex}.}
\fi

Returning to the issue of modular invariance, we find that it simply requires the modular generators to commute with the multiplicity matrix $\multZ_{i\bi}$ of the partition function \eqref{partitionFunctionCharacters},
\begin{equation}
 [\modT^2,\multZ]=0=[\modS,\multZ]
\end{equation}
If the set $\confFam$ of irreducible representations of the symmetry algebra is finite, $|\confFam|<\infty$, we speak of a \emphasize{rational CFT (RCFT)}. Two modular invariant partition functions are immediately identified in this case
\begin{align}
 Z_{\textrm{diag}}(q,\bq)
 &=
 \sum_{i\in\confFam}\chi_i(q)\chi_i(\bq)
 =
 \sum_{i\in\confFam}|\chi_i(q)|^2\label{diagonalModInvariant}\\
 Z_{\textrm{ch.c}}(q,\bq)
 &=
 \sum_{i\in\confFam}\chi_i(q)\chi_{i^+}(\bq)\label{chargeConjModInvariant}
\end{align}
which are called the \emphasize{diagonal modular invariant} and the \emphasize{charge-conjugate modular invariant}, respectively.

As an example consider the Ising CFT. It has three primary fields $\confFam=\{\id,\epsilon,\sigma\}$. All three states have equal holomorphic and anti-holomorphic conformal weights, $h=\bh$. Their values are $h_\id=0$, $h_\epsilon=1/2$ and $h_\sigma=1/16$. The theory has a diagonal partition function
\begin{equation}\label{IsingDiagZ}
 Z(q,\bq)=|\chi_0(q)|^2+|\chi_\varepsilon(q)|^2+|\chi_\sigma(q)|^2
\end{equation}
where as above, we have labeled the vacuum representation by 0. The individual characters are found in \eqref{IsingCharacters}. \new{Many more modular invariants can be found in \cite{cappelli1987modular, cappelli1987ade, kato1987classification, itzykson1986two} as well as chapter 10 of \cite{DiFrancesco} and references therein.}

\newpage
\section{Boundary Conformal Field Theory}\label{secBCFT}
So far we treated systems without boundaries. Yet, clearly, nature is filled with interesting and important systems of finite extend.
\begin{center}
 \textit{How does CFT accommodate boundaries and defects?}
\end{center}
Answering this question lies at the focus of the following few sections. As we will see, the world of conformal boundaries is remarkably rich and has profound applications in various central topics of theoretical physics. Here are a few applications in low-dimensional systems

\subsubsection*{The Kondo Effect}
The Kondo effect describes the \new{screening} of magnetic impurities by conduction electrons in a metal. After its inital description by Kondo, it became the guinea pig for many important techniques we know and love today in theoretical physics; most prominent is the development of the renormalization group by Wilson \cite{Wilson}. In that framework, at high energies the impurity is ignored by the conduction electrons due to their high kinetic energy. Upon lowering the temperature, the conduction electrons begin to notice the presence of the impurity. The conduction electrons then enter a bound state with the impurity with the aim of forming a spin singlet. The latter is not magnetic, and hence the impurity has been screened.

In the \new{nineteen-ninetees}, Affleck and Ludwig realized that the Kondo effect is elegantly described in the framework of boundary conformal field theory (BCFT) \cite{AFFLECK1991641}. By realizing that the physics is dominated by radial s-waves, they reduces the problem to $1+1$ dimensions, where the spatial coordinate is radial distance from the impurity. In this picture, the impurity is seen to impose a conformal boundary condition on the conduction electrons, one in the UV of the renormalization group flow and a different one in the IR. The powerful framework of BCFT allows for an in-depth analysis of the impurity degrees of freedom at both fixed points \cite{AffleckReview}.

\subsubsection*{Entanglement Spectra}
When studying entanglement we assume that Hilbert space factorizes, $\cH=\cH_A\otimes\cH_B$. One popular choice is to associate $\cH_A$ and $\cH_B$ with spatial domains. When dealing with quantum mechanics on a lattice, one easily assembles the local Hilbert spaces of each site into a domain $A$ and $B$. When dealing with quantum field theory, this bipartitioning is actually not straightforward. Indeed, fields are distributions and need to be smeared over space. Hence, cutting fields apart at arbitrarily chosen boundaries between spatial regions $A$ and $B$ is highly problematic.

As argued in \cite{ohmori2015physics}, the solution is to assign boundary conditions to the fields at the entangling region, i.e. at the interface of $A$ and $B$. This is mostly done implicitly, even on the lattice. In the context of two-dimensional CFT, this approach has proven particularly useful. Indeed, while conventional techniques only probe a subsector of $\cH_A$, BCFT grants full, unrestrained access to $\cH_A$; see for instance \cite{DiGiulio:2022jjd,Northe:2023khz} for studies of the entanglement spectrum of $\cH_A$. It turns out that these techniques even capture the universal parts of entanglement spectra belonging to gapped phases adjacent to a quantum phase transition described by a CFT \cite{cho2017universal}.

\subsubsection*{Symmetry-Protected Topological Phases of Matter}
Symmetry-protected topological phases of matter (SPT) have non-trivial topological degeneracies in presence of a symmetry. When restricted to a disk of two spatial dimensions, the SPT harbors critical modes on the edge\footnote{This is reminiscent of the Quantum Hall effect (QHE). However, the edge modes are chiral in the QHE, while they are non-chiral in SPTs.}. While the bulk and edge are both anomalous individually, in combination they form a non-anomalous system. This highly fine-tuned interplay is exploited in \cite{Han:2017hdv}, where is is argued that one cannot cut the edge open while preserving the symmetry of the SPT. This is synonymous with showing that no conformal boundary condition can be found which also preserves the symmetry.

BCFT also plays a role for $(1+1)$-dimensional SPTs. Indeed, there are several non-trivial phases protected by the same symmetry, none of which can be deformed adiabatically into each other. In order to cross over between these phases, one necessarily passes through a quantum critical point, a CFT! Imagine having two distinct SPT phases for the same symmetry on a line and separated by a domain wall. Hallmarks of such transitions are the presence of degenerate degrees of freedom on the domain wall, which are controlled by the symmetry of the SPT. From the point of view of the CFT these topological degrees of freedom, along with their degeneracies, are stored in spectra of BCFTs \cite{cho2017relationship}. 

\subsubsection*{Useful Literature}
\begin{itemize}
 \item The book \quotes{Boundary Conformal Field Theory and the Worldsheet Approach to D-branes} \cite{Recknagel:2013uja} by Recknagel and Schomerus is an excellent resource to learn BCFT, in particular chapter 4, which we are following here for the longest part. All remaining chapters contain interesting applications and advanced tools of BCFT. 
 \item \new{No name is as intertwined — argueably even synonymous — with BCFT as that of John L. Cardy. His groundbreaking contributions span from the early studies of BCFT \cite{cardy1984conformal} to the uncovering of the profound principles governing critical boundaries \cite{cardy1986effect, cardy1989boundary, cardy1991bulk}. This chapter aims to follow in his footsteps by explaining his \emphasize{boundary state formalism} and the resulting \emphasize{Cardy constraint} \cite{cardy1989boundary}. Together with the work of Ishibashi \cite{Ishibashi:1988kg,Onogi:1988qk} and Lewellen \cite{lewellen1992sewing}, one has an excellent introduction to BCFT}.
 \item The lectures by Petkova and Zuber \cite{petkova2001conformal} are a useful resource adding alternative viewpoints to the previous reference. The focus here lies strongly with representation theoretic data of the CFT and their connections to graphs.
 \item Chapter 6 of Blumenhagen and Plauschinn \cite{Blumenhagen} provides a nice introduction to BCFT geared toward string theory, where conformal boundaries describe D-branes. 
\end{itemize}

\subsection{Generalities and Outline}\label{secBCFTgeneralities}

The construction of a boundary Conformal Field Theory (BCFT) starts from a CFT on the full complex plane, which is assumed to be solved or known in sufficient detail, meaning we have access to
\begin{itemize}
 \item[1)] The spectrum on the full plane, as discussed in \secref{secStructureStateSpace}
 \begin{equation}\label{planeSpectrum}
 \cH^{\pl}=\bigoplus_{(i,\bi)\in\ipl }\multZ_{i\bi}\,\cH_i\otimes\cH_{\bi}
\end{equation}
with primary fields $\phi_{i\bi}(z,\bz)$ taken from the set $\ipl\,\subseteq \confFam\times\confFam$ for which $\multZ_{i\bi}\neq0$. As before, $\confFam$ denotes the set of irreducible representations of the chiral symmetry algebra. The superscript $\pl$ is used to indicate the full complex plane throughout this lecture.

 \item[2)] The operator algebra encapsulated by the Operator product expansion (OPE) is known\footnote{We are not concerned with the concrete form of $f$ here, which is found in the literature, and prefer to highlight the structure of the OPE \eqref{bulkOPE}.}
 \begin{equation}\label{bulkOPE}
  \phi_{i,\bi}(z,\bz)\phi_{j,\bj}(w,\bw)
  =
  \sum_{k,\bar{k}}c_{ijk}c_{\bi\bj\bar{k}}\,f(z-w,\bz-\bw,\{h\},\{\bh\})\,\phi_{k,\bar{k}}(w,\bw)
 \end{equation}
 where $c_{ijk}$ ($c_{\bi\bj\bar{k}}$) carry the information of the three-point correlators for the left- (right-) moving degrees of freedom. These are non-trivial dynamical data of a CFT. The sets $\{h\}$ ($\{\bh\}$) are abbreviations for all involved conformal weights. 
\end{itemize}
This CFT on the full plane will be referred to as \emphasize{parent CFT} or \emphasize{bulk CFT}, and in order to construct a BCFT from it, the parent CFT  is now restricted to the upper half-plane
\begin{equation}
 \mathds{H}=\{z\in\C\,|\,\Im z\geq0\}
\end{equation}
The boundary is placed on the real line, so that the lower half-plane is not part of the system\footnote{Most of the discussion here applies to symmetry algebras containing the Virasoro algebra $\vir$ as proper subalgebra. These generalizations are left to the reader to explore in the aforementioned literature however. }.

Local properties are inherited from the parent CFT. In particular, the singularity structure of correlators of fields $\phi_{i\bi}(z,\bz)$ for $\Im(z)$ is determined by the corresponding OPE on the full plane.

A number of things do change, however. First of all, $\cH^{\pl}$ is only an index set for the BCFT and \textit{not the state space of the theory on $\mathds{H}$}. Global properties follow from the boundary condition imposed on the real line. Two kinds of global data specify the BCFT
\begin{itemize}
 \item[a)] \emphasize{Gluing conditions}. These are linear constraints imposed by the boundary condition on the symmetry algebra of the parent CFT. These are discussed in \secref{secGluingConditions}.
 \item[b)] \emphasize{One-point Functions}. In contrast to the complex plane, the presence of a boundary allows non-trivial bulk fields $\phi_{i\bi}$ to have a non-vanishing expectation value. This is possible due to the presence of fields localized at the boundary. This is discussed in \secref{sec1pt}.
\end{itemize}
These new boundary excitations build up the sought-after \emphasize{state space of the BCFT}, as explained in \secref{secStateSpaces}. They constitute the counterpart to \eqref{planeSpectrum} on the upper half-plane $\mathds{H}$. In \secref{secBdyState} we develop a sophisticated toolkit, namely the \emphasize{boundary state formalism}, which allows us to determine the state spaces of boundary fields in \secref{secCardyConstraint} via the \emphasize{Cardy constraint}. 

\subsection{The Conformal Boundary Condition}\label{secGluingConditions}
In this subsection, we discuss the \emphasize{conformal symmetry preserved by the boundary}, we learn about \emphasize{gluing conditions} and have a first encounter with the \emphasize{folding trick}.

The mere presence of a boundary already breaks symmetries. An obvious instance of this is translation symmetry perpendicular to the boundary. In general, imposing boundaries can largely destroy a system's symmetry. On the other hand, there also exist special boundary conditions preserving plenty of symmetry.

We are interested in boundary conditions preserving the maximal amount of conformal symmetry present on the plane. This is given by the subset of conformal transformations $f$ which map the boundary of the upper half-plane into itself, but may otherwise deform the interior of the upper half-plane,
\begin{equation}
 f:\mathds{H}\to\mathds{H},
 \qquad
 f(x)\in\R
 \quad
 \text{ for }
 x\in\R
\end{equation}
They remain symmetries in a BCFT. They include globally defined Möbius transformations $\SL(2,\R)$, and as on the plane, we also include transformations harboring isolated singularities. Taken together, unsurprisingly, these transformations generate a single Virasoro algebra. We emphasize that only \textit{real} analytic functions survive the trip to the upper half-plane, i.e. in complex coordinates $\bar{f}(\bar{z})=f(z)$. Hence, from formerly two independent copies of $\vir$, only the diagonal subalgebra survives.

A more physically intuitive way to see this is to require that no energy-momentum flow can leak across the boundary. Parametrizing the upper half-plane by $z=x+\iu y$ and placing the boundary at $y=0$ this is \new{expressed by} $T_{xy}(x,y=0)=0$. \new{Indeed, recall that this component of the energy-momentum tensor is the momentum density in $y$-direction.} In complex coordinates this \new{demand} is recast as follows:
\begin{tcolorbox}
The \emphasize{gluing condition} imposed on the energy-momentum tensor by a conformal boundary condition is
 \begin{equation}\label{glueT}
 T(z)=\bT(\bz)
 \qquad
 \text{for}
 \qquad 
 z=\bz
 \end{equation} 
 This provides a linear constraint relating the holomorphic and anti-holomorphic copies of Virasoro present in the plane. Only their diagonal subalgebra with $L_n=\bL_n$ survives, providing the single copy of Virasoro governing the system on the boundary. 
\end{tcolorbox}

We anticipate that there may be several solutions to \eqref{glueT} allowing for various distinct conformal boundary conditions\footnote{When extended symmetries are at play, their currents are also glued linearly. Their gluing must respect \eqref{glueT}, if we insist on conformality of the boundary condition.}. We will touch upon this in \secref{secStateSpaces} and become fully concrete about this fact in \secref{secCardyConstraint}.

\begin{figure}
\begin{center}
 \includegraphics[scale=.3]{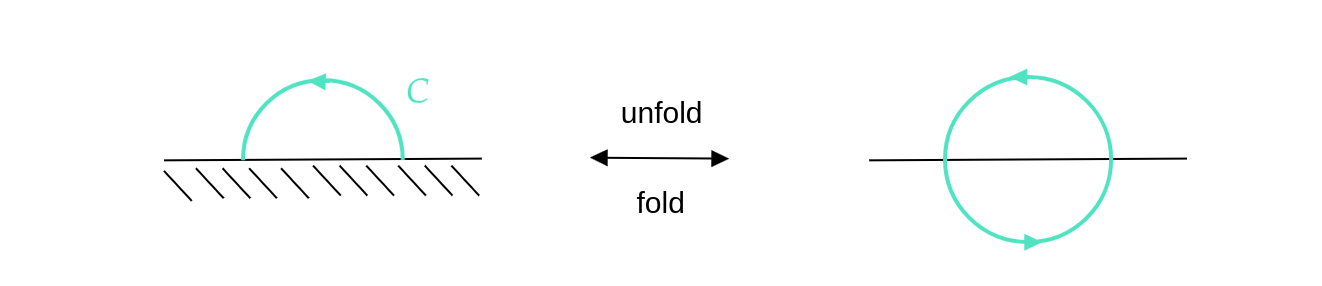}
 \caption{The folding trick. The semicircle traced by the antiholomorphic sector is reflected at the boundary into the lower half-plane. The gluing condition \eqref{glueT} prescribes the analytic continuation justifying this move. }
 \label{figFoldingContour}
\end{center}
 
\end{figure}

Note that the gluing condition \eqref{glueT} may be viewed as prescribing an analytic continuation of the energy-momentum tensor to the lower half-plane,
\begin{equation}
 \mathsf{T}(z)
 =
 \sum_nL_n^{\hp}z^{-n-2}
 =
 \begin{cases}
  T(z),\quad \Im z\geq0\\
  \bT(\bz),\quad \Im z<0
 \end{cases}
\end{equation}
Here and below we use $\hp$ to emphasize that a quantity is specified on the upper half-plane, in contrast to the full plane indicated by $\pl$. The modes of this energy-momentum tensor are 
\begin{equation}\label{VirasoroModeH}
 L_n^{\hp}
 =
 \int_C\frac{dz}{2\pi \iu}z^{n+1}T(z)-\int_C\frac{d\bz}{2\pi \iu}\bz^{n+1}\bT(\bz)
 \overset{\eqref{glueT}}{=}
 \oint\frac{dz}{2\pi \iu}z^{n+1}\mathsf{T}(z) 
\end{equation}
where the integration contour $C$ is a semi-circle around the origin, as in the left-hand side of \figref{figFoldingContour}. In going to the last equality, we have analytically continued the system to the lower half-plane, where we could close the contour as indicated on the right-hand side of \figref{figFoldingContour}. Note that this expression looks entirely as for a single holomorphic copy of Virasoro on the full complex plane, see \eqref{Texpansion}. Clearly, it thus satisfies the Virasoro algebra \eqref{Virasoro}. 

This is our first encounter with the \emphasize{folding trick}. The gluing condition \eqref{glueT} allows us to view the anti-holomorphic piece situated on the upper-half plane as holomorphic continuation to the lower half-plane of the holomorphic sector. This also implies that $\bz$ is to be viewed as a holomorphic coordinated in the unfolded picture.

\subsection{One-point Functions and Boundary Fields}\label{sec1pt}
In this subsection, \emphasize{one-point functions of bulk fields} are derived, \emphasize{boundary fields} are introduced and the \emphasize{coupling between bulk and boundary fields} is explained.

Boundaries are not just vacant spaces for bulk fields to terminate on. In fact many interesting aspects of boundaries stem from the degrees of freedom they harbor. To give a few examples, the degrees of freedom residing on boundaries have been central in understanding phenomena such as the Kondo effect in impurity studies, D-branes in string theory, entanglement spectra in critical systems and topological edge modes in topological phases of matter. The first thing we learn now is how these degrees of freedom allow bulk fields to acquire a one-point function.

\begin{figure}
 \begin{center}
  \includegraphics[scale=.3]{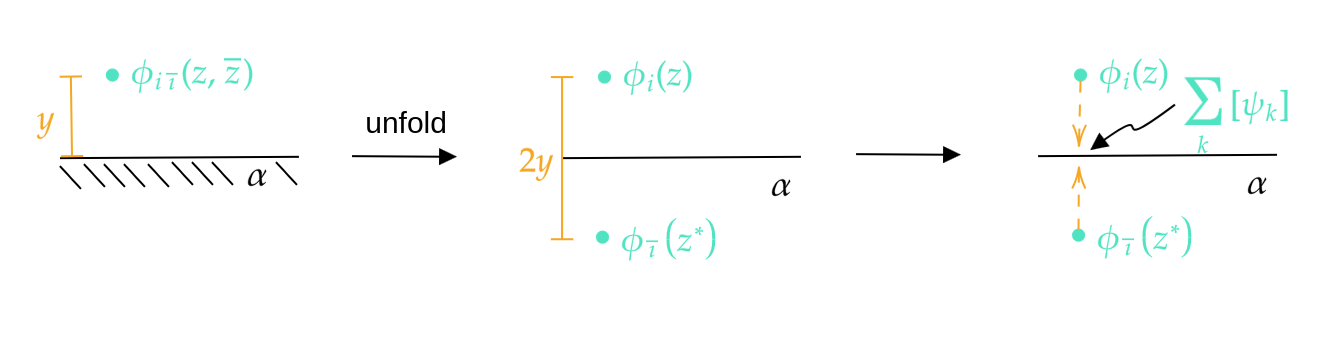}
 \end{center}
\caption{By virtue of the folding trick, a bulk field $\phi_{i\bi}$ can be viewed as two individual chiral fields $\phi_i$ in the upper half-plane and $\phi_{\bi}$ in the lower half-plane. The latter becomes a mirror charge. As shown in the third panel, when approaching the boundary, the two chiral fields can be expanded in terms of operators confined to the boundary, as in \eqref{bulkBdyOPE}.}
\label{figFolding1pt}
\end{figure}


Consider a bulk field $\phi_{i\bi}(z,\bz)$. Using the folding trick, we are allowed to view this non-holomorphic field as two holomorphic fields $\phi_i(z)$ placed at $z$ in the upper half plane, and $\phi_{\bi}(\bz)$ placed at $\bz=z^*$ in the lower half-plane as mirror charge, see \figref{figFolding1pt}. This is in fact justified on a deeper level by conformal Ward identities \cite{Recknagel:2013uja}. Note that we do not introduce new notation for fields in the unfolded picture and only distinguish the fields by double or single indices. This innocent looking gymnastic has tremendous physical implications. Indeed, it turns a one-point function of a non-holomorphic field into a two-point function of holomorphic fields, as on the full plane \eqref{2pt}, so we can access it.
\begin{tcolorbox}
 In the presence of a boundary labeled by $\alpha$, a bulk field of non-trivial conformal dimensions, $(h_i,h_{\bi})\neq(0,0)$, may acquire an expectation value
\begin{equation}\label{1ptBdy}
 \corr{\phi_{i\bi}(z,\bz)}_\alpha
 =
 \corr{\phi_i(z)\phi_{\bi}(\bz)}_\alpha
 =
 \frac{A_{i\bi}^\alpha \, d_{i i^+}}{|z-\bz|^{2h_i}}\delta_{\bi,i^+}
\end{equation}
\end{tcolorbox}

Several remarks are in order here
\begin{itemize}
 \item $\alpha$ distinguishes boundary conditions, such as Dirichlet and Neumann boundary conditions on free fields. The notation $\corr{\,\cdot\,}_\alpha$ signifies an expectation value in presence of the boundary condition $\alpha$.
 \item A two-point function of holomorphic fields can only be non-zero if the fusion of the contributing conformal families contains the identity family, $[\phi_i]\fuse[\phi_{\bi}]=[\id]+\dots$. This happens when the families are conjugate to one another, hence the $\delta_{\bi,i^+}$\footnote{This is a little refinement to what we had stated earlier in \eqref{2pt}, but back then we did not know about conjugate representations, so we could not appreciate this little subtlety.}. We learn that a bulk field cannot acquire an expectation value unless $\bi=i^+$, solidifying the notion of a mirror charge in the lower-half plane. When restricting to minimal models, the representations are always self-conjugate, $i^+=i$. 
 \item  Note that $h_{i^+}=h_i$, which is clearly pointing back at the gluing condition \eqref{glueT}, i.e. a necessary condition for a bulk field to acquire an expectation value in presence of a boundary is that its holomorphic and antiholomorphic parts carry the same energy. The constant $d_{ii^+}$ is the normalization of the bulk field.
 \item The constant $A_{i\bi}^\alpha$ is non-trivial new data arising from the boundary, and should not be naively scaled to unity as that would interfere with the normalizations $d_{i\bi}$ of the bulk fields. $A_{i\bi}^\alpha$ may be seen as a matrix in $(i,\bi)$. More importantly it is a structure constant of the boundary and we turn to it in-depth momentarily. 
 \item Note that the one-point function of the identity field on the plane $\phi_{00}=\id^{\pl}$ is non-trivial a priori, $\corr{\id^{\pl}}=A_{00}^\alpha\neq1$ (On the plane, $\id^{\pl}$ is normalized, $d_{00}=1$). This is the celebrated \emphasize{g-factor}. It has important implications for boundary renormalization group flows, but we will not have time to discuss this. The reader is referred to \cite{Recknagel:2013uja} for details.
 \item  In contrast to the holomorphic two-point function on the plane \eqref{2pt}, there are two reasons for the absolute value to appear on the spacetime dependence. The first is that the only scale in the problem is the distance $(z-\bz)$ of the fields to the boundary (recall $z=x+\iu y$), and it better be real. The second is that on the full plane the holomorphic and antiholomorphic parts complemented each other to yield a real value for the correlator. Here we only have this one piece so it is forced to be real by itself.
 \item The folding trick works for general $n$-point functions of bulk fields, turning them into $2n$-point functions. When extended symmetries are at play, the boundary may induce an automorphism on the symmetry algebra leading to a further action $\omega$ on the representations, $\bi\to\omega(\bi)$, but this is not discussed here; see \cite{Recknagel:2013uja}.
\end{itemize}

\begin{tcolorbox}
\hypersetup{linkcolor=\boxlinkcolor}
Observe the singularity in \eqref{1ptBdy} when $\phi_{i\bi}$ approaches the boundary, $z\to0$, as in the right panel of \figref{figFolding1pt}. This, fellow detectives, is the smoking gun of an excitation localized at the boundary!
 This idea is formalized by the \emphasize{bulk-boundary OPE} \new{\cite{cardy1991bulk}},
\begin{equation}\label{bulkBdyOPE}
 \phi_{i\bi}(z,\bz)=\sum_{k}C_{(i\bi)k}^\alpha|z-\bz|^{h_k-h_i-h_{\bi}}\psi_k(x)
\end{equation}
The fields $\psi_k(x)$ are localized at the boundary and cannot be moved into the interior of the upper half-plane. The boundary fields $\psi_k$ fall into representations of the Virasoro algebra governing the BCFT generated by \eqref{VirasoroModeH}. The structure constants $C^\alpha_{(i\bi)k}$ are non-trivial dynamical data and are not fixed by conformal symmetry.
\end{tcolorbox}
While $\psi_k$ stands for primaries and descendants in \eqref{bulkBdyOPE}, from now on we take it to always mean a primary, unless stated otherwise. To get our hands on the structure constants $C^\alpha_{(i\bi)k}$, we must in general solve non-linear, model dependent constraints such as the \textit{sewing constraint}. Since it is too time-consuming, we will not discuss it here, and refer the reader to the book \cite{Recknagel:2013uja}. There are still a number of insightful conclusions we can draw here with an expendable amount of effort.

First off, we emphasize that the identity field on the boundary $\psi_0=\id^{\hp}$ is \textit{not} the same field as the identity field on the plane $\phi_{00}=\id^{\pl}$. The reason being that $\id^{\pl}$ adheres to two copies of the Virasoro algebra, while $\id^{\hp}$ adheres only to one. More specifically, $\id^{\pl}$ is invariant under the global conformal group on the plane $\SL(2,\C)/\Z_2$, while $\id^{\hp}$ is invariant under the conformal group on the boundary $\SL(2,\R)/\Z_2$. Another way of seeing this is to note that $\id^{\pl}$ gives rise to state in $\cH^{\pl}$, see \eqref{planeSpectrum}, while $\id^{\hp}$ gives rise to a state in the boundary state space  $\cH_\alpha$ that we discuss momentarily. Nevertheless, it is reasonable to demand that $\id^{\pl}$ neatly becomes $\id^{\hp}$ when on the boundary, implying $C^\alpha_{(00)k}=\delta_{k,0}$. While the bulk-boundary OPE \eqref{bulkBdyOPE} thus becomes $\id^{\pl}=\id^{\hp}$, it actually only means that these fields \textit{act the same} in the current setup. 

Second, the conformal families $[\psi_k]$ appearing on the right-hand-side of \eqref{bulkBdyOPE} naturally follow from the fusion of the bulk field's conformal families, 
\begin{equation}
 [\phi_i]\fuse[\phi_{\bi}]=\sum_k\fus_{i\bi}^k\,[\psi_k]
\end{equation}
Clearly, the structure constant $C_{(i\bi)k}^\alpha=0$ whenever its corresponding fusion coefficient $\fus_{i\bi}^k=0$. The concept of fusion and our notation is discussed in \secref{secFusion}.

Third, we make use of the unbroken scale invariance along the boundary. This implies, that the only boundary field $\psi_k$ allowed to have a non-vanishing one-point function is the identity field on the boundary $\psi_0=\id^{\hp}$, $\corr{\psi_k}_\alpha=\Lambda^\alpha\delta_{k,0}$ for some constant $\Lambda^\alpha$. 
\begin{exercise}\label{exBulkBdyOPE1pt}
 Show that the proportionality constant is given by
 \begin{equation}\label{1ptBdyField}
 \corr{\psi_k(x)}_\alpha=A^\alpha_{00} \delta_{k,0}
 \end{equation}
 Use this to relate the bulk field's one-point coefficient to the bulk-boundary structure constants
 \begin{equation}\label{bulkBdyOPE1ptBdy}
  C_{(ii^+)0}^\alpha=\frac{A_{ii^+}^\alpha}{A_{00}^\alpha}
\end{equation}

\end{exercise}
\ifsol
\solution{exBulkBdyOPE1pt}{
\begin{equation}
 \corr{\id^{\pl}}_\alpha
 =
 A_{00}^\alpha d_{00}
 =
 \sum_kC_{(00)k}^\alpha|z-\bz|^{h_k}\corr{\psi_k(x)}_\alpha
 =
 C_{(00)0}^\alpha\Lambda^\alpha
\end{equation}
Using that $d_{00}=C_{(00)0}^\alpha=1$, the claim \eqref{1ptBdyField} follows. Repeat this with a bulk field with non-vanishing one-point function,
\begin{equation}
 \corr{\phi_{ii^+}(z,\bz)}
 =
 \frac{A^\alpha_{ii^+}}{|z-\bz|^{2h_i}}
 =
 \sum_kC_{(ii^+)k}^\alpha|z-\bz|^{h_k-2h_i}\corr{\psi_k(x)}_\alpha
 =
 \frac{C_{(ii^+)0}^\alpha A^\alpha_{00}}{|z-\bz|^{2h_i}}
\end{equation}
Rearranging this formula provides the second claim \eqref{bulkBdyOPE1ptBdy}.
}
\fi
\subsection{Boundary State Spaces and Boundary-Condition Changing Operators}\label{secStateSpaces}
As mentioned above the state space of the CFT on the full plane \eqref{planeSpectrum} is only an index set for the BCFT. We are now in a position to understand what the \emphasize{state space of the BCFT} is. Additionally, we allow for and discuss the possibility that the boundary condition changes along the real line, leading to the concept of a \emphasize{boundary-condition changing operator}.

The gluing condition \eqref{glueT} ensures that the state space describing the boundary falls into irreducible representations of the Virasoro algebra generated by \eqref{VirasoroModeH},
\begin{equation}\label{Halpha}
\cH_{\alpha}=\bigoplus_{i\in\confFam}\cH_i^{\oplus\nia} 
\end{equation}
As explained in detail in \secref{secCardyConstraint}, the boundary condition $\alpha$ together with the bulk spectrum \eqref{planeSpectrum} select which modules $\cH_i$ are allowed to appear in \eqref{Halpha} and with what multiplicity $\nia\in\N_0$. It is stressed however that the bulk fields $\phi_{i\bi}$ do not correspond to the states in $\cH_\alpha$ since they transform under two copies of the Virasoro algebra.

The individual modules $\cH_i$ correspond to the conformal families $[\psi_i]$ of the boundary fields via the operator-state correspondence. To make this more precise, let us assume the presence of an $\SL(2,\R)$ invariant vacuum state $\ket{0}\in\cH_\alpha$. The operator-state correspondence is the statement that for any state $\ket{v}\in\cH_\alpha$ there exists a field operator such that
\begin{equation}
 \lim_{x\to0}\Psi_v(x)\ket{0}=\ket{v}
\end{equation}
Because conformal families are the \quotes{field} version of a Virasoro module, the field operators $\Psi_v$ are indeed the boundary fields $\psi_k$ (and their descendants) encountered in the previous subsection. This entails that the primary states $\ket{i}=\psi_i(0)\ket{0}$ for a primary field $\psi_i$ satisfy the familiar highest weight conditions, 
\begin{equation}
 L^{\hp}_0\ket{i}=h_i\ket{i},
 \qquad
 L_n^{\hp}\ket{i}=0 \quad \text{for } n>0
\end{equation}
Translating from state to operator language, this is the statement that
\begin{equation}
 [L^{\hp}_n,\psi_i(x)]=x^n\left(x\p_x+h(n+1)\right)\psi_i(x)
\end{equation}
analogous to chiral fields on the plane, see \eqref{LnPrimaryCommutator}.
\medskip 

So far we have restricted to the case that only a single boundary condition is present on the real line. While we will be very precise only in \secref{secCardyConstraint}, we anticipate here that the gluing conditions \eqref{glueT} allow for several distinct boundary conditions $\alpha$. To give an idea, think of a free fermion or free boson theory. The free fields naturally admit Neumann (free) and Dirichlet (fixed) boundary conditions. Individually, both types of boundary conditions fall into the framework discussed above, i.e. they are distinct $\alpha$.

\begin{figure}
 \begin{center}
  \includegraphics[scale=.3]{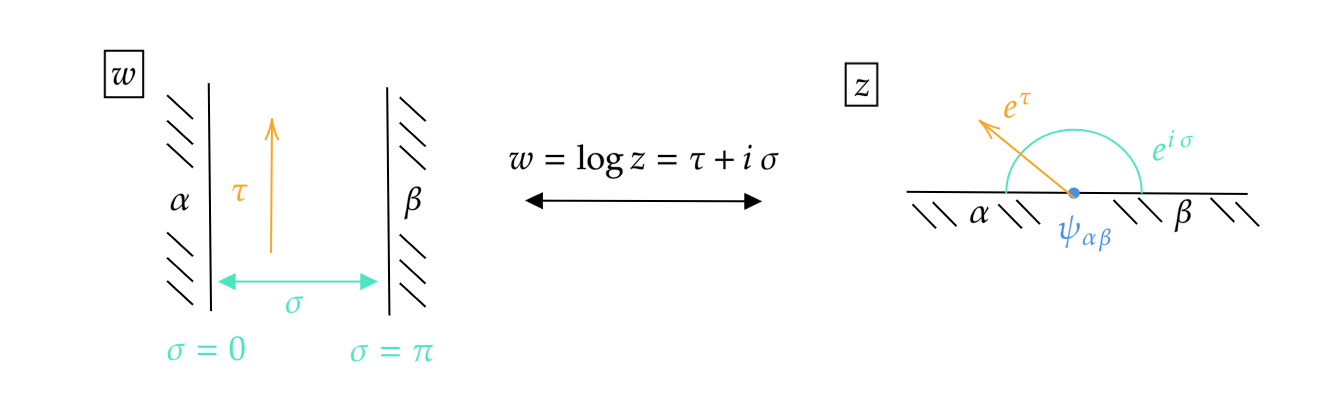}
 \end{center}
\caption{Left: A strip with two distinct boundary conditions $\alpha$ and $\beta$. Right: On the plane we see that this setup is induced by a BCCO $\psi^{\alpha\beta}$ situated on the real line.}
\label{figBCCO}
\end{figure}


Moving forward, it is now perfectly reasonable to ask about situations where the boundary condition changes from one boundary condition $\alpha$ to a distinct boundary condition $\beta$. This is best motivated on the strip, which is reached from the upper half-plane after the conformal map \eqref{ExpMap} where the phase is restricted to the range $\spa\in[0,\pi]$. As depicted on the left of \figref{figBCCO}, we are free to place a boundary condition $\beta$ at $\spa=0$ and a distinct boundary condition $\alpha$ at $\spa=\pi$ \footnote{It should be noted that the two ends of the strip ar secretely connected at $\tau\to-\infty$.}. Mapping this back onto the plane, we find a discontinuity in boundary conditions at $z=0$, with $\alpha$ on $\Re(z)<0$ and $\beta$ on $\Re(z)>0$. This is the hallmark of another kind of boundary field, the \emphasize{boundary-condition changing operator} (BCCO) $\psi^{\alpha\beta}$, mediating between two boundary conditions $\alpha$ and $\beta$. 

The BCCOs can as well be assigned a Hilbert space which we call $\cH_{\alpha\beta}$\footnote{A small word of warning for anybody who wishes to delve deeper into BCFT using the book \cite{Recknagel:2013uja}: what is called $\cH_{\alpha\beta}$ here is $\cH_{\beta\alpha}$ in the book. }. If $\alpha$ and $\beta$ are conformal, i.e. respect \eqref{glueT}, then this Hilbert space decomposes into irreducible Virasoro modules $\cH_i$ with multiplicities $\niab\in\N$,
\begin{equation}\label{bdySpaceAlphaBeta}
 \cH_{\alpha\beta}=\bigoplus_{i\in\confFam}\cH_i^{\oplus\niab} 
\end{equation}
The state space \eqref{Halpha} corresponds to the fields which do not change the boundary condition, or in other words, those fields changing $\alpha$ to $\alpha$, hence $\cH_{\alpha\alpha}\equiv\cH_\alpha$ and $\niaa=\nia$. 

It should be noted at this point that the $\SL(2,\R)$ invariant vacuum state does not appear for $\alpha\neq\beta$, i.e. $\ket{0}\notin\cH_{\alpha\beta}$ . The reason is simple: the vacuum state corresponds to the identity field $\id^{\hp}$, which by definition acts trivially, and hence cannot change the boundary condition. In turn this implies that the conformal weights $h(\psi^{\alpha\beta})>0$.

In passing let us mention a few modern applications of BCFT and BCCOs. First, BCCOs play an important role in symmetry-protected topological phases of matter (SPTs). It is possible to associate specific conformal boundary conditions $\alpha$ an SPT \cite{cho2017relationship}. Imagine a setup, where two (1+1)-dimensional SPTs, one labeled by $\alpha$ and the other by $\beta$, are separated by a domain wall. These domain walls are required to host degenerate edge modes and they are described by BCCOs in $\cH_{\alpha\beta}$. Said degeneracy is in fact a topological marker. Another application comes about in entanglement studies, where one thinks of $\cH_{\alpha\beta}$ as describing distinct ways of factorizing a bulk CFT Hilbert space $\cH\to\cH_A\otimes\cH_B$, where $A$ and $B$ are spatial domains \cite{ohmori2015physics}.  

\begin{figure}
  \includegraphics[scale=.3]{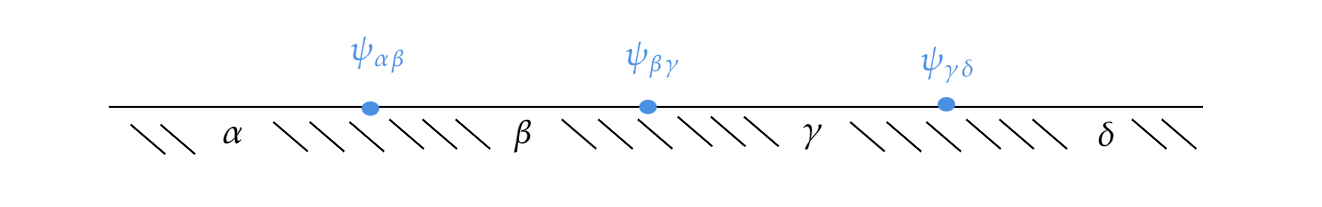}
\caption{A boundary with several BCCOs.}
\label{figSeveralBCCOs}
\end{figure}


Nothing stops us now from considering systems with several jumps in its boundary condition, see for example \figref{figSeveralBCCOs} . This is formalized by the boundary-boundary OPE
\begin{equation}
 \psi_k^{\alpha\beta}(x_1)\psi_l^{\gamma\delta}(x_2)
 =
 \sum_m\delta_{\beta\gamma}\,C^{\alpha\beta\delta}_{klm}(x_1-x_2)^{h_m-h_k-h_l}\psi^{\alpha\delta}_m(x_2)
\end{equation}
Because these fields are restricted to the real line, we need to fix an order, $x_1>x_2$. The Kronecker deltas secures that the boundary conditions can indeed be joined. In this formula we have again been lax about distinguishing primaries from descendants, i.e. the labels $k,l,m$ describe both sorts of fields. 
\subsection{The Boundary State Formalism}\label{secBdyState}
In this section we develop one the most sophisticated tools at our disposal in the study of BCFT. \emphasize{Boundary states} are a rephrasing of the boundary condition in terms of bulk CFT data, allowing us to elegantly constrain the boundary spectrum in \secref{secCardyConstraint}. We anticipate that this does not mean that the new degrees of freedom populating the boundary are secretely already present in the bulk theory; but this will become clearer as we move along. In this section, we encounter the \emphasize{Ishibashi equation} and a specific class of its solutions, the so-called \emphasize{Ishibashi states}.

We set out with the finite temperature definition of bulk field correlators in presence of a boundary $\beta$ for $\Re(z)<0$ and $\alpha$ for $\Re(z)>0$,
\begin{equation}\label{ThermalCorr}
 \corr{\phi_1^{\hp}(z_1,\bz_1)\dots \phi_N^{\hp}(z_N,\bz_N)}^{\beta_0}_{\beta\alpha}
 :=
 \Tr_{\cH_{\beta\alpha}}\left(e^{-\beta_0H^{\hp}}\phi_1^{\hp}(z_1,\bz_1)\dots \phi_N^{\hp}(z_N,\bz_N)\right)
\end{equation}
with inverse temperature $\beta_0=1/T$ and the $\hp$ superscript on the bulk fields indicates that these are situated on the upper half-plane. The Hamiltonian is 
\begin{equation}
 H^{\hp}=L_0^{\hp}-\frac{\cc}{24}
\end{equation}
The loci $z_i$ are radially ordered. For the correlator to be periodic\footnote{For fermions, the correlator is also allowed to be anti-periodic upon traversing the range of $\tau$.} in time $\tau=\log|z|$ up to a scale factor, the bulk fields are assumed to be quasi-primary, $\phi^{\hp}(\lambda z,\bar{\lambda}\bz)=\lambda^{-h}\bar{\lambda}^{-\bh}\phi^{\hp}(z,\bz)$. 

\begin{figure}
 \begin{center}
  \includegraphics[scale=.3]{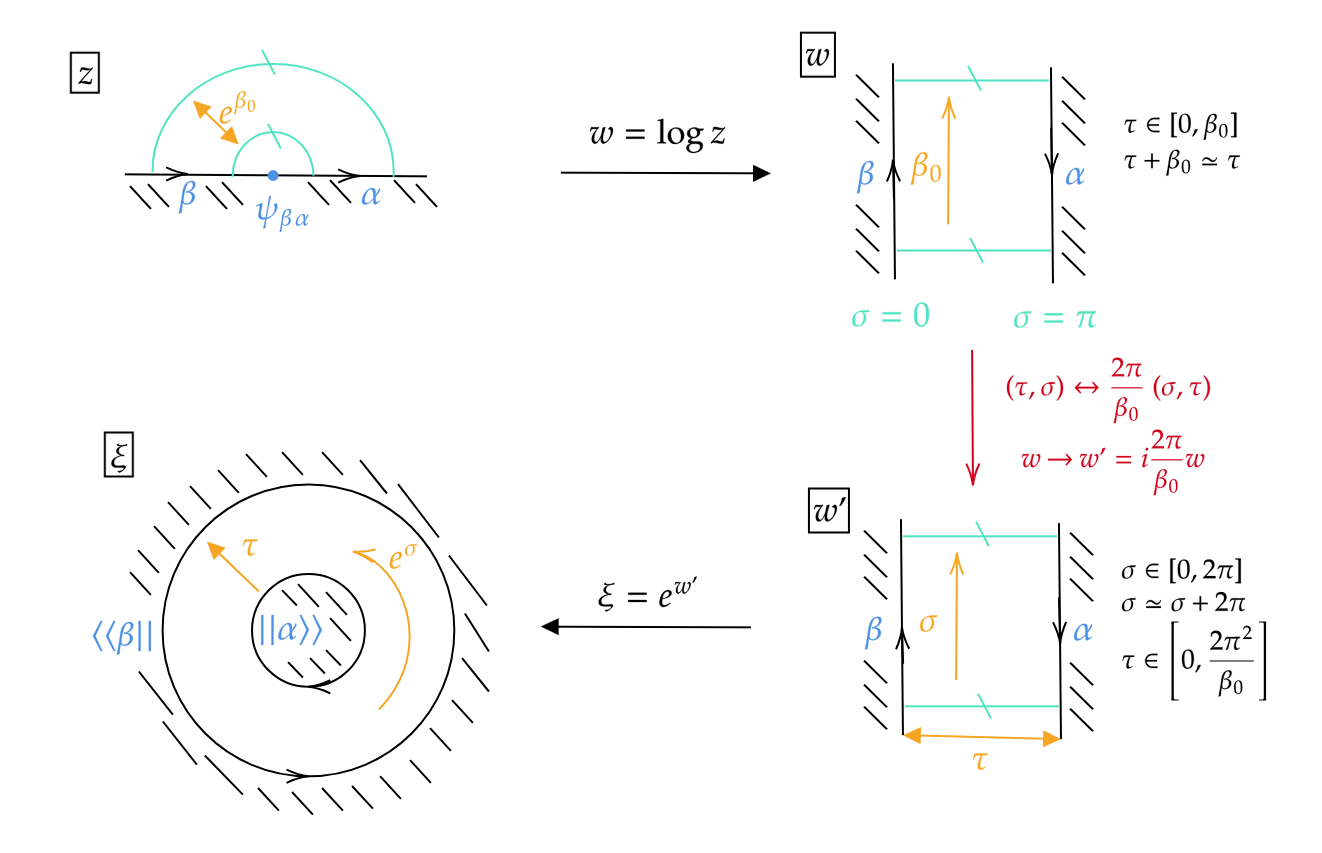}
 \end{center}
\caption{Upper left ($z$-frame): The fundamental domain of a thermal system on the upper half-plane with boundary conditions $\beta$ on the left and $\alpha$ on the right is enclosed by the two green semi-circles. Upper right ($w$-frame): The same thermal setup, now on the infinite strip. The BCCO $\psi^{\beta\alpha}$ has been pushed to $\tau\to-\infty$ and is not depicted any longer. Lower right ($w'$-frame): Time and space are swapped (multiplication with $\iu$.) and rescaled by $2\pi/\beta_0$. Lower left: The setup is exploded onto the full complex plane, forming an annulus, where $\beta$ labels now the outer ring and $\alpha$ the lower ring. Because of the opposing orientation, indicated by arrow tips in all frames, the CPT operator $\CPT$ dresses the outer boundary.}
\label{figAnnulus}
\end{figure}

For $N=0$ the correlator \eqref{ThermalCorr} reduces to the BCFT partition function and we shall focus on this case. On the upper half-plane this is visualized as two semi-circles a radial distance $e^{\beta_0}$ apart, which are identified, see the $z$-frame of \figref{figAnnulus}. We can choose the thermal fundamental domain on the upper half-plane to have its inner semi-circle at $|z|=e^0=1$ and its outer semicircle at $|z|=e^{\beta_0}$. In the following, we perform manipulations of this setup, which will have it appear as an overlap amplitude between states in a bulk CFT. These states are the boundary states encoding a boundary condition. 

The first step is to move to the strip via the exponential map $w=\log z=\tau+\iu\spa$. The spatial coordinate lies within $\spa\in[0,\pi]$, as before, and due to the finite temperature the temporal coordinate is now periodic too, i.e. $\tau\in[\tau_0,\tau_0+\beta_0]$, $\tau_0+\beta_0\simeq\tau_0$. The fundamental domain that we work with is thus a finite cylinder, see the $w$-frame of \figref{figAnnulus}. 

An innocent but crucial observation is that in Euclidean space, time and space are on equal footing, so that we are free to swap $(\tau,\sigma)\leftrightarrow(\spa,\tau)$. To make contact with modular $\modS$ transformations later, we rescale the new coordinate by $2\pi/\beta_0$. In short, we perform $w\to w'=2\pi\iu w/\beta_0$. Here, the $\iu$ is responsible for the swap of space and time. The new coordinates take values in $\spa\in[\spa_0,\spa_0+2\pi]$ with $\spa\simeq \spa_0+2\pi$ and $\tau\in[0,2\pi^2/\beta_0]$; see the $w'$-frame of \figref{figAnnulus}.

This \quotes{new} cylinder can now be exploded back onto the plane via the exponential map
\begin{equation}
 \xi=e^{w'}=\exp\left(\frac{2\pi\iu}{\beta_0}\log z\right),
 \qquad
 \bar{\xi}=e^{\bar{w}'}=\exp\left(-\frac{2\pi\iu}{\beta_0}\log \bz\right)
\end{equation}
as presented in the $\xi$-frame of \figref{figAnnulus}. The boundary $\alpha$, formerly at $\Re(z)>0$, lies now on a circle of radius $|\xi_{\text{min}}|=1$ and the boundary $\beta$, formerly at $\Re(z)<0$, lies now on a circle of radius $|\xi_{\text{max}}|=\exp(2\pi^2/\beta_0)$. More specifically, we find only the segments of the boundaries which lie in the thermal fundamental domain designated above, i.e. for $1\leq|z|\leq e^{\beta_0}$. The resulting setup is an \emphasize{annulus}.

Because the new time coordinate flows radially outward, the annulus reminds us of radial quantization \textit{in the bulk} for the particular case that a \quotes{bulk state} $\bket{\alpha}$ is prepared at the inner circle and another \quotes{bulk state} $\bbra{\CPT\beta}$ is prepared at the outer circle. $\CPT$ is the bulk CPT operator, which accounts for the opposite orientations of the inner and outer circles. The new bra-ket notation is in place to indicate that these are not ordinary states in the bulk. After all, there are no boundaries in the parent theory, so $\cH^{\pl}$ cannot know about the states $\bket{\alpha}$ or $\bbra{\CPT\beta}$, which implement the boundary conditions. Nevertheless, these new states are constructed from states in $\cH^{\pl}$, as we see in due time. For now, 
\begin{tcolorbox}
\begingroup
\hypersetup{linkcolor=\boxlinkcolor}
 We conclude that the boundary states are constructed to paraphrase a correlator in presence of a boundary via a \quotes{bulk overlap amplitude} \cite{cardy1989boundary}\footnotemark,
\begin{align}\label{BdyStatesDef}
 \bbra{\CPT\beta}e^{-\frac{2\pi^2}{\beta_0}H^{\pl}}\phi_1^{\pl}(\xi_1,\bar{\xi}_1)&\dots \phi_N^{\pl}(\xi_N,\bar{\xi}_N)\bket{\alpha}\\
 &:=
 \cJ(\xi,z)\,\Tr_{\cH_{\beta\alpha}}\left(e^{-\beta_0H^{\hp}}\phi_1^{\hp}(z_1,\bz_1)\dots \phi_N^{\hp}(z_N,\bz_N)\right)\notag
\end{align}
\new{where the field insertions lie within the fundamenal domain in the $z$-frame, and thus inside the annulus in the $\xi$-frame.}
\endgroup
\end{tcolorbox}
\footnotetext{\new{The original resource for the boundary state formalism \cite{cardy1989boundary} works without field insertions in \eqref{BdyStatesDef}, so that $\cJ$ is dropped. The arguments leading to this relation, however, permit the presence of bulk fields. While formally possible, inclusions of boundary fields typically lead to inpractical boundary states, at least to the authors knowledge.}}
On the left-hand-side, we have used superscripts $\pl$ to indicate that these fields are on the full plane, i.e. the $\xi$-coordinate frame. Importantly the Hamiltonian 
\begin{equation}\label{HamiltonianXiPlane}
 H^{\pl}=L_0^{\pl}+\bL_0^{\pl}-\frac{\cc}{12}
\end{equation}
is actually the bulk Hamiltonian with $\bar{\cc}=\cc$; recall exercise \ref{exHamiltonianCyl}. The function $\cJ(\xi,z)$ is the product of Jacobians arising from the standard transformations
\begin{align}
 \phi^{\pl}(\xi,\bar{\xi})
 &=
 \left(\frac{\p z}{\p\xi}\right)^{h}
 \left(\frac{\p \bz}{\p\bar{\xi}}\right)^{\bh}\phi^{\hp}(z,\bz)\\
 T^{\pl}(\xi)
 &=
 \left(\frac{\p z}{\p\xi}\right)^{2}
 T^{\hp}(z)+\frac{\cc}{12}S(z,\xi)
\end{align}
where the Schwarzian \eqref{Schwarzian} makes an appearance. 

\subsubsection{Ishibashi States}
So far we have gathered a heuristic understanding of boundary states. Now it is time to become more concrete about their properties, in particular on how these states implement boundary conditions. In other words, boundary states translate the gluing condition \eqref{glueT} from the upper half-plane parameterized by $z$ into the full complex plane parameterized by $\xi$. 

We start by choosing $\phi_N^{\hp}(z_N,\bz_N)=T^{\hp}(z_N)-\bT^{\hp}(\bz_N)$ in \eqref{BdyStatesDef}. By virtue of \eqref{glueT}, the correlator \eqref{BdyStatesDef} vanishes at $\bz=z$ regardless of the remaining field insertions $\phi_r(z_r,\bz_r)$ at the remaining $N-1$ loci. We restrict to real and positive $z$ meaning that we are dealing with the boundary state $\bket{\alpha}$ implemented at the unit circle $|\xi_{\text{min}}|=1$ of the annulus. The case of negative real $z$ is commented on later. The following observations are useful in mapping the energy-momentum tensor from the upper half-plane to the annulus for $z>0$,
\begin{equation}
 \bar{\xi}=\xi^{-1},
 \qquad
 \left(\frac{\p\bar{\xi}}{\p\bz}\right)\left(\frac{\p\xi}{\p z}\right)^{-1}=-\bar{\xi}^2,
 \qquad
 S(\xi,z)=S(\bar{\xi},\bz)
\end{equation}
Translating the right-hand-side of \eqref{BdyStatesDef} to its left-hand-side, the last of these properties has the Schwarzian contribution cancel, leading to
\begin{equation}\label{almostIshibashi}
 \left[
 \left(\frac{\p\xi}{\p z}\right)^{2}T^{\pl}(z)
 -
 \left(\frac{\p\bar{\xi}}{\p\bz}\right)^2\bT^{\pl}(\bar{\xi})
 \right]\bket{\alpha}
 =
 0
\end{equation}
Crucially, this holds only when evaluated on the unit circle, $\bar{\xi}=\xi^{-1}$. We can remove the spacetime dependence altogether by employing the standard mode expansion
\begin{equation}
 T^{\pl}(\xi)=\sum_n\xi^{-n-2}L_n^{\pl},
 \qquad
 \bT^{\pl}(\bar{\xi})=\sum_n\bar{\xi}^{-n-2}\bL_n^{\pl}
\end{equation}
Dividing \eqref{almostIshibashi} by $\left(\frac{\p\xi}{\p z}\right)^{2}$ and ordering by powers of $\xi$, we derive the
\begin{tcolorbox}
 \emphasize{Ishibashi condition}
 \begin{equation}\label{IshibashiCond}
  \left[
 L_n^{\pl}
 -
 \bL_{-n}^{\pl}
 \right]\bket{\alpha}
 =
 0
 \end{equation}
 It endows the boundary state $\bket{\alpha}$, situated on the unit circle of the annulus, $\bar{\xi}=\xi^{-1}$, with the gluing condition $T^{\hp}(z)=\bT^{\hp}(\bz)$ for $z=\bz>0$ on the upper half-plane.
\end{tcolorbox}
This analysis can be repeated at the outer circle of the annulus situated at $|\xi_{\text{max}}|=\exp(2\pi^2/\beta_0)$ leading to $\bbra{\CPT\beta}\left[L_n^{\pl} - |\xi_{\text{max}}|^{2n}\bL_{-n}^{\pl}\right]=0
 $, however no new information is extracted from this condition if $\bket{\beta}$ already satisfies \eqref{IshibashiCond}. This is a linear constraint on a vector. 
 Ishibashi showed the following \cite{Ishibashi:1988kg,Onogi:1988qk}:
\begin{tcolorbox}
\hypersetup{linkcolor=\boxlinkcolor}
 For each irreducible representation $i\in\confFam$ of the chiral algebra, in our case the Virasoro algebra $\vir$, a solution of the Ishibashi equation \eqref{IshibashiCond} can be constructed. These solutions are called \emphasize{Ishibashi states} and take the form
 \begin{equation}\label{IshibashiState}
  \iket{i}=\sum_{N=0}^\infty\ket{i,N}\otimes U\ket{i,N}\in\cH_i\otimes\cH_{i^+}
 \end{equation}
Here, $\ket{i,N},\,N\in\Z$ is an orthonormal basis in the irreducible module $\cH_i$ and $U$ is an anti-unitary operation\footnotemark acting like the chiral CPT operator, in particular $U\ket{i,N}\in\cH_{i^+}$. We choose $\ket{i,0}$ to be the primary state in $\cH_i$. Finally, $U\bL^{\pl}_n=\bL_n^{\pl}U$. Therefore $U$ maps Virasoro primaries into Virasoro primaries.

\end{tcolorbox}
\footnotetext{\new{A reminder on properties of anti-unitaries. Given two vectors $\ket{v},\ket{w}$ in a complex Hilbert space $\cH$ and an operator $A\in \End(\cH)$, they satisfy $\braket{Uv}{Uw}=\braket{v}{w}^*$, and $\bra{Uv}UA\ket{w}=\bra{w}A^\dagger\ket{v}$.}}
\new{Note the discrepancy in ket notation between $\bket{\alpha}$ and $\iket{i}$. The former is reserved for physical solutions, which is explained in \secref{secCardyConstraint}. See for instance equation \eqref{bdystate}.}

Following \cite{petkova2001conformal}, we now present an argument based on the folding trick, which justifies the statements in this box. Indeed, for any state $\ket{a}\in\cH_i\otimes\cH_{\bi^+}$ the folding trick prescribes an operation $X_a\in \Hom(\cH_i,\cH_{\bi})$, which works as follows
\begin{equation}
 \ket{a}
 =
 \sum_{N,\bar{N}=0}^\infty a_{N,\bar{N}}\ket{i,N}\otimes U\ket{\bi,\bar{N}}
 \qquad
 \overset{\textrm{fold}}{\longleftrightarrow}
 \qquad
 X_a
 =
 \sum_{N,\bar{N}=0}^\infty a_{N,\bar{N}}\ketbra{i,N}{\bi,\bar{N}}
\end{equation}
Because folding reverses orientation, the chiral CPT operator $U$ is removed/added upon unfolding/folding. The scalar product in $\cH_{\bi}$, for which $\bL_{-n}=\bL_n^\dagger$, implies that the Ishibashi condition \eqref{IshibashiCond} unfolds into $\pi_i(L_n)X_a=X_a\pi_{\bi}(L_n)$. We have been pedantic about representations $\pi$ here to emphasize that the operator $X_a$ intertwines the action of $\vir$ on the two irreducible modules $\cH_i$ and $\cH_{\bi}$. Schur's lemma thus forces these two representations to be equivalent $\cH_i\sim\cH_{\bi}$. Two consequences arise immediately. First, the labels of the two modules are the same, $\bi=i$. Second, $X_a$ is necessarily the projector $P_i$ onto $\cH_i$ meaning that $a_{N,\bar{N}}=\delta_{N,\bar{N}}$. Because no representations are excluded by the folding trick, we find that indeed, there exists one solution \eqref{IshibashiState} to \eqref{IshibashiCond} for each $i\in\confFam$. This demonstrates the statements in the box.

One can moreover show that the Ishibashi states $\iket{i}$ do not depend on the choice of basis $\ket{i,N}$. In fact, up to an overall phase the Ishibashi states are unique, as demonstrated in \cite{Recknagel:2013uja} (section 4.3.2). They are also linearly independent of each other, simply because distinct Ishibashi states are built from different energy eigenspaces of the bulk Hamiltonian and so one cannot use one Ishibashi state to construct the other.

Even though the Ishibashi state $\eqref{IshibashiState}$ is built from objects in $\cH^{\pl}$ it cannot possibly be inside that Hilbert space, because it does not converge in $\cH^{\pl}$. This is finally the highly anticipated justification alluded to above that boundary states are bulk objects that are not quite in the bulk. They cannot be true bulk objects, since no boundary data exist in the bulk $\cH^{\pl}$ after all and boundary states are built from the boundary condition \eqref{glueT}. Nevertheless, scalar products of states in $\ket{k,M}\otimes\ket{j,L}\in\cH^{\pl}$ of definite energy with $\iket{i}$ are still well-defined, and all such scalar products taken together constitute the information of $\iket{i}$. Therefore, the following exercise presents an alternative, and sufficient, route to show that \eqref{IshibashiState} solves the Ishibashi condition \eqref{IshibashiCond}.

\begin{exercise}\label{exIshibashiProof}
 Show that the Ishibashi state \eqref{IshibashiState} satisfies
 \begin{equation}
  \bra{k,M}\otimes\bra{U(j,L)}(L_n^{\pl}-\bL_{-n}^{\pl})\iket{i}=0
 \end{equation}
for all orthonormal basis vectors $\ket{k,M}\otimes\ket{j,L}\in\cH^{\pl}$. Therefore the Ishibashi condition  \eqref{IshibashiCond} must hold true. The Virasoro modes are to be thought of as $L_n^{\pl}=L_n^{\pl}\otimes \id$ and $\bL_{-n}^{\pl}=\id\otimes\bL_{-n}^{\pl}$.
\end{exercise}
\ifsol
\solution{exIshibashiProof}{The ${\pl}$ superscript on the Virasoro modes is dropped here to avoid clutter, 
\begin{align}
 \bra{k,M}&\otimes\bra{U(j,L)}(L_n-\bL_{-n})\iket{i}\notag\\
 &=
 \sum_{N=0}^\infty\biggl[\bra{k,M}L_n\ket{i,N}\bra{U(j,L)}U\ket{i,N}-\braket{k,M}{i,N}\bra{U(j,L)}\bL_{-n}U\ket{i,N}\biggr]\notag\\
 &=
 \sum_N\biggl[\delta_{ij}\,\delta_{LN}\,\bra{k,M}L_n\ket{i,N}-\delta_{ki}\,\delta_{MN}\,\bra{U(j,L)}U\bL_{-n}\ket{i,N}\biggr]\notag\\
 &=
 \delta_{ij}\,\bra{k,M}L_n\ket{i,L}-\delta_{ki}\,\bra{i,M}\bL_{+n}\ket{j,L}\notag\\
 &=
 \delta_{ij}\,\delta_{ki}\,\bigl(\bra{i,M}L_n\ket{i,L}-\bra{i,M}\bL_{+n}\ket{i,L}\bigr)\notag\\
 &=
 0
\end{align}
The first equality uses \eqref{IshibashiState}, the second that $U\bL_n=\bL_n U$, the third that $\bra{U(v)}UA\ket{w}=\bra{w}A^\dagger\ket{v}$ and that $L_n^\dagger=L_{-n}$ (recall exercise \ref{exVirModeDagger}), the fourth that $L_n$ maps an irreducible module $\cH_i$ into itself. Finally the last equality follows because $\bL_n$ acts on the right-moving module $\cH_i$ just as $L_n$ acts on the left-moving $\cH_i$.}
\fi

Despite the non-normalizability of the $\iket{i}$, it is still possible to introduce an \quotes{inner product} between Ishibashi states by employing a damping factor $q^{L_0-\cc/24}$. In physical situations, the damping factor has thermal origin involving the inverse temperature $\beta_0$, as we see below.
\begin{exercise}\label{exIshibashiOverlap}
 Starting with the $\xi$-frame Hamiltonian \eqref{HamiltonianXiPlane}, show that 
 \begin{equation}\label{IshibashiOverlap}
  \ibra{j}q^{\frac{1}{2}H^{\pl}}\iket{i}
  =
  \ibra{j}q^{L_0^{\pl}-\cc/24}\iket{i}
  =
  \delta_{ji}\,\chi_i(q)
 \end{equation}
where the character \eqref{character} of the conformal family $i$ is employed.
\end{exercise}
\ifsol
\solution{exIshibashiOverlap}{The ${\pl}$ superscript on the Virasoro modes is dropped here to avoid clutter, 
\begin{align}
 \ibra{j}q^{\frac{1}{2}(L_0-\bL_0-\cc/12)}\iket{i}
 &=
 \ibra{j}q^{L_0-\cc/24}\iket{i}\notag\\
 &=
 \sum_{M,N=0}^\infty\bra{j,M}\otimes\bra{U(j,M)}q^{L_0-\cc/24}\ket{i,N}\otimes U\ket{i,N}\notag\\
 &=
 \sum_{M,N=0}^\infty\bra{j,M}q^{L_0-\cc/24}\ket{i,N}\braket{i,N}{j,M}\notag\\
 &=
 \delta_{ij}\sum_{N=0}^\infty\bra{i,N}q^{L_0-\cc/24}\ket{i,N}\notag\\
 &=
 \delta_{ij}\,\tr_{\cH_i}\left[q^{L_0-\cc/24}\right]
\end{align}
The first equality uses \eqref{IshibashiCond}, the second plugged in the Ishibashi state \eqref{IshibashiState}, the third reordered terms and used up the antiunitarity of $U$, the fourth the orthognoality of states and the last recognized the definition of the trace over $\cH_i$, which is the character.
}
\fi

\subsection{The Cardy Constraint: Physical Boundary States}\label{secCardyConstraint}
We have now understood the solution space to the Ishibashi condition \eqref{IshibashiCond}. In this subsection, we learn how to construct a \emphasize{physical boundary state $\bket{\alpha}$} describing the boundary condition $\alpha$ from Ishibashi states by virtue of the \emphasize{Cardy constraint}.

The Ishibashi states form linearly independent solutions to the Ishibashi condition \eqref{IshibashiCond}. Thus they function as basis for any solution of the Ishibashi condition,
\begin{equation}\label{bdystate}
 \bket{\alpha}=\sum_{i\in\ihp}B_\alpha^i\,\iket{i}
\end{equation}
Any such superposition implements the gluing condition $\eqref{glueT}$, however, not all of these are physical. To determine what constitutes a physical boundary condition we need to constrain the coefficients $B_\alpha^i$ so that they reproduce the boundary degrees of freedom discussed in \secref{secStateSpaces}, which we turn to momentarily. Beforehand, we remark
\begin{itemize}
 \item The only bulk fields $\phi_{i\bi}$ corresponding to states in $\cH^{\pl}$ -- recall \eqref{planeSpectrum} -- that survive the placement of the boundary are those acquiring a one-point function \eqref{1ptBdy}. By means of the folding trick these were seen to be the select few in the set 
 \begin{equation}\label{ihp}
  \ihp:=\{i\in\confFam\,|\,(i,i^+)\in\ipl\}
 \end{equation}
 Hence, not all Ishibashi states contribute to \eqref{bdystate}. Equivalently we could have written $\bket{\alpha}=\sum_{i,\bi\in\confFam}B^{i\bi}_\alpha\delta_{\bi,i^+}\iket{i}$.
 \item There is in fact a stronger connection with the one-point functions \eqref{1ptBdy} and the bulk-boundary OPE coefficients of \eqref{bulkBdyOPE}. Indeed one can show that \new{\cite{cardy1991bulk}}\footnote{\new{The paper \cite{cardy1991bulk} uses a slightly altered, yet equivalent definition of boundary states. For a correct matching with the conventions used here see 4.3.3 of \cite{Recknagel:2013uja}.}}
 \begin{equation}\label{1ptCoeffBdyCoeff}
  C_{(ii^+)0}^\alpha=\frac{B^{i^+}_\alpha}{B^{0}_\alpha}
  \qquad
  \overset{\eqref{bulkBdyOPE1ptBdy}}{\Rightarrow}
  \qquad
  B^{i^+}_\alpha=\kappa\, A_{ii^+}^\alpha
 \end{equation}
where $\kappa\neq0$ is a normalization constant. As pointed out by \eqref{1ptBdy}, the one-point coefficients $A_{ii^+}^\alpha$ are fundamental data of the boundary and, as we see here, they determine the boundary states via 
\begin{equation}\label{1ptBdyState}
\bbra{\CPT\alpha}=\sum_{i\in\ihp}B_\alpha^{i^+}\ibra{i}
\qquad
\Rightarrow
\qquad
 \corr{\phi_{ii^+}(z,\bz)}
  =
  \frac{1}{\kappa}\frac{\bbra{\CPT\alpha}\phi_{ii^+}\rangle \, d_{i i^+}}{|z-\bz|^{2h_i}}
\end{equation}
underlining their relevance. Here, $\CPT B^j_\alpha\iket{j}=(B^j_\alpha)^*\iket{j^+}$ has been used. A derivation of this important fact would lead us too far afield, but interested readers are referred to section 4.3.3 of \cite{Recknagel:2013uja}. The overall normalization of boundary states is determined by the Cardy constraint below, which also fixes $\kappa$. While \eqref{1ptBdyState} gives physical justification to the $B_\alpha^i$, for now it only replaces one set of unknowns, namely $A^\alpha_{ii^+}$, by the new set of unkowns $B_\alpha^i$.
\item
In \secref{secBCFTgeneralities} we remarked that a BCFT is governed by two fundamental data, the gluing condition \eqref{glueT} and the one-point functions \eqref{1ptBdy}. Note how the boundary state \eqref{bdystate} elegantly combines these two pieces of information. The gluing condition \eqref{glueT} results in the Ishibashi condition \eqref{IshibashiCond} and the Ishibashi states \eqref{IshibashiState} furnish an $|\confFam|$-dimensional solution space thereof. What is left is to find the physically meaningful points in this solution space. As explained in \eqref{1ptCoeffBdyCoeff}, these physical points are controlled precisely by the one-point functions of the bulk fields. 
\end{itemize}

We now turn to constraining the coefficients $B_\alpha^i$, and thus -- finally -- also the one-point data $A^\alpha_{ii^+}$, by relating them to the boundary state spaces \eqref{bdySpaceAlphaBeta}. As was pointed out in \eqref{bdySpaceAlphaBeta}, a physical boundary state space has non-negative integer multiplicities $\niab\in\N_0$. Preparing the system at finite temperature $\beta_0=-2\pi\iu\tau$ and tracing its thermal density matrix $e^{-\beta_0 H^{\hp}}=q^{L_0-\cc/24}$ over $\cH_{\alpha\beta}$ leads to the system's partition function
\begin{equation}\label{bdyZAlphaBeta}
 Z_{\alpha\beta}(q)
 =
 \sum_{i\in\confFam}\niab\,\chi_i(q) 
\end{equation}
where $q=e^{-\beta_0}=e^{2\pi\iu\tau}$ and we re-encounter the characters \eqref{character} of a conformal family $i$. This object counts boundary degrees of freedom $\psi_{\alpha\beta}$ propagating in a loop of compactified time $\tau\in[0,\beta_0]$, as sketched in the $w$-frame of \figref{figAnnulus}. By means of \eqref{BdyStatesDef} and $\tq=e^{-4\pi^2/\beta_0}=e^{-2\pi\iu/\tau}$, this is the same as\footnote{String theorists call \eqref{bdyZAlphaBeta} the open string picture, whereas they refer to \eqref{bulkZAlphaBeta} as the closed string picture. The fact that these two expressions are the same is termed open-closed worldsheet duality. In order to not estrange non-string theorists, we \new{do} not use this terminology.}
\begin{equation}\label{bulkZAlphaBeta}
 Z_{\alpha\beta}(q)
 =
 \bbra{\CPT\alpha}
 \tq^{\frac{1}{2}\left(L_0^{\pl}+\bL_0^{\pl}-\frac{\cc}{12}\right)}\bket{\beta}
 =
\sum_{i\in\ipl} B_\alpha^{i^+}B^i_\beta\,\chi_i(\tq)
\end{equation}
In order to reach the last equality, the \quotes{overlap} \eqref{IshibashiOverlap}, the left-hand side of \eqref{1ptBdyState} and that $i^+$ appears amongst the left-movers in $\cH^{\pl}$ whenever $i$ does were used. This way of looking at the partition function describes bulk degrees of freedom propagating from the boundary $\beta$ to the boundary $\alpha$; this corresponds to the $w'$-frame of \figref{figAnnulus}, albeit after swapping $\alpha\leftrightarrow\beta$. As emphasized in our discussion in \secref{secBdyState}, and evident here, this rephrasing of the partition function amounts to a modular $\modS$ transformation \eqref{modStransformation}, which interchanges $q\leftrightarrow\tq$. Combining everything, we end up with the following non-linear constraint.
\begin{tcolorbox}
 The boundary state coefficients $B^i_\alpha$ and multiplicities $\niab\in\N_0$ are subject to the \emphasize{Cardy constraint} \new{\cite{cardy1989boundary}}
 \begin{subequations}\label{CardyConstraint}
  \begin{align}
  \sum_{i\in\confFam}\sum_{j\in\ihp} B_\alpha^{j^+}B^j_\beta\,\modS_{ji\,}\chi_i(q)
  &=
  \sum_{i\in\confFam}\niab\,\chi_i(q) \label{CardyConstraintA}\\
  \Rightarrow\sum_{j\in\ihp} B_\alpha^{j^+}B^j_\beta\,\modS_{ji\,}
  &=
  \niab\,\in\,\N_0 \label{CardyConstraintB}
 \end{align}
 \end{subequations}
 
Solutions $B_\alpha^i,\,B_\beta^i$ to this equation yield a physical boundary spectrum $\niab$. For every parent CFT, we introduce a formal set $\cB$ accounting for all solutions $\alpha,\beta$ to this constraint, i.e. $\alpha,\beta\in\cB$. They are the sought after \emphasize{physical boundary states}.
\end{tcolorbox}
Some remarks are in order:
\begin{itemize}
 \item As it stands, equation \eqref{CardyConstraintB} does not follow directly from \eqref{CardyConstraintA} since characters are not necessarily linearly independent. In $\su(2)_k$ Wess-Zumino-Witten models, for instance, one finds for non self-conjugate sectors $\chi_{i^+}=\chi_i$. In most models, where such issues occur one has access to an extended symmetry algebra, with \quotes{charged} characters (called \quotes{unspecialized} in the mathematics literature), which are indeed linearly independent. Thus one can repeat our analysis here with these and at the very end extract \eqref{CardyConstraintB}.
 \item When $\beta=\alpha$, a solution to the Cardy constraint is called a \emphasize{self-consistent} boundary condition. The resulting $\niaa\equiv\nia$ provide the boundary fields on $\alpha$. Given a solution for $\beta\neq\alpha$, the two boundaries are called \emphasize{mutually consistent}. The $\niab$ describe the boundary condition-changing operators. Recall our discussion in \secref{secStateSpaces}.
 \item Single Ishibashi states $\iket{i}$ typically do not satisfy this equation, as they would have $B_\alpha^i=\delta_\alpha^i$. We say that they do not provide a consistent set of boundary fields. This little example of a non-solution assumes that the cardinality of $\cB$ agrees with that of $\confFam$, allowing us to identify the two sets. This is not generally true, but below we restrict to an example where this is the case and see what happens in detail. 
 \item It follows immediately from \eqref{CardyConstraint} that it is not allowed to multiply a physical boundary state $\bket{\alpha}$ by a non-integer overall factor. The solutions to \eqref{CardyConstraint} form rather a lattice over the non-negative integers. There are situations where $\niab$ is allowed to be a negative integer, for instance when fermions are around.
 \item Boundary conditions $\alpha$ with a unique vacuum, i.e. $\niaa=1$, are called \emphasize{elementary, fundamental} or \emphasize{simple}. The set of elementary boundary states generate the remaining boundary states through positive integer linear combinations.
 \item Viewing $\niab=(\mathsf{n^i)_{\alpha\beta}}$ as $|\cB|\times|\cB|$ matrix for fixed $i$, the Cardy constraint \eqref{CardyConstraintB} may be read as its decomposition into a number $|\ihp|$ of $|\cB|$-dimensional eigenvectors $(B^i)_\alpha=B_\alpha^i$ and its eigenvalues $\modS_{ji}$. 
 \item Under the assumption of an orthonormal and complete set of boundary conditions, discussed in \cite{petkova2001conformal, Recknagel:2013uja}, it can be shown that the matrices $\mathsf{n}^i$ furnish a representation of the fusion algebra,
 \begin{equation}
  \mathsf{n}^i\mathsf{n}^j=\sum_{k}\fus_{ij}^k\mathsf{n}^k
 \end{equation}
 These are called \emphasize{non-negative integer representations (NIM-rep)}. Due to $\fus_{ij}^k=\fus_{ji}^k$, the $\mathsf{n}^i$ must be mutually commuting matrices.
\end{itemize}

\begin{exercise}\label{exMultChargeConjugation}
 Show that $\mathsf{n}_{\alpha\beta}^{i^+}=\mathsf{n}^i_{\beta\alpha}$.
\end{exercise}
\ifsol
\solution{exMultChargeConjugation}{Using that $\modS_{ji^+}=\modS_{j^+i}$, one must only charge conjugate the summation index $j\to j^+$ in \eqref{CardyConstraintB} to show the claim.}
\fi

\subsubsection*{\new{A Remark on Boundary Sewing Relations}}\label{remBdySewing}
\new{An important question to mention before closing our general discussion of the Cardy constraint is whether its solutions are a fully consistent conformal boundary condition. Lewellen turned to address this issue in \cite{lewellen1992sewing}, where he categorized the Cardy constraint \eqref{CardyConstraint} as one of four \emphasize{sewing relations}, which tie the boundary and plane structure constants together.} 

\new{The additional sewing relations are derived from correlators combining bulk and boundary fields on the upper-half plane. In contrast to the sewing relations on the plane mentioned on page \pageref{remPlaneSewing}, on the boundary, locality of fields is lost. The reason is that boundary fields cannot be passed across each other because they are confined to a single dimension. In consequence their correlators must not be analytical and can have branch cuts. Hence the sewing relations on the upper-half plane derive solely from associativity of the OPE. For more references, the reader may consult section 4.4.2 of \cite{Recknagel:2013uja}. }

\new{The four sewing relations form a complete set of constraints on a BCFT. It can happen that a solution to the Cardy constraint fails to satisfy other sewing relations, and in principle, all of them must be checked. Typically, these sewing relations are difficult to handle. As a result, in practice one evaluates the bulk field one-point functions \eqref{1ptBdy} and assumes that they can be extended consistently to all sewing relations. Fortunately, in certain situations, one relation proves surprisingly practical. It is known as the \emphasize{cluster condition} and has been successfully employed to derive boundary states for the $\SU(2)$ Wess-Zumino-Witten model in \cite{gaberdiel2002conformal}.}

\subsubsection{An Example: The Cardy Case}\label{secCardyCase}
An example will clarify the preceeding concepts further. We study the case originally studied by Cardy, hence the name: \emphasize{the Cardy case}. 

Consider a rational parent CFT, i.e. one for which $|\confFam|<\infty$, with a charge-conjugate modular invariant,
\begin{equation}
 Z_{\textrm{ch.c}}(q,\bq)
 =
 \sum_{i\in\confFam}\chi_i(q)\chi_{i^+}(\bq)
\end{equation}
Evidently, all of these fields satisfy the requirements in \eqref{ihp} so that $\ihp=\confFam$. Cardy derived the following boundary states, as solution to the Cardy constraint \eqref{CardyConstraint}:

\begin{tcolorbox}
 The \emphasize{Cardy states} are \cite{cardy1989boundary}
\begin{equation}\label{CardyState}
 \bket{a}=\sum_{i\in\confFam}\frac{\modS_{ai}}{\sqrt{\modS_{0i}}}\,\iket{i}, 
 \qquad
 a\in\confFam
\end{equation}
\end{tcolorbox}
Since the modular $\modS$ matrix is an $|\confFam|\times|\confFam|$ matrix, we have a boundary state for any $a\in\confFam$. In other words, here $\cB=\confFam$. We have refrained to label these states by the greek letter $\alpha$ to indicate that this is a special case. 

\begin{exercise}\label{exCardySpectra}
 Use the Verlinde formula \eqref{Verlinde} to show that the boundary spectra \eqref{bdyZAlphaBeta} between two Cardy states $\bket{a}$ and $\bket{b}$ are controlled by the fusion rules
 \begin{equation}
  Z_{ab}(q)=\sum_{i\in\confFam}\fus_{ab^+}^i\chi_i(q)
 \end{equation}
 \ifsol
 \solution{exCardySpectra}{Recall \eqref{modSidentities} and \eqref{FusionChargeConjugation} and plug into the Cardy constraint \eqref{CardyConstraintA}
 \begin{equation}
  \sum_{i,j\in\confFam}
  \frac{\modS_{aj^+}}{\sqrt{\modS_{0j^+}}}
  \frac{\modS_{bj}}{\sqrt{\modS_{0j}}}
  \modS_{ji}\chi_i(q)
  =
  \sum_{i,j\in\confFam}\frac{\modS_{a^+j}\modS_{bj}\modS_{ji^+}^*}{\modS_{0j}}\chi_i(q)
  =
  \sum_{i\in\confFam}\fus_{a^+b}^{i^+}\chi_i(q)
  =
  \sum_{i\in\confFam}\fus_{ab^+}^{i}\chi_i(q)
 \end{equation}
}
\fi
\end{exercise}
The fusion rule coefficients are indeed integers, confirming the consistency of the Cardy states. Because $\fus_{aa^+}^0=1$, Cardy states are elementary. Note that Cardy's solution \eqref{CardyState} does not really make use of the Virasoro algebra. Instead it only requires the existence of a modular $\modS$ matrix. Thus any extended symmetry algebra allowing for rational models, such as in Wess-Zumino-Witten models or supersymmetric models, also has Cardy states.  

\subsubsection*{The Ising model}
As an example consider the Ising CFT. It has three primary fields $\confFam=\{\id,\epsilon,\sigma\}$. All three states have equal holomorphic and anti-holomorphic conformal weights, $h=\bh$. Their values are $h_\id=0$, $h_\epsilon=1/2$ and $h_\sigma=1/16$. All of its fields are self-conjugate so that the diagonal modular invariant \eqref{IsingDiagZ} is also charge-conjugate; hence we apply our recent discussion to it. The model's modular $\modS$ matrix is found in \eqref{IsingModS} and repeated here for convenience,
\begin{equation}
 \modS=
 \begin{pmatrix}
        \frac{1}{2} & \frac{1}{2} & \frac{1}{\sqrt{2}} \\
        \frac{1}{2} & \frac{1}{2} & -\frac{1}{\sqrt{2}} \\
        \frac{1}{\sqrt{2}} & -\frac{1}{\sqrt{2}} & 0 \\
       \end{pmatrix}
\end{equation}
where rows and columns are ordered as $\{\id,\epsilon,\sigma\}$. The resulting Cardy states are
\begin{subequations}\label{IsingCardyStates}
\begin{align}
 \bket{\id}
 &=
 \frac{1}{\sqrt{2}}\iket{\id}+\frac{1}{\sqrt{2}}\iket{\epsilon}+\frac{1}{2^{1/4}}\iket{\sigma}\\
 \bket{\epsilon}
 &=
 \frac{1}{\sqrt{2}}\iket{\id}+\frac{1}{\sqrt{2}}\iket{\epsilon}-\frac{1}{2^{1/4}}\iket{\sigma}\\
 \bket{\sigma}
 &=
 \iket{\id}-\iket{\epsilon}
\end{align}
\end{subequations}
Each of these boundary states prescribes a boundary condition on the bulk fields. In the Ising chain, the field $\sigma$ measures the spin. Three boundary conditions are well known: at the boundary, the spin may be either
\begin{itemize}
 \item fixed up, meaning $\corr{\sigma}_\alpha>0$, called \quotes{+},
 \item fixed down, meaning $\corr{\sigma}_\alpha<0$, called \quotes{-},
 \item unfixed, meaning $\corr{\sigma}_\alpha=0$, called \quotes{free}.
\end{itemize}
Taking the overlap with $\bra{\sigma}$, and recalling \eqref{1ptBdyState}, as well as \eqref{bulkBdyOPE1ptBdy}\footnote{We assume $\kappa>0$. Flipping the sign of $\kappa$ would only swap the assignments of $\bket{+}$ and $\bket{-}$.}, we see that we can identify $\bket{\id}=\bket{+},\,\bket{\epsilon}=\bket{-}$ and $\bket{\sigma}=\bket{\textrm{free}}$.

\newpage

\section{Topological Defects and Interfaces}\label{secDCFT}
So far we have discussed CFTs on the plane and have gained an understanding of the implications of adding a boundary to the system. We are now in a position to add another element of structure to a CFT, namely \emphasize{topological interfaces} and \emphasize{topological defects}. Traditionally, these have appeared in statistical and condensed matter systems as disorder lines; an example is present in the classic work \cite{Oshikawa:1996dj} on the Ising model. In modern research topological defects play an important role in the study of symmetries, including the modern subject of non-invertible symmetries; see for instance \cite{shao2023s} or Ho Tat Lam's lectures at this school and references therein. 

After a brief exposition of \emphasize{conformal interfaces} in \secref{secConfInterfaces}, we focus on one of their subsets, namely the \emphasize{topological interfaces}. First we develop a mathematical framework for these in \secref{secGeneralTopInterface}, which we employ in \secref{secPetkovaZuberConstraint} to derive the analog of the Cardy constraint for defects, namely the \emphasize{Petkova-Zuber constraint}. The simplest solution to said constraint are \emphasize{Verlinde lines}, which are discussed in \secref{secVerlindeLines}. We also use this as an opportunity to develop \emphasize{graphical calculus}, which is a valuable asset in working with interfaces, as it circumvents tedious calculation while showcasing the intrinsic structure of the theory. 

\subsubsection*{Useful Literature}
\begin{itemize}
 \item The main reference for the first part of this section is the seminal paper \cite{Petkova:2000ip} by Petkova and Zuber. The reader will find that we are following the narrative of this paper closely and generalize their discussion slightly here and there. The authors lectured on the same topic \cite{petkova2001conformal}. This reference provides another and leads into the mathematical field of graphs, which is not done in the present text.
 \item The topic of defects and interfaces aroused much interest in the mathematical physics community at the start of the millenium. In a Herculean feat, Fuchs, Runkel and Schweigert almost single-handedly developed the mathematical framework for defects \cite{Fuchs:2002cm, Fuchs:2003id, Fuchs:2004dz, Fuchs:2004xi, Fjelstad:2005ua}, which is being applied vigorously in contemporary research. They employ elegant formulations based in category theory, and mathematically inclined readers will enjoy their exposition.
 \item The notions of group-like and duality defects were introduced in \cite{Frohlich:2006ch}, and our presentation here is a minimalistic excerpt on some core features of these defects. The power of this formalism is demonstrated in \cite{Frohlich:2004ef}, which embeds the well-known Kramers-Wannier duality of the Ising model in this framework, and elegantly rederives many lessons, which are hard-won with other tools.
 \item To see how defects are manipulated to achieve results of various types, the reader can consult \cite{Runkel:2010ym, Kojita:2016jwe, Konechny:2019wff} and references therein. Deriving defects is generally difficult, yet for two free boson theories whose radii are multiple integers of the self-dual radius topological interfaces have been classified \cite{Fuchs:2007tx}.
\end{itemize}

\subsection{Conformal Interfaces}\label{secConfInterfaces}
\begin{figure}
 \begin{center}
  \includegraphics[scale=.3]{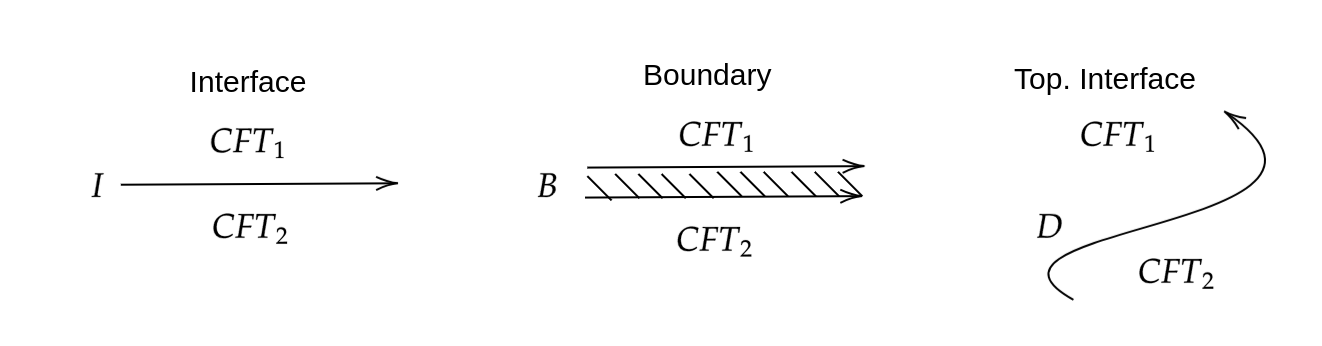}
 \end{center}
\caption{Left: An interface $I$ separating CFT$_1$ from CFT$_2$. Center: A totally reflective interface is a boundary. Right: A totally transmissive interface is a topological interface and may be deformed. Arrow tips indicate an orientation of a defect.}
\label{figInterface}
\end{figure}
To begin, we discuss the notion of an \emphasize{interface} or \emphasize{domain wall}. Imagine having two possibly distinct CFTs on the complex plane, CFT$_1$ one on the upper half-plane, and CFT$_2$ on the other the lower half-plane. This is depicted in the left panel of \figref{figInterface}. For this system to be consistent, some non-local operator must be spread on the real line gluing the two CFTs together. This is the interface! It is said to be a \emphasize{conformal interface} if it glues the energy-momentum tensors of the two CFTs according to
\begin{equation}\label{glueTdef}
 T^{(1)}(x)-\bT^{(1)}(x)=T^{(2)}(x)-\bT^{(2)}(x)\,,
 \qquad
 \text{ for all }x\in\R
\end{equation}
They are called conformal since upon folding this constraint reduces to the familiar conformal boundary condition \eqref{glueT} for the product theory CFT$_1\otimes$CFT$_2$. The interface is implemented by an operator $\defect:\cH_{(1)}\to\cH_{(2)}$ mapping between the CFT Hilbert spaces
 \begin{equation}\label{planeSpectra}
 \cH_{(1)}=\bigoplus_{(i,\bi)\in\ipl_{(1)} }\multZ_{i\bi}^{(1)}\,\cH_i\otimes\cH_{\bi}\,,
 \qquad
 \cH_{(2)}=\bigoplus_{(i,\bi)\in\ipl_{(2)} }\multZ_{i\bi}^{(2)}\,\cH_i\otimes\cH_{\bi}
\end{equation}
Because the central charges of the CFTs may differ, the sets of representations $\confFam$ may differ for CFT$_1$ and CFT$_2$ \footnote{It would be cleaner to decorate the sets $\confFam$ with an index for the central charge, but we will not dwell on this for long, so we are cavalier here.}. We write thus $\ipl_{(1,2)}\subseteq\confFam_{(1,2)}\times\confFam_{(1,2)}$, c.f. \eqref{planeSpectrum}. Note that we have refrained from placing a $\pl$ superscript on the $\cH_{(1,2)}$ as this distinction will not be necessary in this section. The interface with the opposite orientation similarly defines a map $\bar{\defect}:\cH_{(2)}\to\cH_{(1)}$. As we will see below, Hermitian adjungation inverts orientation of a defect.

The condition \eqref{glueTdef} states that energy is partly transmitted and reflected at the interface. \new{A model independent approach for the construction of conformal interfaces is difficult}\footnote{\new{Given a particular model, conformal interfaces have been constructed however. See for instance \cite{bachas2008fusion} for conformal interfaces between free boson theories, see \cite{Becker:2017zai} for interfaces between a free boson and its $\Z_2$ orbifold, see \cite{brehm2015entanglement} for conformal interfaces in the free fermion or see \cite{gaiotto2012domain} for conformal interfaces implementing RG flows in Virasoro minimal models.}}, but there are two obvious limiting cases in which progress can be made. First, if no energy passes through the real line, the interface is \emphasize{totally reflective}. This means that each side of \eqref{glueTdef} vanishes on its own and the real line presents an actual boundary \eqref{glueT} for each CFT; the two CFTs are disconnected, as depicted in the center panel of \figref{figInterface}. We have already learned in previous sections how to treat such a setup. The opposite limiting case is that of
 
\begin{tcolorbox}
\hypersetup{linkcolor=\boxlinkcolor}
 A \emphasize{totally transmissive} interface is characterized by a continuous energy-momentum tensor across the real line
 \begin{equation}\label{GlueTtopDef}
  T^{(1)}(x)=T^{(2)}(x),\qquad \bT^{(1)}(x)=\bT^{(2)}(x)\,,
 \qquad
 \text{ for all }x\in\R
 \end{equation}
 In operator language this implies \quotes{Virasoro blindness} of the interface operator\footnotemark
 \begin{equation}\label{topdef}
 [L_n,\defect]=0=[\bL_n,\defect]
\end{equation}
Since the Virasoro generators generate infinitesimal diffeomorphisms on the complex plane, this means one is allowed to deform the interface $\defect$ away from the real line and into any form for free, so long as we do not cross any field insertions or other objects; see the right panel of \figref{figInterface}. This means that the defect $\defect$ is characterized by the homotopy class of the contour it is following. Hence, these operators are called \emphasize{topological interfaces}.
\end{tcolorbox}
\footnotetext{\new{Because of \eqref{GlueTtopDef}, the Virasoro modes to either side of the defect coincide and we omit superscripts $L_n^{(1)}=L_n^{(2)}=L_n$ and similarly for $\bL_n$.}}

We speak of \emphasize{topological defects} when the two CFTs coincide, CFT$_1=$CFT$_2$. In particular, $\confFam_{(1)}=\confFam_{(2)}=\confFam$ in \eqref{planeSpectra} and the multiplicity matrices agree, $\multZ_{i\bi}^{(1)}=\multZ_{i\bi}^{(2)}$. Sometimes, this distinction is taken a bit more lightly in the literature however, in that the multiplicity matrices are allowed to be distinct while keeping the central charges identical, and one still speaks of a defect.

\subsection{The Most General Topological Interface}\label{secGeneralTopInterface}
As in the case with boundary states, we first analyze the solution space to \eqref{topdef} and ask about physically meaningful solutions afterwards. Consider for a start a mapping between $D:\cH_i\otimes\cH_{\bi}\to\cH_j\otimes\cH_{\bj}$. The irreducibility of the modules forces any operator $\defect$ satisfying \eqref{topdef} to be trivial unless $j=i$ and $\bj=\bi$. Moreover, the operator $\defect$ must act like the projector $\proj{i\bi}:=P_i\otimes P_{\bi}$. Applying this to the case at hand, we have $\multZ_{i\bi}^{1,2}$ copies enumerated by $r=1,\dots \multZ_{i\bi}^{(1)}$ and $s=1,\dots \multZ_{i\bi}^{(2)}$. Then the $\defect$ are interwiners between these copies
\begin{subequations}\label{IshibashiIntertwiners}
\begin{align}
 \proj{i\bi}^{sr}&:\, \left(\cH_i\otimes\cH_{\bi}\right)^r
 \,\to\,
 \left(\cH_i\otimes\cH_{\bi}\right)^s\\
 \proj{i\bi}^{sr}
 =&
 \sum_{N,\bar{N}}
 \left(\ket{i,N}\otimes\ket{\bi,\bar{N}}\right)^s
 \left(\bra{i,N}\otimes\bra{\bi,\bar{N}}\right)^r
\end{align}
\end{subequations}
where $\ket{i,N}\otimes\ket{\bi,\bar{N}}$ denotes an orthonormal basis of $\cH_i\otimes\cH_{\bi}$. In a first reading, newcommers are encouraged to consider the case of the same diagonal modular invariant for both CFTs, i.e. $\multZ^{(1)}=\multZ^{(2)}=\delta_{i\bi}$; the notion of modular invariants is discussed in \secref{secStructureStateSpace}. This avoids clutter by the indices $r,s$ and still showcases the underlying structure all the same.

The similarity of the intertwiners with the Ishibashi states \eqref{IshibashiState} on the upper half-plane is not accidental; indeed the $\proj{i\bi}^{sr}$ take on their role in the current setup. Our new operators intertwine the Virasoro algebra to either side of the defect,
\begin{equation}
 L_n^{(1)}\proj{i\bi}^{sr}=\proj{i\bi}^{sr}L_n^{(2)}\,
 \qquad
 \bL_n^{(1)}\proj{i\bi}^{sr}=\proj{i\bi}^{sr}\bL_n^{(2)}
\end{equation}
and satisfy 
\begin{equation}\label{intertwinerProperties}
 \proj{i\bi}^{sr}\proj{j\bj}^{s'r'}
 =
 \delta_{ij}\,\delta_{\bi\bj}\,\delta_{s'r}\,\proj{i\bi}^{sr'}\,,
 \qquad
 \left(\proj{i\bi}^{sr}\right)^\dagger=\proj{i\bi}^{rs}
\end{equation}
Evidently, the Hermitian conjugate maps $\cH_{(2)}\to\cH_{(1)}$.

In total there are thus $\sum_{i\bi}\multZ_{i\bi}^{(1)}\multZ_{i\bi}^{(2)}$ linearly independent solutions to \eqref{topdef}. The most general topological interface operator is then constructed as a superposition
\begin{equation}\label{topDefect}
 \defect_A:\cH_{(1)}\to\cH_{(2)}\,,
 \qquad
 \defect_A
 =
 \sum_{i,\bi}\sum_{r,s}D_A^{(i\bi,rs)}\,\proj{i\bi}^{sr}\,,
 \quad
 A\in\cF
\end{equation}
where the set $\cF$ has $\Lambda=\sum_{i\bi}\multZ_{i\bi}^{(1)}\multZ_{i\bi}^{(2)}$ elements. This is the analog of the set $\cB$ that conformal boundary conditions are taken from, see \eqref{CardyConstraint}. The coefficients $\defect_A^{(i\bi,rs)}$ furnish  complex $\Lambda\times\Lambda$ matrices. The identity operation clearly satisfies \eqref{topdef} and is thus amongst our interface operators. We denote it by $A=0$,
\begin{equation}\label{identityInterface}
 D_0=\id=\sum_{i\bi}\proj{i\bi}^{rr}\,,
 \qquad
 D_0^{(i\bi,rs)}=\delta_{rs}
\end{equation}
Strictly speaking, the identity only appears when the two CFT Hilbert spaces in \eqref{planeSpectra} coincide. We can still view the identity interface operator \eqref{identityInterface} as the one that transports all fields existing in both theories simultaneously across the interface without modification. 

%
 Using the properties in \eqref{intertwinerProperties}, the action of two topological interface operators can be composed,
\begin{align}
 \defect_A^\dagger\circ\defect_B
 =
 \sum_{i\bi}\sum_{r,r',s}\left(\defect_A^{(i\bi,rs)}\right)^*\defect_B^{(i\bi,r's)}\proj{i\bi}^{rr'}
\end{align}
Because Hermitian adjungation inverts orientation, this sequence defines a new interface operator $\defect_{A^\dagger B}:\cH_{(1)}\to\cH_{(1)}$,
\begin{equation}\label{fusedDefect}
\defect_{A^\dagger B}
 =
 \sum_{i\bi}\sum_{r,r'}\defect_{A^\dagger B}^{(i\bi,r'r)}\proj{i\bi}^{rr'}\,,
 \qquad
 \defect_{A^\dagger B}^{(i\bi,r'r)}
 =
 \sum_{s}\left(\defect_A^{(i\bi,rs)}\right)^*\defect_B^{(i\bi,r's)}
\end{equation}
We will see in exercise \ref{exVerlindeLineFusion} that for certain topological interfaces, this map composition is in fact fusion in the sense of \eqref{fusionRules}. We shall be a cavalier about details and always refer to composed interfaces as fused ones.

\subsection{The Petkova-Zuber Constraint}\label{secPetkovaZuberConstraint}
In this subsection we develop the analog of the Cardy constraint for topological defects. It is not quite standard to refer to this as the Petkova-Zuber constraint, but we refer to it as such nevertheless, paying homage to the original paper \cite{Petkova:2000ip} that is followed here.  Just as the Cardy constraint did for conformal boundaries, the Petkova-Zuber constraint constrains the coefficients $\defect_A^{(i\bi,rs)}$ to provide physically consistent interface operators.

From now on we restrict to the case CFT$_1=$ CFT$_2$ and we write $\cH_{(1)}=\cH_{(2)}=\cH$ as well as $\multZ^{(1)}=\multZ^{(2)}=\multZ$ so that the set $\cF$ now constitutes $\Lambda=\sum_{i\bi}(\multZ_{i\bi})^2$ values. All the interfaces in this section are thus defects and the trivial defect \eqref{identityInterface} is now the honest trivial map on $\cH$. Following the terminology that the trivial defect is \emphasize{invisible}, we sometimes refer to non-trivial defects as \emphasize{visible}. Be warned that these somewhat colloquial terminologies are not standard in the literature.

We turn again to the cylinder and imagine inserting a defect operator $\defect_A$ at constant time $\tau$. Afterward we close the cylinder into a torus. The resulting object is computed by
\begin{equation}\label{defectInTorus}
 Z_A=\tr_{\cH}\left[\defect_A\,\tq^{L_0-\cc/24}\tilde\bq^{L_0-\bar{\cc}/24}\right]
\end{equation}
Using the modular nome $\tq$ instead of $q$ is just a matter of choice, which turns out to be convenient when comparing with the Cardy constraint \eqref{CardyConstraint} shortly. Using 
\begin{equation}
 \tr_{\cH}\left[\proj{i\bi}^{rr'}\,\tq^{L_0-\cc/24}\tilde\bq^{\bL_0-\bar{\cc}/24}\right]
 =
 \chi_i(\tq)\,\chi_{\bi}(\tilde\tq)\,\delta_{rr'}
\end{equation}
traces of defects with opposite orientation, or equivalently composed defects \eqref{fusedDefect}, are easily evaluated
\begin{align}
 Z_{A^\dagger B}
 &=
 \tr_{\cH}\left[\defect_A^\dagger\defect_B\,\tq^{L_0-\cc/24}\tilde\bq^{\bL_0-\bar{\cc}/24}\right]\label{ZAB}\\
 &=
 \sum_{i\bi}\sum_{r}\defect_{A^\dagger B}^{(i\bi,rr)}\,\chi_i(\tq)\,\chi_{\bi}(\tilde\bq)\notag\\
 &=
 \sum_{i\bi}\sum_{rs}\left(\defect_A^{(i\bi,rs)}\right)^*\defect_B^{(i\bi,rs)}\,\chi_i(\tq)\,\chi_{\bi}(\tilde\bq)\label{ZABconcrete}
\end{align}
\begin{figure}
 \begin{center}
  \includegraphics[scale=0.3]{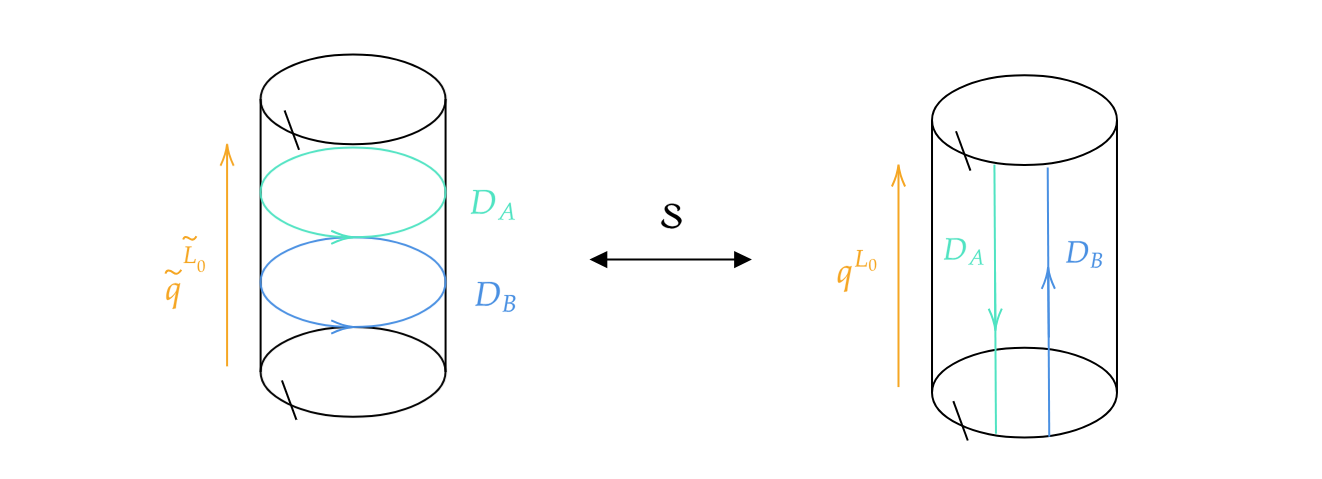}
 \end{center}
\caption{Left: Time evolution runs orthogonal to the defects in the $\tq$ frame, and is generated by a Hamiltonian $\tilde L_0+\tilde \bL_0$ (only one chirality is drawn). This is the analog of $H^{\pl}$ in the boundary case. The green defect has reversed orientation and is implemented mathematically by $\defect_A^\dagger$. Right: A modular $\modS$ transformation results in the $q$ frame, where time evolution runs parallel to the defects and is generated by $L_0+\bL_0$ (only one chirality is drawn). This is the analog of $H^{\hp}$ in the boundary case. }
\label{figPetkovaZuber}
\end{figure}
This situation describes the states of $\cH$, evolved along a temporal coordinate $\tau$ running orthogonal to the defects until they reach $\defect_B$, and past $\defect_{A}^\dagger$, both extended along a slice of constant time $\tau$, as seen in the left panel of \figref{figPetkovaZuber}. It is reassuring that the modular invariant of the CFT is reproduced when $A=B=0$,
\begin{equation}\label{ZtrivialTwist}
 Z_{00}=\sum_{i\bi,r}\chi_i(\tq)\,\chi_{\bi}(\tilde\tq)=\sum_{i\bi}\multZ_{i\bi}\chi_i(\tq)\,\chi_{\bi}(\tilde\tq)=Z(q,\bq)
\end{equation}
where the second equality uses that $r=1,\dots,\multZ_{i\bi}$ and modular invariance \eqref{modInvariance} establishes the last equality.

As in the case of boundaries, due to the Euclidean spacetime, we can study the same setup where time and space are swapped $\tau\leftrightarrow\spa$. This is easily accomplished by a modular $\modS$ transformation. This transformation exchanges the two cycles of the torus, so that now the defects run parallel to the new time coordinate, see the right panel of \figref{figPetkovaZuber}. While we are still looking at the same object as \eqref{ZAB}, the interpretation is now very different. Indeed, in this modular frame, we are looking at traces
\begin{equation}
 Z_{A^\dagger B}(q,\bq)
 =
 \tr_{\cH_{A^\dagger B}}\left[q^{L_0-\cc/24}\tilde\bq^{\bL_0-\bar{\cc}/24}\right]
 =
 \sum_{i\bi}\multZtwist{i\bi}{A^\dagger}{B}\,
 \chi_i(q)\,\chi_{\bi}(\bq)\label{twistSpaceZ}
\end{equation}
over a Hilbert space that has been \emphasize{twisted} by $\defect_A^\dagger$ and $\defect_B$, or equivalently $\defect_{A^\dagger B}$,
\begin{align}
 \cH_{A^\dagger B}
 &=
 \bigoplus_{(i,\bi)\in\confFam\times\confFam}\multZtwist{i\bi}{A^\dagger}{B}\,
 \cH_i\otimes\cH_{\bi}\,,
 \qquad
 \multZtwist{i\bi}{A^\dagger}{B}\in\N_0\label{twistSpace}
 %
\end{align}
Fields corresponding to the states in twisted Hilbert spaces are called \emphasize{defect fields}. Four types are distinguished \cite{Frohlich:2006ch}
\begin{enumerate}
 \item The most general class comes about when inserting at least three defects into the trace \eqref{defectInTorus} (which we haven't done explicitely, but let us mention this straightforward generalization nevertheless here). Such fields are called \emphasize{junction fields} and map one defect into at least two other defects and vice versa.
 \item The second class are \emphasize{defect-changing fields} and are captured by \eqref{twistSpace} when $A^\dagger$ and $B$ are arbitrary defects. These fields map the defect $A$ (no dagger!) into the defect $B$.  
 These fields are junction fields for the case that $A$ and $B$ are arbitrary and the remaining ones are invisible.
 \item The third class are \emphasize{disorder fields} or \emphasize{twist fields}. These are fields which terminate a defect $A$, i.e. they are defect-changing fields mapping $A$ into the trivial defect. These are the fields that have been discussed traditionally in the context of orbifolds or disorder physics.
 \item The last class are the standard \emphasize{bulk fields}, sometimes also called \emphasize{order fields}. These fields are disorder fields for which the arbitrary defect $A$ is now also invisible, i.e. no defect lines are attached to this type of field.
\end{enumerate}

These fields either terminate a defect line or change one defect line into another. The requirement $\multZtwist{i\bi}{A^\dagger}{B}\in\N_0$ secures the meaningfulness of this Hilbert space, making it the analog of \eqref{bdySpaceAlphaBeta} in the boundary case; the partition function \eqref{twistSpaceZ} is the analog of \eqref{bdyZAlphaBeta}. The untwisted partition function \eqref{ZtrivialTwist}, implies $\multZtwist{i\bi}{0}{0}=\multZ_{i\bi}$. 

We have now gathered every ingredient to present the main dish of this section by comparing the trace of \eqref{twistSpaceZ} to \eqref{ZABconcrete} (and pretending that we have used unspecialized characters from the start; see the first comment below \eqref{CardyConstraint})
\begin{tcolorbox}
 The \emphasize{Petkova-Zuber constraint} \new{\cite{Petkova:2000ip}} relates the defect coefficients $\defect^A_{(i\bi,rs)}$ to the twist multiplicities $\multZtwist{i\bi}{A^\dagger}{B}$ non-linearly ,
  \begin{align}\label{PetkovaZuber}
   %
   \sum_{j\bj}\sum_{rs}\left(\defect_A^{(j\bj,rs)}\right)^*\defect_B^{(j\bj,rs)}\modS_{ji}\modS_{\bj\bi}
   &=
   \multZtwist{i\bi}{A^\dagger}{B}\,\in\N_0
  \end{align}
 where $A,B\in\cF$ and $i\bi\in\confFam$.

\end{tcolorbox}

The resemblance of our entire approach to that in deriving the Cardy constraint \eqref{CardyConstraint} is evident. Let us make a number of similar remarks as back then:
\begin{itemize}
 \item Just as the $\niaa$ specify the spectra harbored on a boundary with boundary conditions $\alpha$, the $\multZtwist{i\bi}{A^\dagger}{A}\in\N_0$ specify the fields living on a defect $\defect_A$. Note that these transform under two copies of the chiral algebra, as indicated by the upper indices on $\multZtwist{i\bi}{A^\dagger}{A}$. Similarly, while the $\niab$ determine boundary condition-changing operators, the $\multZtwist{i\bi}{A^\dagger}{B}$ are defect-changing operators mapping between $\defect_A$ and $\defect_B$. 
 \item Any topological defect which cannot be decomposed into a superposition of other interfaces with positive coefficients is called \emphasize{elementary, fundamental} or \emphasize{simple}. Elementary topological defects furnish a basis for all topological defects and form a subset $\cE\subset\cF$. Evidently, a defect cannot be decomposed when at least one $\multZtwist{i\bi}{A^\dagger}{A}=1$, since $\multZtwist{i\bi}{A^\dagger}{B}\geq1$ (if representations $(i\bi)$ occur). Furthermore, an elementary defect always has a unique vacuum sector, $\multZtwist{00}{A^\dagger}{A}=1$. Unless $A=0$, this is not the vacuum of the parent CFT! Because $h=\bh=0$, this is a topological degree of freedom confined to the defect.
 \item The numbers $\multZtwist{i\bi}{A^\dagger}{B}$ may be viewed either as $|\confFam|\times|\confFam|$ matrices $\mathsf{n}_{A^\dagger B}$ with $A,B\in\cF$ or as $|\cF|\times|\cF|$ matrices $\mathsf{n}_{i\bi}$ with $i,\bi\in\confFam$.
 \item The Petkova-Zuber constraint \eqref{PetkovaZuber} may be regarded as spectral decomposition of the matrices $\mathsf{n}^{i\bi}$ into their $\cF$-dimensional eigenvectors $\defect^{(i\bi,rs)}$ with entries $(\defect^{(i\bi,rs)})_A=\defect^{(i\bi,rs)}_A$ and eigenvalues $\modS_{ji}\modS_{\bj\bi}$.
 \item Under the assumption that the coefficients $\defect^{(i\bi,rs)}_A$ form a unitary (orthonormal and complete) change of basis from the intertwiners \eqref{IshibashiIntertwiners} to the defects \eqref{topDefect} It can be shown that the matrices $\mathsf{n}_{i\bi}$ furnish a representation of the tensor product of two copies of the fusion algebra
 \begin{equation}
  \mathsf{n}_{i\bi}\mathsf{n}_{j\bj}=\sum_{k,\bar{k}}\fus_{ij}^k\fus_{\bi\bj}^{\bar{k}}\, \mathsf{n}_{k\bar{k}}
 \end{equation}
This is a NIM-rep. This is explained in the last comment below the Cardy constraint \eqref{CardyConstraint}.
\end{itemize}

\subsubsection*{\new{A Remark on Consistent Defects}}
\new{As with solutions to the Cardy constraint \eqref{CardyConstraint}, we may ask if solutions to the Petkova-Zuber constraint are fully consistent topological defects. To the authors knowledge, no complete list of sewing relations has been written down. By virtue of the folding trick, these sewing relation must however be an extension of the corresponding relations on the upper half-plane; see the remark on page \pageref{remBdySewing}. }

\new{There exist nevertheless powerful techniques imported from \emphasize{topological field theory (TFT)}, which provide a complementary angle for the construction of topological defects for RCFTs. These are guaranteed to provide a consistent set of topological defects. They are not necessarily all that exist however. An upper bound on the number of all consistent topological defects can be found by combining these techniques with the Petkova-Zuber constraint -- which is to be viewed as a necessary condition -- and further relations obtained by deforming defect lines in the presence of bulk fields. These ideas are successfully employed to construct topological defects for the free boson theory in \cite{Fuchs:2007tx}.}

\subsection{Diagonal Theories and Verlinde Lines}\label{secVerlindeLines}
This section presents the analog of the Cardy case studied in \secref{secCardyCase}. \new{Its solution are the \emphasize{Verlinde lines}, which lead us to a powerful \emphasize{graphical calculus}.}

We focus once more on diagonal theories, i.e. CFT$_1=$ CFT$_2$, with $\multZ_{i\bi}=\delta_{i\bi}$, in which case we can drop the multiplicity labels $r,s$, and we relax our notation for projectors by $\proj{ii}\to\proj{i}$. A defect operator then takes the shape
\begin{equation}
 \defect_a=\sum_{i\in\confFam}\defect_a^i\,\proj{i}
\end{equation}
As in the Cardy case in \secref{secCardyCase}, we have identified the set of defect labels $\cF$ with the set of chiral representations $\confFam$, and labeled these by lower case latin letters, $a,b,\dots$. 

\begin{exercise}\label{exPetkovaZuberSimple}
 Retrace the steps taken above to derive the simplified Petkova-Zuber constraint for defects in diagonal theories
 \begin{equation}\label{PetkovaZuberDiag}
  \sum_{j}(\defect_a^j)^*\defect_b^j\modS_{ji}\modS_{j\bi}
  =
  \multZtwist{i\bi}{a^\dagger}{b}\in\N_0
 \end{equation}
\end{exercise}
\ifsol
\solution{exPetkovaZuberSimple}{See the lecture.}
\fi
\begin{tcolorbox}
\hypersetup{linkcolor=\boxlinkcolor}
The so-called \emphasize{Verlinde lines} \new{\cite{Petkova:2000ip}},
 \begin{equation}\label{VerlindeLines}
  \defect_a=\sum_{i\in\confFam}\frac{\modS_{ai}}{\modS_{0i}}\,\proj{i}\,
  \qquad
  \defect_a^\dagger=\defect_{a^+}
 \end{equation} 
 furnish a solution to the Petkova-Zuber constraint \eqref{PetkovaZuberDiag}. Note that the defect coefficients are the eigenvalues of the fusion matrices $\fus_a$ on the eigenvector $\mathbf{v}_i$.
\end{tcolorbox}
Because of $\defect_a^\dagger=\defect_{a^+}$, $\multZtwist{i\bi}{a^\dagger}{b}\to\multZtwist{i\bi}{a^+}{b}$ can be specified, so that the spectra of Verlinde lines are written as follows
\begin{equation}
 Z_{a^+b}(q,\bq)
 =
 \sum_{i\bi}\multZtwist{i\bi}{a^+}{b}\,
 \chi_i(q)\,\chi_{\bi}(\bq)
\end{equation}
\begin{exercise}\label{exTwistMultiplicities}
Show that Verlinde lines solve \eqref{PetkovaZuberDiag} by deriving their matrix $\mathsf{n}^{i\bi}$. You are seeking an expression in terms of the fusion matrices $\fus_i$. Are Verlinde lines elementary?
\end{exercise}
\ifsol
\solution{exTwistMultiplicities}{
\begin{align}
      \multZtwist{i\bi}{a^+}{b}
      =
      \sum_j\frac{\modS_{aj}^*}{\modS_{0j}^*}\frac{\modS_{bj}}{\modS_{0j}}\modS_{ji}\modS_{j\bi}
      &=
      \sum_{j,k}
      \frac{\modS_{bj}\modS_{ji}}{\modS_{0j}}\delta_{jk}\frac{\modS_{k\bi}\modS_{ak}^*}{\modS_{0k}}\notag\\
      &=
      \sum_{j,k,l}
      \frac{\modS_{bj}\modS_{ji}}{\modS_{0j}}\modS_{jl}^*\modS_{lk}\frac{\modS_{k\bi}\modS_{ak}^*}{\modS_{0k}}\notag\\
      &=
      \sum_l\fus_{bi}^l\fus_{l\bi}^a\notag\\
      &=
      \sum_l\fus_{ib}^l\fus_{\bi l}^a\notag\\
      &=
      \sum_l(\fus_i)_{bl}(\fus_{\bi})_{la}
      =
      (\fus_i\fus_{\bi})_{ba}
     \end{align}
To see elementarity observe $\multZtwist{00}{a^+}{a}=(\fus_0\fus_0)_{aa}=(\mathbb{1})_{aa}=1$.}
\fi
\begin{figure}
 \begin{center}
  \includegraphics[scale=.3]{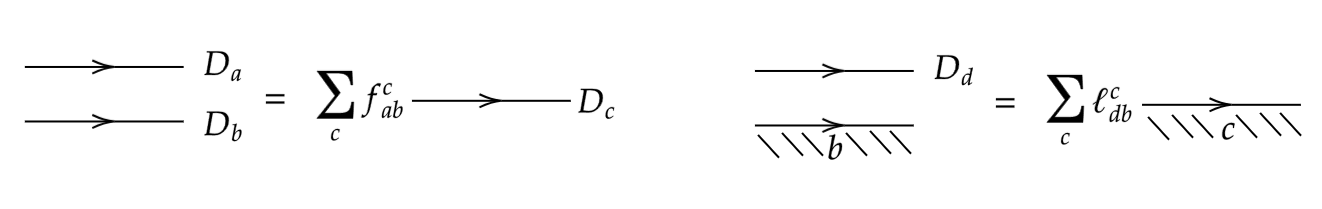}
 \end{center}
\caption{Left: Defects naturally satisfy an algebra. Right: The action of defects on boundaries closes in boundaries, meaning that a new boundary is formed after application of a defect.}
\label{figDefectAlgebra}
\end{figure}
\begin{exercise}\label{exVerlindeLineFusion}
Verlinde lines earn their name on basis of their algebra. Compute the structure constants $f_{ab}^c$ of the algebra of Verlinde lines 
 \begin{equation}
  \defect_a\defect_b=\sum_cf_{ab}^c\defect_c
 \end{equation}
 Topological defects may be applied to boundary states. Compute the action of a Verlinde line on a Cardy state \eqref{CardyState}, 
 \begin{equation}
  \defect_a\bket{b}
  =
  \sum_c\ell_{ab}^c\bket{c}
 \end{equation}
 Both of these actions are conveniently represented pictorially in \figref{figDefectAlgebra}. Are the new defects and boundaries elementary?
\end{exercise}
\ifsol
\solution{exVerlindeLineFusion}{
\begin{align}
 \defect_a\defect_b
 &=
 \sum_{ij}\frac{\modS_{ai}}{\modS_{0i}}\frac{\modS_{bj}}{\modS_{0j}}\delta_{ij}\proj{j}\notag\\
 &=
 \sum_{ijc}\frac{\modS_{ai}\modS_{bi}}{\modS_{0i}}\frac{\modS_{ic}^*\modS_{cj}}{\modS_{0j}}\proj{j}\notag\\
 &=
 \sum_{cj}\fus_{ab}^c\frac{\modS_{cj}}{\modS_{0j}}\proj{j}
 \quad
 =
 \sum_c\fus_{ab}^c\defect_c\label{DefectAlgebra}\\
 \defect_d\bket{b}
 &=
 \sum_{ij}\frac{\modS_{di}}{\modS_{0i}}\frac{\modS_{bj}}{\sqrt{\modS_{0j}}}\delta_{ij}\iket{j}\notag\\
 &=
 \sum_{ijc}\frac{\modS_{di}\modS_{bi}}{\modS_{0i}}\frac{\modS_{ic}^*\modS_{cj}}{\sqrt{\modS_{0j}}}\iket{j}\notag\\
 &=
 \sum_{cj}\fus_{db}^c\frac{\modS_{cj}}{\sqrt{\modS_{0j}}}\iket{j}
 \quad
 =
 \sum_c\fus_{db}^c\bket{c}
\end{align}
The new defects and boundaries are only elementary if the fusion coefficient is equivalent to a permutation of the label, i.e. if the result of the fusion is only one conformal family. 
}
\fi
\begin{figure}
 \begin{center}
  \includegraphics[scale=.3]{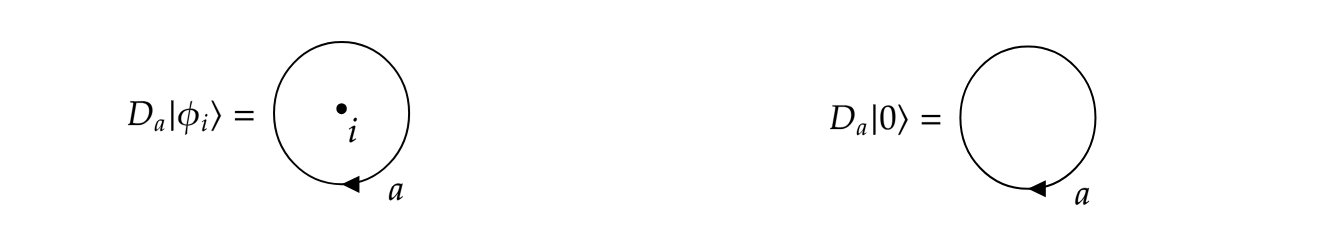}
 \end{center}
\caption{Left: Defect action on a generic field/state. Right: The vacuum is invisible so that the defect action on it is just an empty bubble. Due to tranlation invariance of the vacuum, this empty bubble can be inserted anywhere in the 2d space.}
\label{figDefectOnPrimary}
\end{figure}
\begin{exercise}\label{exDefectOnPrimary}
 Evaluate the action of a defect on a primary state, $\defect_a\ket{\phi_{i}}$ (it has $\bi=i$). What happens when evaluating the defect on a descendant of this primary state? This action can also be represented pictorially, see \figref{figDefectOnPrimary}. Because the vacuum state is \quotes{invisible}, the action of $\defect_a$ thereon is simply a bubble, which is called the \emphasize{defect's g-factor}. What physical quantity is it in case of a Verlinde line?
\end{exercise}
\ifsol
\solution{exDefectOnPrimary}{
\begin{align}
 \defect_a\ket{\phi_i}
 =
 \defect_a^i\ket{\phi_i}
 =
 \frac{\modS_{ai}}{\modS_{0i}}\ket{\phi_i}\,,
 \qquad
 \defect_a\ket{0}
 =
 \frac{\modS_{a0}}{\modS_{00}}\ket{\phi_i}
 =
 \qdim_a\ket{0}
\end{align}
Where the quantum dimension \eqref{qdimModS} was re-encountered; see \figref{figDefectOnPrimarySolution}.
\begin{figure}
 \begin{center}
  \includegraphics[scale=.3]{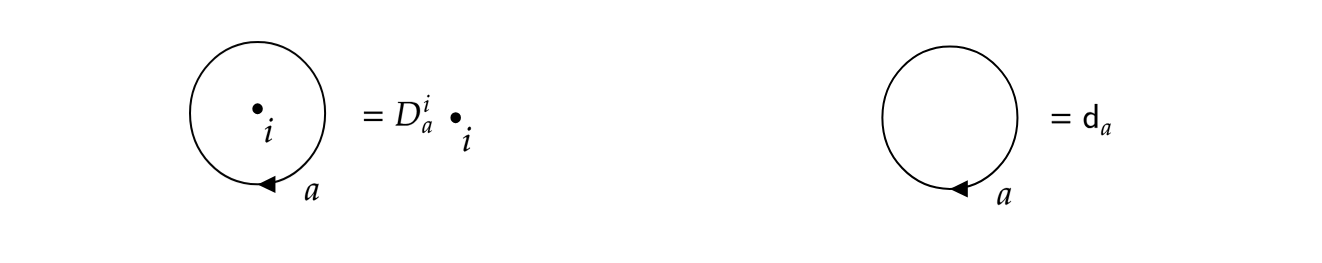}
 \end{center}
\caption{Left: Defect action on a generic field/state. Right: The vacuum is invisible so that the defect action on it is just an empty bubble. Due to tranlation invariance of the vacuum, this empty bubble can be inserted anywhere in the 2d space.}
\label{figDefectOnPrimarySolution}
\end{figure}
}
\fi

\begin{exercise}\label{exTwistFieldIdentity}
 Show that 
 \begin{equation}\label{ZabFromZ0c}
  Z_{a^+ b}(q,\bq)=\sum_c\fus_{a^+b}^cZ_{0c}(q,\bq)
 \end{equation}
It is useful to first solve exercise \ref{exTwistMultiplicities}. It is furthermore recommended to solve exercise \ref{exVerlindeLineFusion} first.
\end{exercise}
\ifsol
\solution{exTwistFieldIdentity}{
 The simplest way is to utilize the defect algebra \eqref{DefectAlgebra} and stick this into the partition funtion of the CFT according to \eqref{defectInTorus}, i.e.
 \begin{align}
 Z_{a^+b}
 &=
 \tr\left[\defect_{a^+}\defect_b\,q^{L_0-\cc/24}q^{\bL_0-\bar{\cc}/24}\right]\notag\\
 &=
 \sum_c\fus_{a^+b}^c\tr\left[\defect_c\,q^{L_0-\cc/24}q^{\bL_0-\bar{\cc}/24}\right]\notag\\
 &=
 \sum_c\fus_{a^+b}^cZ_{0c}
 =
 \sum_c\fus_{a^+b}^cZ_{c^+0}
 \end{align}
The claim can also be demonstrated without having solved exercise \ref{exVerlindeLineFusion} first. This is a good exercise to get acquainted with the fusion matrices $\fus_i$. Begin with
\begin{subequations}
\begin{align}
 Z_{0c}
 &=
 \sum_{i\bi}\left(\fus_i\fus_{\bi}\right)_{c0}\chi_i(q)\chi_{\bi}(\bq)\label{Z0cA}\\
 &=
 \sum_{i\bi k}\fus_{ic}^k\fus_{\bi k}^0\chi_i(q)\chi_{\bi}(\bq)\notag\\
 &=
 \sum_{i\bi}\left(\fus_c\right)_{i\bi^+}\chi_i(q)\chi_{\bi}(\bq)\label{Z0cB}
\end{align}
\end{subequations}
Moreover, $(\fus_0)_{ka}=\fus_{0k}^a=\fus_{k^+a}^0=\fus_{a^+k}^0=(\fus_{a^+)_{k0}}$. This allows us the following manipulations
\begin{align}
 (\fus_i\fus_{\bi})_{ba}
 &=
 (\fus_i\fus_{\bi}\fus_0)_{ba}\notag\\
 &=
 \sum_k(\fus_i\fus_{\bi})_{bk}(\fus_0)_{ka}\notag\\
 &=
 \sum_k(\fus_i\fus_{\bi})_{bk}(\fus_{a^+)_{k0}}\notag\\
 &=
 (\fus_i\fus_{\bi}\fus_{a^+})_{b0}\notag\\
 &=
 (\fus_{a^+}\fus_i\fus_{\bi})_{b0}\notag\\
 &=
 \sum_c\fus_{a^+b}^c(\fus_i\fus_{\bi})_{c0}
\end{align}
The fifth line uses commutativity of the fusion matrices. Finally,
\begin{align}
 Z_{a^+b}
 =
 \sum_{i\bi}(\fus_i\fus_{\bi})_{ba}\chi_i\chi_{\bi}
 =
 \sum_{i\bi c}\fus_{a^+b}^c(\fus_i\fus_{\bi})_{c0}\,\chi_i\chi_{\bi}
 =
 \sum_{c}\fus_{a^+b}^c\,Z_{0c}
\end{align}
In going to the last line we recognized \eqref{Z0cA}.}
\fi

\begin{exercise}\label{exIsingTwistFields}
 Evaluate all twisted partition functions \eqref{twistSpaceZ} for the Ising model. The Verlinde lines are labeled by the three conformal families $\cF=\confFam={\id,\epsilon,\sigma}$. Their non-trivial fusion rules are 
 \begin{equation}\label{IsingFusionTwistEx}
  [\epsilon]\fuse[\epsilon]=[\id],
  \qquad 
  [\epsilon]\fuse[\sigma]=[\sigma],
  \qquad
  [\sigma]\fuse[\sigma]=[\id]+[\epsilon]
 \end{equation}
All fields are self-conjugate. It is useful to first solve exercise \ref{exTwistMultiplicities}. 
\end{exercise}
\ifsol
\solution{exIsingTwistFields}{
The Ising model's fusion matrices are 
\begin{equation}
 \fus_0
 =
 \begin{pmatrix}
 \,1 & 0 & 0\, \\
 0 & 1 & 0 \\
 0 & 0 & 1 \\
 \end{pmatrix}
\qquad
\fus_\varepsilon
 =
 \begin{pmatrix}
 \,0 & 1 & 0\, \\
 1 & 0 & 0 \\
 0 & 0 & 1 \\
 \end{pmatrix}
\qquad
\fus_\sigma
 =
 \begin{pmatrix}
 \,0 & 0 & 1\, \\
 0 & 0 & 1 \\
 1 & 1 & 0 \\
 \end{pmatrix}
\end{equation}
Working systematically with \eqref{ZabFromZ0c} we first construct the non-trivial single-twist partition functions. To that end employ \eqref{Z0cB} and that families are self-conjugate in the Ising model, i.e. $\bi^+=\bi$,
\begin{subequations}\label{IsingTwistFields}
\begin{align}
 Z_{0\varepsilon}(q,\bq)
 &=
 \chi_{1/2}(q)\bar\chi_0(\bq)
 +
 \chi_{0}(q)\bar\chi_{1/2}(\bq)
 +
 \chi_{1/16}(q)\bar\chi_{1/16}(\bq)\label{epsilonTwist}\\
 Z_{0\sigma}(q,\bq)
 &=
 \left(\chi_0(q)+\chi_{1/2}(q)\right)\bar\chi_{1/16}(\bq)
 +
 \chi_{1/16}(q)\left(\bar\chi_{0}(\bq)+\bar\chi_{1/2}(\bq)\right)
\end{align}
\end{subequations}
The twist fields in $Z_{0\varepsilon}$ terminate a $\defect_\varepsilon$ defect,  while the twist fields in $Z_{0\sigma}$ terminate a $\defect_\sigma$ defect. We have chosen to label the characters by conformal dimensions, rather than the conformal family as before. This is because these fields are fundamentally different from the fields $\id,\,\varepsilon,\,\sigma$. Indeed, the fields in \eqref{epsilonTwist} are from left to right: the holomorphic fermion $\psi(z)$, the anti-holomorphic fermion $\bar{\psi}(\bz)$ and the disorder field $\mu(z,\bz)$. The latter is Kramers-Wannier dual to $\sigma(z,\bz)$, hence the
coinciding conformal dimensions.
The partition functions \eqref{IsingTwistFields} are not modular invariant. The defect changing operators are now found by fusion by virtue of \eqref{ZabFromZ0c}. Looking back \eqref{IsingFusionTwistEx} these are
\begin{align}
 Z_{\varepsilon\varepsilon}(q,\bq)
 &=
 Z_{00}(q,\bq)\\
 Z_{\varepsilon\sigma}(q,\bq)
 &=
 Z_{0\sigma}(q,\bq)\\
 Z_{\sigma\sigma}(q,\bq)
 &=
 Z_{00}(q,\bq)+Z_{0\varepsilon}(q,\bq)
\end{align}
Recall that $Z_{00}(q,\bq)=Z(q,\bq)$ is the bulk's modular invariant partition function. While only the last one of these spectra is new, the interpretation of these twist fields is different. They are defect-changing operators. For instance the fields in $Z_{\varepsilon\sigma}$ turn a defect $\defect_\varepsilon$ into a defect $\defect_\sigma$ and we learn that they have the same weights as the fields in $Z_{0\sigma}$.
}
\fi

\bibliographystyle{JHEP}
\bibliography{Bibliography.bib}

\end{document}